\documentclass[11pt,oneside]{article}

\usepackage[natbibapa]{apacite}
\usepackage{graphicx}
\graphicspath{ {./images/} }
\usepackage{subfig}
\usepackage[subfigure]{tocloft}
\usepackage{caption}

\usepackage{multirow}
\usepackage[table,xcdraw]{xcolor}

\usepackage{soul}
\usepackage{amsmath, amssymb}
\usepackage{calrsfs}
\DeclareMathAlphabet{\pazocal}{OMS}{zplm}{m}{n}
\usepackage[shortlabels]{enumitem}
\usepackage{stmaryrd}			
\usepackage{bm}    				
\usepackage{amsthm}
\usepackage[nottoc]{tocbibind}
\usepackage{tocloft}
\newcommand{\R}{\mathbb{R}}
\newcommand{\N}{\mathbb{N}}
\newcommand{\U}{\mathbb{U}}

\usepackage{bbm}					

\numberwithin{equation}{section}

\newtheorem{thm}{Theorem}[section]

\newtheorem{prop}{Proposition}[section]
\newtheorem{cor}{Corollary}[section]

\theoremstyle{definition}
\newtheorem{defn}{Definition}[section]

\theoremstyle{remark}
\newtheorem{example}{Example}[section]

\theoremstyle{remark}
\newtheorem{rem}{Remark}[section]

\theoremstyle{theorem}

\title{Quantifying Dimensional Change \\ in Stochastic Portfolio Theory}

\vspace{2mm}

\author{  
	\textsc{Erhan Bayraktar} 
	\thanks{
		Department of Mathematics, University of Michigan, Ann Arbor, MI, USA (E-mail: {\it erhan@umich.edu}). 
	}  
	\and
	\textsc{Donghan Kim} 
	\thanks{ 
		Department of Mathematics, University of Michigan, Ann Arbor, MI, USA (E-mail: {\it donghank@umich.edu}).
	}
	\and
	\textsc{Abhishek Tilva}
	\thanks{ 
		Department of Statistics, Columbia University, New York, NY, USA (E-mail: {\it akt2143@columbia.edu}).
	}
}

\begin{document}
	
	\maketitle
	
	\begin{abstract}
		\noindent
		In this paper, we develop the theory of functional generation of portfolios in an equity market with changing dimension. By introducing dimensional jumps in the market, as well as jumps in stock capitalization between the dimensional jumps, we construct different types of self-financing stock portfolios (additive, multiplicative, and rank-based) in a very general setting. Our study explains how a dimensional change caused by a listing or delisting event of a stock, and unexpected shocks in the market, affect portfolio return. We also provide empirical analyses of some classical portfolios, quantifying the impact of dimensional change in portfolio performance relative to the market.
	\end{abstract}
	
	\smallskip
	
	
	
	\smallskip
	
	\setcounter{tocdepth}{2}
	\tableofcontents
	
	\input amssym.def
	\input amssym
	
	\section{Introduction}
	\label{sec: introduction}
	In this paper, we study an equity market model of stochastic dimension. This is a continuation of our previous work \citep{BKT:arbitrage}, where a complete version of the fundamental theorem of asset pricing is developed in a stock market with a changing number of companies, and access to a money market. On the other hand, this paper focuses solely on the stock market of fluctuating dimension, without access to the money market, and develops the theory of functional generation of portfolios which only invest in stocks.
	
	The theory of functional generation of portfolios, introduced by \cite{F_generating} over $20$ years ago, is the central part of the stochastic portfolio theory \citep{Fe}~(see also \cite{FK_survey} for a brief overview). It explains how to construct a class of self-financing portfolios from a function depending on the companies' relative capitalization. It also derives an explicit decomposition of their log relative return with respect to the entire stock market, as the sum of the log of generating function and a drift process. Such decomposition enables us to find conditions on the portfolio-generating function, such that the drift process is increasing in time and the generated portfolio outperforms the market over the long run. More recently, \cite{Karatzas:Ruf:2017} introduced a new construction of portfolios which they call `additive' generation, whereas Fernholz's original construction is called `multiplicative'. This new method provides a simpler formulation of conditions for portfolios to outperform the market over appropriate time horizons.
	
	All of these previous works assume a fixed number of stocks in the market. Our primary purpose is to remove this assumption on the market dimension when constructing self-financing portfolios. Concretely, we shall use a piecewise semimartingale of stochastic dimension, introduced by \cite{Strong2}, to model the capitalization dynamics of stocks; the moments of dimensional changes are given by a sequence $(\tau_k)_{k=0}^{\infty}$ of stopping times, and the stock capitalization process is assumed to be a semimartingale of some fixed dimension, during each `epoch' of the market between those dimensional jumps. Using this concept, our equity market model incorporates all types of listing/delisting events such as IPO, splits, mergers, bankruptcies, etc.
	
	The main difficulty is then to make the (functionally-generated) portfolios self-financing whenever the market undergoes a dimensional change. In order to handle this, we shall normalize our portfolio-generating function at the starting moment $\tau_k$ of each epoch by an $\mathcal{F}_{\tau_{k}}$-measurable random variable, and construct portfolios in a recursive manner for each epoch of the market. Now that we have two different methods of generating portfolios, both additive and multiplicative normalizations will be applied accordingly. The introduction of such normalizations yields an extra correction term in the decomposition of the relative return of the generated portfolios with respect to the market, and this correction term quantifies precisely the impact of dimensional change in the performance of the portfolios relative to the market.
	
	The correction term arises at each dimensional change, when a new stock enters or an old stock exits the market, due to two factors; a jump in total market capitalization and a jump in the portfolio-generating function. The former is an uncontrollable factor, whereas the latter can be controlled by choosing an appropriate generating function. For multiplicatively generated portfolios, these two factors are nicely separated so that we can quantify them individually in portfolio returns with respect to the market. However, in the case of additively generated portfolios, there is no such simple uncoupling between these two correction factors.
	
	Moreover, the aforementioned portfolio returns measure the wealth of generated portfolios with respect to the total market capitalization~(total market index fund); at the time of dimensional change, the latter quantity jumps, whereas the wealth of the portfolio remains the same, as it is designed to be self-financing. Thus, we introduce a new notion of a `self-financing' market portfolio, which follows the total market index fund as a buy-and-hold strategy between dimensional jumps, but remains self-financed at times of dimensional change, as an alternative baseline for comparing the wealth of functionally-generated portfolios. Especially for multiplicatively generated portfolios, the relative wealth with respect to the new baseline only contains the second factor of the correction term. Since this second factor will be shown to offset exactly the jump in the portfolio-generating function at moments of dimensional change, we recover Fernholz's original decomposition of log relative wealth, as the sum of the log of generating function and a drift term accumulated between two consecutive dimensional changes, with respect to the self-financing market portfolio.
	
	The second contribution of this paper is that our model relaxes the continuity assumption on the capitalization process between dimensional jumps, as it is modeled by an RCLL semimartingale of some fixed dimension between dimensional jumps. Allowing such left-discontinuities can help explain not only how certain unexpected shocks from the market between dimensional jumps affect the portfolios, but also how a delisting event of a stock influences the portfolio returns~(see Remark~\ref{rem : delisting event} for more details).
	
	Therefore, this equity market model using piecewise RCLL semimartingales~(see Definition~\ref{Def : piecewise semimartingale} below) is the most general model considered so far in the realm of stochastic portfolio theory. The method of constructing functionally generated portfolios developed in this paper generalizes the previous theories in both continuous-time and discrete-time market models, and is also easily applicable to rank-based portfolio generation in the stock market of stochastic dimension.
	
	Finally, we provide empirical analyses for classical examples of (multiplicatively generated) portfolios in stochastic portfolio theory. Using a stock dataset over $40$ years on two U.S. stock exchanges (NYSE and AMEX), we present the explicit decompositions~(see Section~\ref{subsec : empirical results}) of the relative wealth processes of functionally-generated portfolios. When we use the original total market index as baseline, the first uncontrollable correction factor significantly influences the relative performance of the functionally-generated portfolios; when the self-financing market portfolio is used as baseline, the drift process~(so-called excess growth) mainly contributes to the outperformance of generated portfolios over long-run, very much in accordance with Fernholz's original theory. 
	
	\smallskip
	
	\textit{Preview: } This paper is organized as follows. Section~\ref{sec: preliminaries} reviews the concept of piecewise semimartingale and describes our equity market model of stochastic dimension. Section~\ref{sec : FGP} explores how to construct functionally-generated portfolios under the market model. In Section~\ref{sec : empirical}, we provide a discrete-time version of the results developed in previous sections and present empirical analyses of classical portfolios using real stock data. Section~\ref{sec : conclusion} contains some concluding remarks, and Appendix~\ref{sec : rank FGP} illustrates how to construct rank-based portfolios.

	\bigskip
	
	\section{Equity market of stochastic dimension}
	\label{sec: preliminaries}
	
	We provide in this section some preliminary concepts and definitions and describe an equity market model of stochastic dimension, which we shall use throughout the paper.
	
	\subsection{Piecewise semimartingales}	\label{subsec : piecewise semimartingales}
	
	We first recall the notion of piecewise semimartingales of stochastic dimension, which was originally introduced by \cite{Strong2}. The same definitions and notations were provided in Section~2 of our earlier paper \citep{BKT:arbitrage}, but we include them in this subsection for the completeness of the paper.
	
	Let us denote a state space $\U:= \cup_{n=1}^{\infty} \R^n$, equipped with the topology generated by the union of the standard topologies of $\R^n$. Besides the additive identity element $0^{(n)}$, the $n$-dimensional vector of zeros, in $\R^n$ for each $n \in \N$, we define an additive identity element $\odot$, a topologically isolated point in $\hat{\U} := \U \cup \{\odot\}$ satisfying $\odot + x = x + \odot = x$ and $\odot x = x \odot = \odot$ for each $x \in \hat{\U}$. We define the modified indicator
	\begin{equation}	\label{def : modified indicator}
		\hat{\mathbbm{1}}_A(t, \omega) :=
		\begin{cases}
			1 \in \R \quad &\text{for } (t, \omega) \in A \subset [0, \infty) \times \Omega,
			\\
			\odot &\text{otherwise},
		\end{cases}
	\end{equation}
	which will be used for dissecting $\U$-valued stochastic processes. We shall sometimes add the zero vector $0^{(n)}$ with an appropriate dimension $n \in \mathbb{N}$ to some expressions involving the modified indicator $\hat{\mathbbm{1}}$, to make sure that the resulting expression has the correct dimension in $\U$. We denote $\mathbbm{1}_A$ the usual indicator function for a set $A$. 
	
	We use the notations $\R_+ := [0, \infty)$, $[n] := \{1, \cdots, n\}$ for every $n \in \N$, and $B^{\top}$ the transpose of a matrix $B$.
	
	All relationships among random variables are understood to hold almost surely. On a filtered probability space $(\Omega, \mathcal{F}, (\mathcal{F}_t)_{t \ge 0}, \mathbb{P})$ satisfying the usual conditions, let $X$ be a $\U$-valued progressive process having paths with left and right limits at all times, and denote $N := \dim (X)$ the dimension process of $X$. The following definition characterizes time instants of dimensional jumps for a given $\U$-valued process $X$, as a sequence of stopping times.
	
	\begin{defn} [Reset sequence]   \label{Def : reset sequence}
		A sequence of stopping times $(\tau_k)_{k=0}^{\infty}$ is called a \textit{reset sequence} for a progressive $\U$-valued process $X$, if the following hold for $\mathbb{P}$-a.e. $\omega$:
		\begin{enumerate} [(1)]
			\item $\tau_0(\omega) = 0$, $\tau_{k-1}(\omega) \le \tau_k(\omega)$ for all $k \in \N$, and $\lim_{k \rightarrow \infty} \tau_k(\omega) = \infty$;
			\item $N(t, \omega) = N(\tau_{k-1}+, \omega)$ for every $t \in (\tau_{k-1}(\omega), \tau_k(\omega)]$ and $k \in \N$;
			\item $t \mapsto X(t, \omega)$ is right-continuous on $(\tau_{k-1}(\omega), \tau_k(
			\omega))$ for every $k \in \N$.
		\end{enumerate}
	\end{defn}
	
	When $X$ has a reset sequence $(\tau_k)_{k=0}^{\infty}$, we shall always consider the minimal one $(\hat{\tau}_k)_{k=0}^{\infty}$ in the sense of the fewest resets by a given time:
	\begin{equation*}
		\hat{\tau}_0 := 0, \qquad \hat{\tau}_k := \inf\{t > \hat{\tau}_{k-1} \, \vert \, X(t+) \neq X(t) \}, \quad k \in \N,
	\end{equation*}
	and assume that the initial dimension is deterministic, i.e., $\dim(X(0)) = N_0 \in \N$.
	
	In what follows, we fix such $\U$-valued process $X$ with the reset sequence $(\tau_k)_{k = 0}^{\infty}$, and define the \textit{dissections} of $\Omega$ and $X$
	\begin{alignat}{3}
		&\Omega^{k, n} := \{ \tau_{k-1} < \infty, ~ N(\tau_{k-1}+) = n \} \subset \Omega, \qquad &&\forall (k, n) \in \N^2,	\label{def : Omega dissect}
		\\
		&X^{k, n} := \big(X^{\tau_k} - X(\tau_{k-1}+)\big) \hat{\mathbbm{1}}_{\rrbracket \tau_{k-1}, \infty \llbracket \cap (\R_+ \times \Omega^{k, n})} + 0^{(n)}, \qquad &&\forall (k, n) \in \N^2.		\label{def : X dissect}
	\end{alignat}
	
	\begin{defn} [Piecewise semimartingale]	\label{Def : piecewise semimartingale}
		A \textit{piecewise semimartingale} $X$ is a $\U$-valued progressive process having paths with left and right limits at all times, and possessing a reset sequence $(\tau_k)_{k=0}^{\infty}$ such that $X^{k, n}$ is an $\R^n$-valued semimartingale for every $(k, n) \in \N^2$.
		
		A piecewise semimartingale $X$ is called \textit{piecewise continuous~(RCLL) semimartingale}, if each dissection $X^{k, n}$ is an $n$-dimensional continuous~(RCLL, i.e., right continuous with left limits, respectively) semimartingale for every $(k, n) \in \N^2$.
	\end{defn}
	We note that the $(k, n)$-dissection $X^{k, n}$ of $X$, defined in \eqref{def : X dissect}, will be used when $X$ plays the role of an integrator. For integrands, a different version of dissection is necessary.
	
	\begin{defn} [Stochastic integral]
		For a piecewise semimartingale $X$ and its reset sequence $(\tau_k)_{k=0}^{\infty}$, let $H$ be a $\U$-valued predictable process satisfying $\dim (H) = \dim (X)$. We dissect $H$ in the following manner
		\begin{equation}	\label{def : H dissect}
			H^{(k, n)} := H\hat{\mathbbm{1}}_{\rrbracket \tau_{k-1}, \tau_k \rrbracket \cap (\R_+ \times \Omega^{k, n})} + 0^{(n)}, \qquad \forall (k, n) \in \N^2,
		\end{equation}
		and define
		\begin{align}
			&\mathcal{L}(X) := \{ H \text{ predictable} \, \vert \, \dim (H) = \dim (X) \text{ and } H^{(k, n)} \text{ is } X^{k, n} \text{-integrable, } \forall (k, n) \in \N^2 \},	\label{def : space of integrand}
			\\
			&\mathcal{L}_0(X) := \{ H \in \mathcal{L}(X) \, \vert \, H_0 = 0^{(N_0)} \}. 	\nonumber
		\end{align}
		For $H \in \mathcal{L}(X)$, the stochastic integral $H \cdot X$ is defined as 
		\begin{equation}	\label{def : stochastic integral}
			H \cdot X := H_0^{\top}X_0 + \sum_{k=1}^{\infty} \sum_{n=1}^{\infty} (H^{(k, n)} \cdot X^{k, n})
			= H_0^{\top}X_0 + \sum_{k=1}^{\infty} \sum_{n=1}^{\infty} \int \sum_{i=1}^n H^{(k, n)}_i dX^{k, n}_i.
		\end{equation}
	\end{defn}
	
	Note that each dissection $H^{(k, n)}$ of \eqref{def : H dissect} is predictable, since the process $H$ and the $(k, n$)-dissection set $\rrbracket \tau_{k-1}, \tau_k \rrbracket \cap (\R_+ \times \Omega^{k, n})$ are predictable. 
	
	The stochastic integral $H \cdot X$ in \eqref{def : stochastic integral} is an $\R$-valued semimartingale, and it generalizes the usual $\R^n$-valued semimartingale stochastic integration since any sequence $\tau_k \uparrow \infty$ of stopping times is a reset sequence. We also note that $\mathcal{L}(X)$ is a vector space, i.e., $H \cdot X + G \cdot X = (H+G) \cdot X$ holds for $H, G \in \mathcal{L}(X)$.
	
	\medskip
	
	\subsection{Capitalization and market weight processes}	\label{subsec : basic processes}
	
	With the concept of piecewise semimartingale, we now describe a model of an equity market having a finite, but unbounded, stochastic number of investable assets.
	
	We first define a $\U$-valued capitalization process $S$ allowing left-discontinuities between two consecutive dimensional jumps. In order to handle discontinuities, we shall denote $\Delta X(t) := X(t) - X(t-)$ and $X_-(t) := X(t-)$ the jump and the left-continuous process of a RCLL semimartingale $X$ at time $t$, respectively, for any $t \ge 0$.
	
	\begin{defn} [Capitalization process]	\label{Def : price process RCLL}
		A $\U$-valued piecewise RCLL semimartingale $S$ with reset sequence $(\tau_k)_{k=0}^{\infty}$ is called a \textit{capitalization process}, if every $n$ component of $S$ is nonnegative, but at least one component is strictly positive on each dissection set $\rrbracket \tau_{k-1}, \tau_k \rrbracket \cap (\R_+ \times \Omega^{k, n})$ for every $(k, n) \in \N^2$. 
	\end{defn}
	
	The dimension process $N = \dim(S)$ of $S$ represents the number of companies present in the market, and the $n$ components of $S$ on the $(k, n)$-dissection set, i.e., $\rrbracket \tau_{k-1}, \tau_k \rrbracket \cap (\R_+ \times \Omega^{k, n})$, represent the capitalizations of the $n$ extant companies, for every $(k, n) \in \N^2$. For simplicity of the model, we assume without loss of generality that every stock has a single outstanding share, thus the capitalization of a stock is equal to its price. By definition, this model allows individual stock prices to hit zero, but the market's total capitalization should always remain strictly positive.
	
	\begin{example}	\label{ex : market models}
		This model, which adopts a piecewise RCLL semimartingale of variable dimension as a capitalization process, describes a more realistic equity market in the most flexible way. It includes several other models depicting a market with a changing number of assets. We present here a few such examples.
		\begin{enumerate} [(i)]
			\item The entrance of a new company and the exit of an existing company are modeled by a birth-death process; for each $n \in \N$, whenever the dimension of the market is $n$, a new company enters the market according to the exponential arrival rate of $\lambda_n$, and one of the companies exits the market with the exponential departure rate of $\mu_n$. The dimension process $N = \dim(S)$ is then the birth-death process with birth and death rates $(\lambda_n)_{n \in \N}$ and $(\mu_n)_{n \in \N}$, respectively. The reset sequence $(\tau_k)_{k=0}^{\infty}$ describes the time of either birth or death of a company, and the size of each dimension change is always one, i.e., $\vert N(\tau_{k-1}+) - N(\tau_{k-1}) \vert = 1$ for each $k \in \N$.
			\item The diverse market model introduced in \cite{Karatzas:Sarantsev} studies a fluctuating number of companies by describing a certain form of splits and mergers. When the market weight~(relative capitalization with respect to total market capitalization) of the largest company exceeds a fixed threshold value between $0$ and $1$, the company is split into two companies of random size, modeling a regulatory breakup. On the other hand, any two of the existent companies excluding the largest one, can merge at random times, whenever the exponential clock rings. The reset sequence $(\tau_k)_{k=0}^{\infty}$ then depicts the time of either split of the largest company or merger between two small companies, and the size of each dimension change is again equal to one at each stopping time $\tau_k$.
			\item We can even consider a combination of (i) and (ii) above, i.e., a stock market allowing all kinds of events including the entrance, exit, splits, and mergers of companies, whenever each stopping time $\tau_k$ describes such dimensional change.
		\end{enumerate}
	\end{example}
	
	\smallskip
	
	For a given capitalization process $S$ in Definition~\ref{Def : price process RCLL}, we define another $\U$-valued process whose components represent the relative capitalizations of individual stocks.
	
	\begin{defn} [Market weight]	\label{Def : market weights}
		We call a $\U$-valued process $\mu$ the \textit{market weight process}, if it is defined via dissection
		\begin{equation*}
			\mu(\cdot) := \mu^{(0)}(\cdot) \hat{\mathbbm{1}}_{\{0\}\times \Omega}  + \sum_{k=1}^{\infty} \sum_{n=1}^{\infty} \mu^{(k, n)}(\cdot) \hat{\mathbbm{1}}_{\rrbracket \tau_{k-1}, \tau_k \rrbracket \cap (\R_+ \times \Omega^{k, n})},
		\end{equation*}
		where the components of $\mu^{(0)}$ and $\mu^{(k, n)}$ for $(k, n) \in \N^2$ are given by
		\begin{alignat}{3}
			\mu^{(0)}_i(t) &:= \frac{S_i(0)}{\sum_{j=1}^{N_0} S_j(0)} \hat{\mathbbm{1}}_{\{0\}\times \Omega}(t) + 0, \qquad && t \ge 0, \quad i = 1, \cdots, N_0,			\nonumber
			\\
			\mu^{(k, n)}_i(t) &:= \frac{S_i(t)}{\sum_{j=1}^n S_j(t)} \hat{\mathbbm{1}}_{\rrbracket \tau_{k-1}, \tau_k \rrbracket \cap (\R_+ \times \Omega^{k, n})}(t) + 0, \qquad && t \ge 0, \quad i = 1, \cdots, n.	\label{def : relative capitalization}
		\end{alignat}
	\end{defn}
	
	Note that the $(k, n)$-dissection $\mu^{k, n}$ of the market weight process $\mu$, in the sense of \eqref{def : X dissect}, is different from $\mu^{(k, n)}$ of \eqref{def : relative capitalization} for each $(k, n) \in \N^2$. Due to its construction, the quantity $\mu^{(k, n)}_i(t)$ is equal to zero when $t > \tau_k$, and has an important property
	\begin{equation}	\label{eq : sum is equal to 1}
		\sum_{i=1}^n \mu^{(k, n)}_i(t) = 1
	\end{equation}
	on the dissection set $\rrbracket \tau_{k-1}, \tau_k \rrbracket \cap (\R_+ \times \Omega^{k, n})$; whereas the dissection $\mu^{k, n}_i(t)$ of $\mu$ is not necessarily equal to zero when $t > \tau_k$, and does not have such property. However, since each $\mu^{k, n}$ is a RCLL semimartingale, $\mu$ is a $\U$-valued piecewise RCLL semimartingale~(Definition~\ref{Def : piecewise semimartingale}), thus the dissection $\mu^{k, n}$ shall play the role of integrators. When $\mu$ does not appear as integrators, we shall use $\mu^{(k, n)}$ instead of $\mu^{k, n}$, in order to exploit the property \eqref{eq : sum is equal to 1}. We finally note that on each $(k, n)$-dissection set, their increments coincide, i.e.,
	\begin{equation}	\label{eq : same increments}
		d\mu^{k, n}_i(t) \equiv d\mu^{(k, n)}_i(t), \qquad i = 1, \cdots, n,
	\end{equation}
	and thus also from \eqref{eq : sum is equal to 1}
	\begin{equation}	\label{eq : increments add up to zero}
		\sum_{i=1}^n d\mu^{k, n}_i(t) \equiv 0.
	\end{equation}
	
	\begin{rem} \label{rem : two dissections}
		Though we shall use two different dissections $\mu^{(k, n)}$ (for integrands) and $\mu^{k, n}$ (for integrators) of the market weight process $\mu$, every integral in this paper having $d\mu^{k, n}$ as an integrator (see, e.g. \eqref{def : integral I} and \eqref{eq : Ito with jumps}) will only be considered on the $(k, n)$-dissection set, therefore the integrator $d\mu^{k, n}$ can be replaced by $d\mu^{(k, n)}$, thanks to \eqref{eq : same increments}. However, we shall keep using $d\mu^{k, n}$ as the integrator in the following to maintain the same notation as in our earlier paper \citep{BKT:arbitrage}.
	\end{rem}
	
	\begin{example}
		Figure~\ref{fig: example} shows an example of a trajectory of $\U$-valued (daily) capitalization process $S$ and corresponding market weight process $\mu$ during 10 years~(2515 trading days). For this trajectory, the reset sequence is $(\tau_1, \tau_2, \tau_3) = (501, 753, 1508)$ and `active' dissections of $\Omega$ are $\Omega^{1, 2}, \Omega^{2, 4}, \Omega^{3, 3}$, and $\Omega^{4, 4}$. Figure~\ref{fig: example}(c) shows a trajectory of the dissection $\mu^{(3, 3)}$ defined in \eqref{def : relative capitalization}, whereas Figure~\ref{fig: example}(d) illustrates a trajectory of the other dissection $\mu^{3, 3}$.
		\begin{figure}[!htb]
			\centering	
			\subfloat[Market capitalization $S$~(in 1 Million \$)]
			{\includegraphics[width=.49\linewidth]{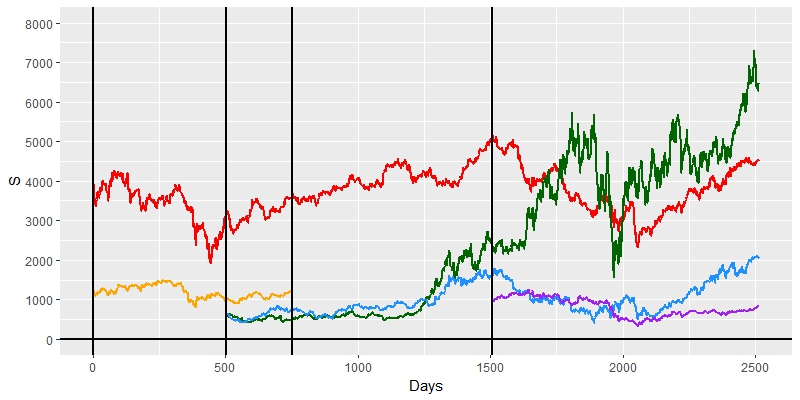}}
			\subfloat[Market weight $\mu$]
			{\includegraphics[width=.49\linewidth]{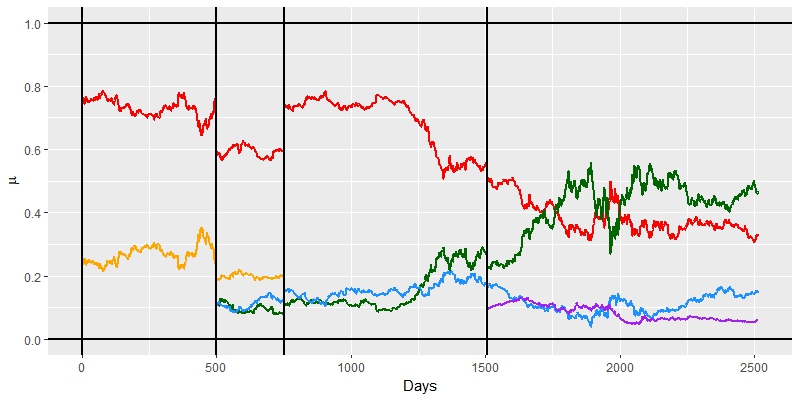}}
			\\
			\subfloat[$\mu^{(3, 3)}$]
			{\includegraphics[width=.49\linewidth]{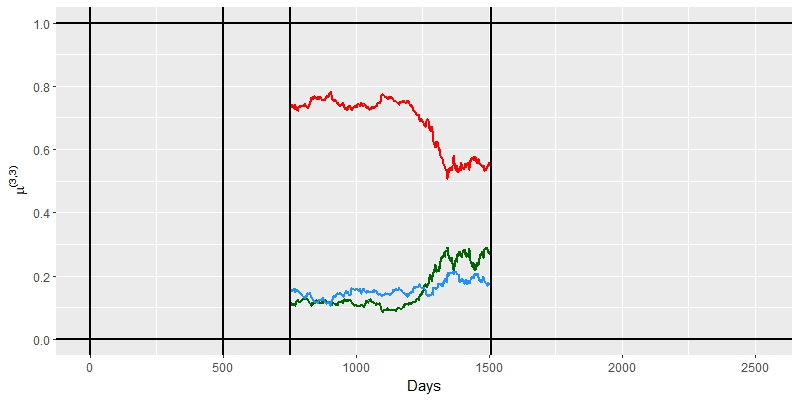}}
			\subfloat[$\mu^{3, 3}$]
			{\includegraphics[width=.49\linewidth]{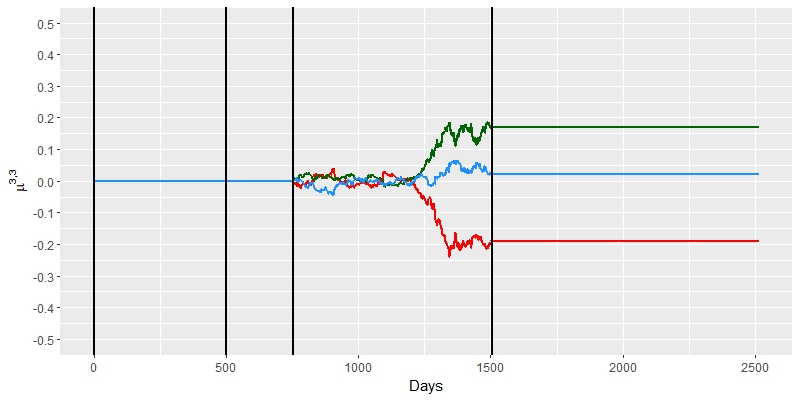}}
			\caption{A trajectory of $S$, $\mu$, and dissections of $\mu$}
			\label{fig: example}
		\end{figure}
	\end{example}
	
	\smallskip
	
	Though the trajectories shown in Figure~\ref{fig: example} are continuous between the dimensional changes, we emphasize here that the capitalization process $S$ and the market weight process $\mu$, defined in Definitions~\ref{Def : price process RCLL} and \ref{Def : market weights}, have RCLL paths between two consecutive dimensional jumps. Any left-discontinuity \textit{between} the dimensional jumps models an unexpected change in the price of an individual stock~(e.g. a company announces/distributes dividends, etc), and such left-discontinuity of $S$ does affect the stochastic integration $H \cdot S$, which represents the capital gains of holding $H$ shares in the stocks.
	
	On the other hand, the dimensional jumps \textit{at} the stopping times $\tau_0, \tau_1, \cdots$ of $S$ are mandated to occur only as right-discontinuities such that the dimensional jumps do not affect the integral $H \cdot S$ for any integrand $H \in \mathcal{L}(S)$. The stochastic integration $H \cdot S$ just stops at each jump and resumes a new piece just afterward. Since each piece $S^{k, n}$ has RCLL paths, the integral $H \cdot S$ remains right-continuous at all times.
	
	This shows how different types of jumps~(\textit{between} and \textit{at} the stopping times $(\tau_k)_{k=0}^{\infty}$) of the capitalization process $S$ are interpreted in the setting of piecewise RCLL semimartingale. We refer to Section~1.1.2 and Remark 3.1 of \cite{Strong2} for more details on the aforementioned discontinuity properties of the piecewise RCLL process $S$ and the integration $H \cdot S$ with their financial interpretation. The market weight process $\mu$ and the stochastic integration $\vartheta \cdot \mu$ for some $\vartheta \in \mathcal{L}(\mu)$ also have the same discontinuity properties as $S$ and $H \cdot S$.
	
	We conclude this subsection with a remark providing a more detailed description of a delisting event of stock from the market~(e.g. due to bankruptcy).
	
	\begin{rem} [Delisting event] \label{rem : delisting event}
		If the $i$-th stock leaves the $(k, n)$-dissected market at the stopping time $\tau_k$, there are two possible scenarios: (i) the component $S_i$ is left-continuous at the moment $\tau_k$ of delisting, i.e., $S_i(\tau_{k}-) = S_i(\tau_k)$; (ii) $S_i$ has a left-discontinuity at $\tau_k$, i.e., $S_i(\tau_{k}-) \neq S_i(\tau_k)$. In case (i), since the dissection $S^{k, n}_i$ remains continuous~(both left- and right-continuous) at $\tau_k$, the exit of the $i$-th company does not affect $H \cdot S$ at $\tau_k$, which means the investor doesn't lose any money from the exit of the $i$-th company. In other words, we have a chance to liquidate the shares holding in the $i$-th stock at the final moment of its death. However, in case (ii), the left jump $S_i(\tau_k) - S_i(\tau_k-)$ at $\tau_k$ does affect the integral $H \cdot S$ such that we may lose some of~(all) the money holding in the $i$-th stock from its delisting, if the final capitalization $S_i(\tau_k)$ is close~(equal, respectively) to zero.
	\end{rem}
	
	\medskip
	
	\subsection{Trading strategies and portfolios}
	
	Recalling the market weight process $\mu$ of Definition~\ref{Def : market weights} and the class $\mathcal{L}(\mu)$ of the integrands for $\mu$ in \eqref{def : space of integrand}, we present the following notion of trading strategy. Here and in what follows, we shall use the notation for stochastic integral which starts from $\tau_{k-1}+$ of any $\mu^{k, n}$-integrable $n$-dimensional process $\vartheta^{(k, n)}$:
	\begin{equation}    \label{def : integral I}
		I_{\mu^{k, n}}(\vartheta^{(k, n)})(t) := \int_{\tau_{k-1}+}^t \sum_{i=1}^n \vartheta^{(k, n)}_i(s) \, d\mu^{k, n}_i(s)
	\end{equation}
	for every $t > \tau_{k-1}$ and $(k, n) \in \N^2$.
	
	\begin{defn} [Trading strategy]	\label{Def : TS}
		A $\U$-valued process $\vartheta$ is called \textit{trading strategy}, if $\vartheta \in \mathcal{L}(\mu)$, and $\vartheta$ satisfies the following two self-financing conditions on each dissection set $\rrbracket \tau_{k-1}, \tau_k \rrbracket \cap (\R_+ \times \Omega^{k, n})$ for every $(k, n) \in \N^2$:
		\begin{equation}	\label{eq : self-financing}
			\sum_{i=1}^n \vartheta^{(k, n)}_i(t) \mu^{(k, n)}_i(t)
			= \sum_{i=1}^n \vartheta^{(k, n)}_i(\tau_{k-1}+) \mu^{(k, n)}_i(\tau_{k-1}+)
			+ I_{\mu^{k, n}}(\vartheta^{(k, n)})(t),
		\end{equation}
		and
		\begin{equation}	\label{eq : self-financing at dimensional change}
			\sum_{i=1}^n \vartheta^{(k, n)}_i(\tau_{k-1}+)S_i(\tau_{k-1}+) = \sum_{i=1}^{N_{\tau_{k-1}}} \vartheta^{(k-1, N_{\tau_{k-1}})}_i(\tau_{k-1})S_i(\tau_{k-1}),
		\end{equation}
		where $\vartheta^{(0, N_0)} \equiv \vartheta^{(0)}$.
	\end{defn}
	
	Here, the quantity $\vartheta^{(k, n)}_i(t)$ represents the number of shares held in the $i$-th stock at time $t$ in the $(k, n)$-dissected market. The \textit{self-financing} property means that there should be neither withdrawals nor
	injections of capital at all times; gains are reinvested, and losses are absorbed. It is straightforward to notice such property from the second identity \eqref{eq : self-financing at dimensional change}; the absolute wealth~(not relative wealth with respect to the market) of $\vartheta$ is maintained, at the moment of dimensional change $\tau_{k-1}$, on every $(k, n)$-dissection set. 
	
	On the other hand, for the first self-financing condition \eqref{eq : self-financing}, we shall need a more detailed explanation. We first denote the left-hand side of \eqref{eq : self-financing}
	\begin{equation}    \label{def : relative wealth}
		V^{\vartheta, k, n}(t) := \sum_{i=1}^n \vartheta^{(k, n)}_i(t) \mu^{(k, n)}_i(t) = \frac{\sum_{i=1}^n \vartheta^{(k, n)}_i(t) S_i(t)}{\sum_{j=1}^n S_j(t)}
	\end{equation}
	where the last identity follows from the definition \eqref{def : relative capitalization}.
	Thus, $V^{\vartheta, k, n}$ represents the ratio of the wealth of the trading strategy $\vartheta$ to the total capitalization of the $(k, n)$-dissected market. This shows the self-financing property of $\vartheta^{(k, n)}$ with respect to the relative capitalization $\mu$ on the interval $(\tau_{k-1}, t]$, but it also implies the self-financing property of $\vartheta^{(k, n)}$ with respect to the capitalization vector $S$ on the same interval, since the self-financing property is num\'eraire independent~(see Lemma~2.9 of \cite{Herdegen:2015}).
	
	Using the definition~\eqref{def : relative capitalization}, we can combine the two self-financing conditions \eqref{eq : self-financing} and \eqref{eq : self-financing at dimensional change} by
	\begin{align}
		V^{\vartheta, k, n}(t) &= \Bigg(\sum_{i=1}^{N_{\tau_{k-1}}} \vartheta^{(k-1, N_{\tau_{k-1}})}_i(\tau_{k-1}) \mu^{(k-1, N_{\tau_{k-1}})}_i(\tau_{k-1}) \Bigg) 
		\Bigg( \frac{ \sum_{j=1}^{N_{\tau_{k-1}}} S_j(\tau_{k-1})}{\sum_{j=1}^n S_j(\tau_{k-1}+)} \Bigg) + I_{\mu^{k, n}}(\vartheta^{(k, n)})(t)	\nonumber
		\\
		&= V^{\vartheta, k-1, N_{\tau_{k-1}}}(\tau_{k-1}) \Bigg( \frac{ \sum_{j=1}^{N_{\tau_{k-1}}} S_j(\tau_{k-1})}{\sum_{j=1}^n S_j(\tau_{k-1}+)} \Bigg) 
		+ I_{\mu^{k, n}}(\vartheta^{(k, n)})(t)	\label{eq : combined self-financing}
	\end{align}
	on the $(k, n)$-dissection set. Thanks to the usual condition on the filtration, we note that the first term on the right-hand side is $\mathcal{F}_{\tau_{k-1}+} = \mathcal{F}_{\tau_{k-1}}$-measurable, thus also $\mathcal{F}_t$-measurable, since $t > \tau_{k-1}$ on the $(k, n)$-dissection set~(see Problem 2.22 of \cite{KS1}).
	
	Collecting the relative wealth on each dissection set, we shall denote
	\begin{align}
		V^{\vartheta} := &\sum_{i=1}^{N_0} \vartheta^{(0)}_i \mu^{(0)}_i \hat{\mathbbm{1}}_{\{0\}\times \Omega} + \sum_{k=1}^{\infty} \sum_{n=1}^{\infty} V^{\vartheta, k, n} \hat{\mathbbm{1}}_{\rrbracket \tau_{k-1}, \tau_k \rrbracket \cap (\R_+ \times \Omega^{k, n})}						\nonumber
		\\
		=&\sum_{i=1}^{N_0} \vartheta^{(0)}_i\mu^{(0)}_i \hat{\mathbbm{1}}_{\{0\}\times \Omega} + \sum_{k=1}^{\infty} \sum_{n=1}^{\infty} \Big[ \sum_{i=1}^n \vartheta^{(k, n)}_i \mu^{(k, n)}_i \Big]	\hat{\mathbbm{1}}_{\rrbracket \tau_{k-1}, \tau_k \rrbracket \cap (\R_+ \times \Omega^{k, n})}		\label{def : relative wealth w.r.t market}
	\end{align}
	the relative wealth process of $\vartheta$ with respect to the (undissected) market.
	
	We conclude this subsection by defining a portfolio $\pi \equiv \pi^{\vartheta}$ corresponding to a trading strategy $\vartheta$ via dissection
	\begin{equation*}
		\pi:= \pi^{(0)} \hat{\mathbbm{1}}_{\{0\}\times \Omega}  + \sum_{k=1}^{\infty} \sum_{n=1}^{\infty} \pi^{(k, n)} \hat{\mathbbm{1}}_{\rrbracket \tau_{k-1}, \tau_k \rrbracket \cap (\R_+ \times \Omega^{k, n})},
	\end{equation*}
	where for each $t \ge 0$
	\begin{alignat}{3}	
		\pi^{(0)}_i(t) &:= \frac{\vartheta^{(0)}_i(0) \mu^{(0)}_i(0)}{\sum_{j=1}^n \vartheta^{(0)}_j(0) \mu^{(0)}_j(0)}, \qquad && i = 1, \cdots, N_0,			\nonumber
		\\
		\pi^{(k, n)}_i(t) &:= \frac{\vartheta^{(k, n)}_i(t) \mu^{(k, n)}_i(t-)}{\sum_{j=1}^n \vartheta^{(k, n)}_j(t) \mu^{(k, n)}_j(t-)}, \qquad && i = 1, \cdots, n, \quad (k, n) \in \N^2.	\label{def : portfolio weights of TS}
	\end{alignat}
	The component $\pi^{(k, n)}_i(t)$ is interpreted as the proportion of wealth invested in the $i$-th stock among $n$ stocks at time $t-$, on the $(k, n)$-dissection set. Every trading strategy $\vartheta$ we construct in the later sections will have left-continuous components $\vartheta^{(k, n)}_i$, i.e., $\vartheta^{(k, n)}_i(t) = \vartheta^{(k, n)}_i(t-)$ for $i = 1, \cdots, n$, on every $(k, n)$-dissection set. Thus, all components $\pi^{(k, n)}_i(t)$ of the corresponding portfolio $\pi^{\vartheta}$ are also left-continuous on every dissection set.
	
	\bigskip
	
	\section{Functional generation of portfolios}   \label{sec : FGP}
	
	Under the stock market model of stochastic dimension described in Section~\ref{sec: preliminaries}, we develop the theory of functional generation of portfolios in this section. After defining the measurable family of generating functions, we describe how to construct additively and multiplicatively generated portfolios from this family. We then decompose their (log-) relative wealth process with respect to the market into 3 terms; the generating function, the excess growth, and the correction term arising due to dimensional jumps.
	
	\medskip
	
	\subsection{Measurable family of generating functions}
	
	Our goal is to construct trading strategies~(and corresponding portfolios) from a function of the market weight process $\mu$. Since the process $\mu$ has changing dimensions on each epoch, we shall need a family of generating functions with different dimensions of domains. When $X$ is a $\U$-valued semimartingale with a reset sequence $(\tau_k)_{k=0}^{\infty}$, we introduce the following notion of piecewise function for $X$. For notational simplicity, we shall use the two notations $X_t \equiv X(t)$ interchangeably.
	
	\begin{defn} [Piecewise function]	\label{def : piecewise function}
		We call $f : \U \rightarrow \R$ a \textit{piecewise function} of $X$, if there exists a family of functions $\{f^0\} \cup \{f^{k, n}\}_{(k, n) \in \N^2}$ such that $f^0 : \R^{N_0} \rightarrow \R$, $f^{k, n}  : \R^n \rightarrow \R$, and the function values are defined by
		\begin{align}
			f(X_0) &:= f^0(X_0),
			\\
			f(X_t) &:= f^{k, n}(X_t) \text{ on each dissection set } \rrbracket \tau_{k-1}, \tau_k \rrbracket \cap (\R_+ \times \Omega^{k, n}), \quad \text{for } (k, n) \in \N^2, ~ t > 0.		\nonumber
		\end{align}
		A piecewise function $f$ of $X$ is said to be \textit{$\ell$-times continuously differentiable}, if every element of the family is $\ell$-times continuously differentiable, i.e., $f^0 \in C^{\ell}(\R^{N_0})$, $f^{k, n} \in C^{\ell}(\R^n)$ for every $(k, n) \in \N^2$. We denote $C^{\ell}(\U)$ the collection of $\ell$-times continuously differentiable piecewise functions.
	\end{defn}
	
	Given a family of functions $f^0 : \R^{N_0} \rightarrow \R$ and $\{f^{k, n} : \R^n \rightarrow \R \}_{(k, n) \in \N^2}$, we can construct a piecewise function $f$ of $X$:
	\begin{equation}	\label{eq : f(X_t)}
		f(X_t) = f^0(X_t) \hat{\mathbbm{1}}_{\{0\} \times \Omega} + \sum_{k=1}^{\infty} \sum_{n=1}^{\infty} f^{k, n}(X_t) \hat{\mathbbm{1}}_{\rrbracket \tau_{k-1}, \tau_k \rrbracket \cap (\R_+ \times \Omega^{k, n})}, \qquad t \ge 0.
	\end{equation}
	Moreover, for any $f \in C^1(\U)$, we consider the vectors of partial derivatives of the family having different dimensions
	\begin{equation}	\label{def : gradients of f}
		\nabla f^0 := (\partial_i f^0 )_{i \in N_0}, \qquad
		\nabla f^{k, n} := (\partial_i f^{k, n})_{i \in n}, \quad \forall \, (k, n) \in \N^2,
	\end{equation}
	and call the $\U$-valued process $\nabla f(X)$, defined by
	\begin{equation*}
		\nabla f(X_t) := \nabla f^0(X_0) \hat{\mathbbm{1}}_{\{0\} \times \Omega} +
		\sum_{k=1}^{\infty} \sum_{n=1}^{\infty} \nabla f^{k, n}(X_t) \hat{\mathbbm{1}}_{\rrbracket \tau_{k-1}, \tau_k \rrbracket \cap (\R_+ \times \Omega^{k, n})}
	\end{equation*}
	the \textit{gradient} of $f(X)$. It is straightforward to check from \eqref{def : H dissect} that dissection of the process $\nabla f(X)$ is given by the standard $n$-dimensional gradient of $f^{k, n}$ in \eqref{def : gradients of f}, i.e., $\big(\nabla f(X)\big)^{(k, n)} = \nabla f^{k, n}(X)$ for all $(k, n) \in \N^2$, and $\nabla f(X) \in \mathcal{L}(X)$.
	
	From each piece $G^{k, n} : [0, 1]^n \rightarrow \R$ of a piecewise function $G$ of $\mu$, we shall generate the corresponding piece, or the dissection $\vartheta^{(k, n)}$, of trading strategy $\vartheta$ in a self-financing way. Then, the initial value $\vartheta^{(k, n)}(\tau_{k-1}+)$ on the interval $\rrbracket \tau_{k-1}, \tau_k \rrbracket$ should depend on the right-hand side of \eqref{eq : self-financing at dimensional change}, which is $\mathcal{F}_{\tau_{k-1}}$-measurable. This implies that the generating function $G^{k, n}$ should also depend on an $\mathcal{F}_{\tau_{k-1}}$-measurable random variable, in other words, $G^{k, n}$ should be determined at the moment $\tau_{k-1}$ of each dimensional change to satisfy the second self-financing identity \eqref{eq : self-financing at dimensional change}. To this end, we introduce the following definition.
	
	\begin{defn} [Measurable modifications]
		For every $(k, n) \in \N^2$ and any function $f : \R^n \rightarrow \R$, we call $\widetilde{f} : \Omega \times \R^n \rightarrow \R$ an \textit{additive $\mathcal{F}_{\tau_{k-1}}$-measurable modification} of $f$, if $\widetilde{f}$ can be represented as $\widetilde{f} = \gamma^{k}+f$ for some $\mathcal{F}_{\tau_{k-1}}$-measurable random variable $\gamma^k$. Similarly, we call $\widehat{f} : \Omega \times \R^n \rightarrow \R$ a \textit{multiplicative $\mathcal{F}_{\tau_{k-1}}$-measurable modification} of $f$, if $\widehat{f}$ can be represented as $\widehat{f} = \delta^{k}f$ for some $\mathcal{F}_{\tau_{k-1}}$-measurable random variable $\delta^k$.
		
		Moreover, given a family $\mathcal{G} := \{G^0\} \cup \{G^{k, n}\}_{(k, n) \in \N^2}$ of functions corresponding to a piecewise function $G$ of $\mu$, piecewise random function $\widetilde{G}$~($\widehat{G}$) of $\mu$ is called a \textit{additive~(multiplicative) measurable modification} of $G$, if the corresponding collection $\widetilde{\mathcal{G}} := \{\widetilde{G}^0\} \cup \{\widetilde{G}^{k, n}\}_{(k, n) \in \N^2}$ of $\widetilde{G}$~($\widehat{\mathcal{G}} := \{\widehat{G}^0\} \cup \{\widehat{G}^{k, n}\}_{(k, n) \in \N^2}$ of $\widehat{G}$) satisfies that $\widetilde{G}^0 \equiv G^0$~($\widehat{G}^0 \equiv G^0$) and each $\widetilde{G}^{k, n}$~($\widehat{G}^{k, n}$) is an additive~(multiplicative, respectively)  $\mathcal{F}_{\tau_{k-1}}$-measurable modification of $G^{k, n}$ for every $(k, n) \in \N^2$.
	\end{defn}
	
	With this notion, we shall handle the second self-financing condition \eqref{eq : self-financing at dimensional change} by constructing trading strategies in a recursive manner. Given a family $\mathcal{G} := \{G^0\} \cup \{G^{k, n}\}_{(k, n) \in \N^2}$ of generating functions and a dissection $\vartheta^{(k-1, N_{\tau_{k-1}})}$ of trading strategy $\vartheta$ on the previous `epoch' of the market, a $\mathcal{F}_{\tau_{k-1}}$-measurable modification~(either additive or multiplicative) $\widetilde{G}^{k, n}$ of $G^{k, n}$ satisfying \eqref{eq : self-financing at dimensional change}, is determined at the moment $\tau_{k-1}$, and $\widetilde{G}^{k, n}$ shall be used to generate the next piece $\vartheta^{(k, n)}$ of $\vartheta$ on each $(k, n)$-dissection set, for $k = 1, 2, \cdots$.
	
	\medskip
	
	\subsection{Functionally generated portfolios}  \label{subsec : FGP}
	
	In this subsection, we shall construct two types of functionally generated portfolios from a piecewise generating function $G \in C^2(\U)$ with its family $\{G^0\} \cup \{G^{k, n}\}_{(k, n) \in \N^2}$, using the market weight process $\mu$ of Definition~\ref{Def : market weights} as its input. Here and in what follows, we assume without loss of generality that $G^0(\mu_0) = 1$ holds, after some normalization~(we can consider $\mathcal{F}_0$-measurable modification satisfying $\widehat{G}^0(\mu_0)=1$ and rename $\widehat{G}^0$ to $G^0$).
	
	\subsubsection{Additive generation}	\label{subsubsec : AGTS}
	Since we have $\nabla G(\mu_-) \in \mathcal{L}(\mu)$, standard $n$-dimensional It\^o's formula with jumps yields
	\begin{align}	
		G^{k, n}(\mu_t) &= G^{k, n}(\mu_{\tau_{k-1}+}) + I_{\mu^{k, n}}\big(\nabla G^{k, n}(\mu_-)\big)(t)  \label{eq : Ito with jumps}
		\\
		& \quad + \frac{1}{2} \int_{\tau_{k-1}+}^t \sum_{i=1}^n \sum_{j=1}^n \partial^2_{i, j} G^{k, n}(\mu_{s-}) \, d[\mu^{k, n, c}_i, \mu^{k, n, c}_j](s)
		+ \sum_{\tau_{k-1} < s \le t} d_{B, G^{k, n}}\big( \mu_s, \, \mu_{s-} \big),	\nonumber
	\end{align}
	on the $(k, n)$-dissection set, where $\mu^{k, n, c}_i$ is the continuous part of $\mu^{k, n}_i$, and $d_{B, G}(x, y)$ denotes the Bregman divergence associated with $G$ for points $x, y \in \Delta_n := \{(x_1, \cdots, x_n) \in [0, 1]^n : \sum_{i=1}^n x_i = 1\}$
	\begin{equation}	\label{def : Bregman}
		d_{B, G}(x, y) := G(x) - G(y) - \sum_{i=1}^n \partial_i G(y)(x_i - y_i).
	\end{equation}
	Let us define
	\begin{equation}	\label{def : Gamma k, n}
		\Gamma^{G, k, n}(t) := 0 + \bigg[ G^{k, n}(\mu_{\tau_{k-1}+}) - G^{k, n}(\mu_t) + I_{\mu^{k, n}}\big(\nabla G^{k, n}(\mu_-)\big)(t) \bigg] \hat{\mathbbm{1}}_{\rrbracket \tau_{k-1}, \tau_k \rrbracket \cap (\R_+ \times \Omega^{k, n})}
	\end{equation}
	the \textit{Gamma process} of $G$ on each $(k, n)$-dissection set. From \eqref{eq : Ito with jumps}, each $(k, n)$-Gamma process admits a different formulation
	\begin{equation}	\label{eq : Gamma k, n}
		\Gamma^{G, k, n}(t) = -\frac{1}{2} \int_{\tau_{k-1}+}^t \sum_{i, j=1}^n \partial^2_{i, j} G^{k, n}(\mu_{s-}) \, d[\mu^{k, n, c}_i, \mu^{k, n, c}_j](s)
		- \sum_{\tau_{k-1} < s \le t} d_{B, G^{k, n}}\big( \mu_s, \, \mu_{s-} \big)
	\end{equation}
	on the $(k, n)$-dissection set, and it is of finite variation. In particular, if $G^{k, n}$ is a concave function, then the Bregman divergence is nonpositive, i.e., $d_{B, G}(x, y) \le 0$ for every $x, y \in \Delta_n$~(by Jensen's inequality), thus the $(k, n)$-Gamma process $\Gamma^{G, k, n}(\cdot)$ is nondecreasing on the $(k, n)$-dissection set.
	
	We could now take the integrand $\vartheta^{(k, n)} := \nabla G^{k, n}(\mu_-)$ of the stochastic integral in \eqref{eq : Ito with jumps} as a candidate for a trading strategy, but we need to make it self-financing. In this effort, we start from defining a dissection $\varphi^{(0)} = (\varphi^{(0)}_1, \cdots, \varphi^{(0)}_{N_0})$ of a trading strategy $\varphi$ at time $t = \tau_0 = 0$
	\begin{align}
		\vartheta^{(0)}(\cdot) &:= 0^{(N_0)} + \nabla G^0(\mu_{\cdot}) \hat{\mathbbm{1}}_{\{0\}\times \Omega}, 		\label{def : vartheta 0}
		\\
		\varphi^{(0)}_i(\cdot) &:= 0 + \big(\vartheta^{(0)}_i(\cdot) - C^{G, \vartheta, 0} \big) \hat{\mathbbm{1}}_{\{0\}\times \Omega}, \qquad i = 1, \cdots, N_0,		\label{def : varphi 0}
	\end{align}
	where $C^{G, \vartheta, 0} := \sum_{i=1}^{N_0} \vartheta^{(0)}_i(0) \mu_i(0) - G^0(\mu_0)$ is called a \textit{defect of balance} at time $0$~(see Definition~\ref{Def : balance} below). In order to construct the next dissections $\varphi^{(k, n)}$ satisfying the first self-financing condition \eqref{eq : self-financing} on the interval $\rrbracket \tau_{k-1}, \tau_k \rrbracket$, recursively in $k = 1, 2, \cdots$, we shall subtract the so-called \textit{defect of self-financibility} $Q^{\vartheta, \mu, k, n}(t)$ from the `basis' $\vartheta^{(k, n)}(t)$~(see the definition~\eqref{def : varphi k, n} below), following the work of \cite{Karatzas:Ruf:2017}. Here, the basis is defined as
	\begin{equation}	\label{def : vartheta k, n}
		\vartheta^{(k, n)}(t) := 0^{(n)} + \nabla G^{k, n} (\mu_{t-}) \hat{\mathbbm{1}}_{\rrbracket \tau_{k-1}, \tau_k \rrbracket \cap (\R_+ \times \Omega^{k, n})},
	\end{equation}
	and the defect of self-financibility is given by
	\begin{align}
		Q^{\vartheta, \mu, k, n}(t)
		:= 0+ \bigg[ \sum_{i=1}^n \vartheta^{(k, n)}_i(t) \mu^{(k, n)}_i(t-)
		&- \sum_{i=1}^n \vartheta^{(k, n)}_i(\tau_{k-1}+) \mu^{(k, n)}_i(\tau_{k-1}+)		\label{def : Q dissection}
		\\
		&- I_{\mu^{k, n}}(\vartheta^{(k, n)})(t-) \bigg] \hat{\mathbbm{1}}_{\rrbracket \tau_{k-1}, \tau_k \rrbracket \cap (\R_+ \times \Omega^{k, n})}.	\nonumber
	\end{align}
	Note that we have $Q^{\vartheta, \mu, k, n}(\tau_{k-1}+) = 0$ by definition. We also consider an additive measurable modification of $G$:
	\begin{equation}	\label{eq : G modification}
		\widetilde{G}^0 = G^0, \qquad \widetilde{G}^{k, n} = \gamma^{G, \varphi, k, n} + G^{k, n}, \qquad \forall \, (k, n) \in \N^2,
	\end{equation}
	where $\gamma^{G, \varphi, k, n}$ is a $\mathcal{F}_{\tau_{k-1}}$-measurable random variable
	\begin{equation}	\label{def : gamma}
		\gamma^{G, \varphi, k, n} := V^{\varphi, k-1, N_{\tau_{k-1}}}(\tau_{k-1}) \cdot \frac{\sum_{j=1}^{N_{\tau_{k-1}}}S_j(\tau_{k-1})}{\sum_{j=1}^n S_j(\tau_{k-1}+)} - G^{k, n}(\mu_{\tau_{k-1}+}).
	\end{equation}
	We next define a $\mathcal{F}_{\tau_{k-1}}$-measurable random variable $C^{G, \vartheta, k, n}$, and an $n$-dimensional vector process $\varphi^{(k, n)}$ by
	\begin{equation}	\label{def : C, G, mu, k, n}
		C^{\widetilde{G}, \vartheta, k, n} := \sum_{i=1}^n \vartheta^{(k, n)}_i(\tau_{k-1}+)\mu^{(k, n)}_i(\tau_{k-1}+) - \widetilde{G}^{k, n}(\mu_{\tau_{k-1}+}),
	\end{equation}
	\begin{equation}	\label{def : varphi k, n}
		\varphi^{(k, n)}_i(t) := 0 + \Big( \vartheta^{(k, n)}_i(t) - Q^{\vartheta, \mu, k, n}(t) - C^{\widetilde{G}, \vartheta, k, n} \Big) \hat{\mathbbm{1}}_{\rrbracket \tau_{k-1}, \tau_k \rrbracket \cap (\R_+ \times \Omega^{k, n})} , \quad i = 1, \cdots, n.
	\end{equation}
	We emphasize again that $\varphi^{(k, n)}$ is defined in a recursive manner for $k = 1, 2, \cdots$; from the initial configuration \eqref{def : vartheta 0} and \eqref{def : varphi 0}, $\gamma^{G, \varphi, k, n}$ of \eqref{def : gamma} depends on the last value $\varphi^{(k-1, N_{\tau_{k-1}})}(\tau_{k-1})$ of the previous epoch, thus, $\varphi^{(k, n)}$ defined in \eqref{def : varphi k, n}, also depends on the quantity $\varphi^{(k-1, N_{\tau_{k-1}})}(\tau_{k-1})$.
	
	Finally, we construct a $\U$-valued process $\varphi$ collecting the dissections
	\begin{equation}	\label{def : varphi}
		\varphi(\cdot):= \varphi^{(0)}(\cdot) \hat{\mathbbm{1}}_{\{0\}\times \Omega}  + \sum_{k=1}^{\infty} \sum_{n=1}^{\infty} \varphi^{(k, n)}(\cdot) \hat{\mathbbm{1}}_{\rrbracket \tau_{k-1}, \tau_k \rrbracket \cap (\R_+ \times \Omega^{k, n})}.
	\end{equation}
	We show that this process $\varphi$ is indeed a trading strategy, and it is called an \textit{additively generated trading strategy} from the piecewise generating function $G \in C^2(\U)$.
	
	\smallskip
	
	\begin{prop} [Additive generation]	\label{prop : AGTS}
		The process $\varphi$ in \eqref{def : varphi} is a trading strategy and its relative wealth process $V^{\varphi}$ is given by
		\begin{equation}	\label{eq : varphi wealth}
			V^{\varphi}(t) 
			= \widetilde{G}(\mu_t) + \sum_{k=1}^{\infty} \sum_{n=1}^{\infty} \, \Gamma^{G, k, n}(t) \, \hat{\mathbbm{1}}_{\rrbracket \tau_{k-1}, \tau_k \rrbracket \cap (\R_+ \times \Omega^{k, n})}, \qquad t \ge 0.
		\end{equation}
	\end{prop}
	
	\begin{proof}
		Since $\nabla G^{k, n}(\mu_-)$ is $\mu^{k, n}$-integrable and the $(k, n)$-dissection set $\rrbracket \tau_{k-1}, \tau_k \rrbracket \cap (\R_+ \times \Omega^{k, n})$ is predictable, $\vartheta^{(k, n)}$ is also predictable for every $(k, n) \in \N^2$. Moreover, $Q^{\vartheta, \mu, k, n}$ is also predictable and $\mu^{k, n}$-integrable~(Lemma~4.13 of \cite{Shiryaev_vector}) for every $(k, n) \in \N^2$. We also have the predictability of $\mathcal{F}_{\tau_{k-1}}$-measurable random variable $C^{\widetilde{G}, \vartheta, k, n}$ on the dissection set. This shows that $\varphi \in \mathcal{L}(\mu)$.
		
		Next, we check that $\varphi$ satisfies the combined self-financing identity \eqref{eq : combined self-financing} for an arbitrary pair $(k, n) \in \N^2$. We recall the properties \eqref{eq : sum is equal to 1} - \eqref{eq : increments add up to zero}, along with the identity
		\begin{equation}    \label{eq : jump relationships}
			\sum_{i=1}^n \vartheta^{(k, n)}_i(\cdot) \Big( \Delta \mu^{(k, n)}_i(\cdot) \Big) = \Big(\Delta I_{\mu^{k, n}}(\vartheta^{(k, n)})\Big)(\cdot),
		\end{equation}
		from Theorem~II.13 of \cite{Protter}, to derive the series of identities on the $(k, n)$-dissection set:
		\begin{align}
			V^{\varphi, k, n}(t) &= \sum_{i=1}^n \varphi^{(k, n)}_i(t)\mu^{(k, n)}_i(t)
			= \sum_{i=1}^n \vartheta^{(k, n)}_i(t)\mu^{(k, n)}_i(t) - Q^{\vartheta, \mu, k, n}(t) - C^{\widetilde{G}, \vartheta, k, n}	\nonumber
			\\
			&= \sum_{i=1}^n \vartheta^{(k, n)}_i(t) \Big( \Delta \mu^{(k, n)}_i(t) \Big) + \widetilde{G}^{k, n}(\mu_{\tau_{k-1}+}) + I_{\mu^{k, n}}(\vartheta^{(k, n)})(t-)	\nonumber
			\\
			&= \widetilde{G}^{k, n}(\mu_{\tau_{k-1}+}) + I_{\mu^{k, n}}(\vartheta^{(k, n)})(t)				
			= \widetilde{G}^{k, n}(\mu_{\tau_{k-1}+}) + I_{\mu^{k, n}}(\varphi^{(k, n)})(t)				\label{eq : additive ito}
			\\
			&= V^{\varphi, k-1, N_{\tau_{k-1}}}(\tau_{k-1}) \Bigg( \frac{ \sum_{j=1}^{N_{\tau_{k-1}}} S_j(\tau_{k-1})}{\sum_{j=1}^n S_j(\tau_{k-1}+)} \Bigg) 
			+ I_{\mu^{k, n}}(\varphi^{(k, n)})(t).	\nonumber
		\end{align}
		Thus, the self-financing condition \eqref{eq : combined self-financing} for $\varphi$ holds, and $\varphi$ is indeed a trading strategy.
		
		For the relative wealth process of $\varphi$, the identities \eqref{eq : additive ito} and \eqref{def : Gamma k, n} yield
		\begin{align}
			V^{\varphi, k, n}(t) 
			&= \widetilde{G}^{k, n}(\mu_{\tau_{k-1}+}) + I_{\mu^{k, n}}(\vartheta^{(k, n)})(t)
			= \gamma^{G, \varphi, k, n} + G^{k, n}(\mu_{\tau_{k-1}+}) + I_{\mu^{k, n}}\big(\nabla G^{k, n}(\mu_-)\big)(t)						\nonumber
			\\
			&= \gamma^{G, \varphi, k, n} + G^{k, n}(\mu_t) + \Gamma^{G, k, n}(t)									\label{eq : V varphi, k, n}
		\end{align}
		on each $(k, n)$-dissection set. Plugging the last equality into \eqref{def : relative wealth w.r.t market}, together with \eqref{eq : f(X_t)}, yields the result \eqref{eq : varphi wealth} for $t > 0$; when $t = 0$, it is easy to check the identity $V^{\varphi}(0) = G^0(\mu_0) = \widetilde{G}^0(\mu_0)$.
	\end{proof}
	
	Proposition~\ref{prop : AGTS} generalizes the earlier construction introduced in \cite{Karatzas:Ruf:2017} for each \textit{piece of the dissected market} of a fixed dimension and patches these pieces seamlessly in the sense that the second self-financing identity \eqref{eq : self-financing at dimensional change} is satisfied by introducing a modified version of generating function at the beginning of each piece.
	
	When computing the relative wealth $V^{\varphi}(t)$ of $\varphi$ in Proposition~\ref{prop : AGTS}, we need to use the identity \eqref{eq : varphi wealth} in a recursive manner, since $\widetilde{G}^{k, n}(\mu_t)$ depends on the quantity $\gamma^{G, \varphi, k, n}$, which in turn depends on the last relative wealth $V^{\varphi, k-1, N_{\tau_{k-1}}}(\tau_{k-1})$ of the previous epoch. This is due to the recursive construction of the trading strategy $\varphi$, however, the following result provides an alternative representation of the relative wealth process $V^{\varphi}(t)$ for any given $t > 0$.
	
	In order to simplify some of the notations in the following, we denote
	\begin{equation}	\label{def : sigma k, n}
		\sigma^{k, n} := \frac{\sum_{j=1}^{N_{\tau_{k-1}}} S_j(\tau_{k-1})}{\sum_{j=1}^n S_j(\tau_{k-1}+)}
	\end{equation}
	the ratio of total capitalization at $\tau_{k-1}$ to total capitalization right afterward for the $(k, n)$-dissection set, and write their products from the $i$-th epoch to the $k$-th epoch for any $1 \le i \le k$ by
	\begin{equation}	\label{def : sigma k, n, i}
		\sigma^{k, n}_{i, N_{\tau_i}} := \prod_{\ell = i}^k \sigma^{\ell, N_{\tau_{\ell}}}
		= \frac{\sum_{j=1}^{N_{\tau_{i-1}}} S_j(\tau_{i-1})}{\sum_{j=1}^{N_{\tau_i}} S_j(\tau_{i-1}+)} \times \cdots \times \frac{\sum_{j=1}^{N_{\tau_{k-1}}} S_j(\tau_{k-1})}{\sum_{j=1}^n S_j(\tau_{k-1}+)}.
	\end{equation}
	
	\smallskip
	
	\begin{thm} [Additive generation] \label{thm : AGTS alternative}
		For an arbitrary pair $(t, \omega)$ with $t > 0$, belonging to a $(k, n)$-dissection set for $(k, n) \in \N^2$, the relative wealth $V^{\varphi}(t)$ of the additively generated trading strategy $\varphi$ in Proposition~\ref{prop : AGTS} can be decomposed as the sum of the generating function, the excess growth~(EG), and the correction term~(C) which arises from dimensional jumps up to time $t$:
		\begin{equation}    \label{eq : V varphi decomposition}
			V^{\varphi}(t) = G^{k, n}(\mu_t) + EG(t) + C(t), \qquad \text{where}
		\end{equation}
		\begin{align}
			EG(t) &:= \sum_{\ell=1}^{k-1} \Gamma^{G, \ell, N_{\tau_{\ell}}}(\tau_{\ell}) \sigma^{k, n}_{\ell+1, N_{\tau_{\ell+1}}} + \Gamma^{G, k, n}(t),	\label{eq : GR add}
			\\
			C(t) &:= \sum_{\ell=1}^k \bigg[ \sigma^{\ell, N_{\tau_{\ell}}} G^{\ell-1, N_{\tau_{\ell-1}}}(\mu_{\tau_{\ell-1}}) - G^{\ell, N_{\tau_{\ell}}}(\mu_{\tau_{\ell-1}+}) \bigg] \, \sigma^{k, n}_{\ell+1, N_{\tau_{\ell+1}}},		\label{eq : C add}
		\end{align}
		with the notation $\sigma^{k, n}_{k+1, N_{\tau_{k+1}}} = 1$.
	\end{thm}
	
	\begin{proof}
		For the given pair $(t, \omega)$~(omitting $\omega$ in the following notations), we repeatedly apply \eqref{eq : V varphi, k, n} and \eqref{def : gamma} to derive
		\begin{align*}
			V^{\varphi}(t) &=  V^{\varphi, k, n}(t) 
			= \gamma^{G, \varphi, k, n} + G^{k, n}(\mu_t) + \Gamma^{G, k, n}(t)
			\\
			&= V^{\varphi, k-1, N_{\tau_{k-1}}}(\tau_{k-1}) \sigma^{k, n} - G^{k, n}(\mu_{\tau_{k-1}+}) + G^{k, n}(\mu_t) + \Gamma^{G, k, n}(t)
			\\
			&= \Big[ \gamma^{G, \varphi, k-1, N_{\tau_{k-1}}} + G^{k-1, N_{\tau_{k-1}}}(\mu_{\tau_{k-1}}) + \Gamma^{G, k-1, N_{\tau_{k-1}}}(\tau_{k-1}) \Big] \sigma^{k, n} 
			\\
			& \qquad \qquad \qquad \qquad \qquad \qquad \qquad \qquad \qquad \qquad - G^{k, n}(\mu_{\tau_{k-1}+}) + G^{k, n}(\mu_t) + \Gamma^{G, k, n}(t) 
			\\
			&= \, \cdots
			\\
			&= \Bigg[ V^{\varphi, 0, N_0}(0) \sigma^{k, n}_{1, N_{\tau_1}} + \sum_{\ell=1}^{k-1} \Big[ G^{\ell, N_{\tau_{\ell}}}(\mu_{\tau_{\ell}}) - G^{\ell, N_{\tau_{\ell}}}(\mu_{\tau_{\ell-1}+}) \Big] \sigma^{k, n}_{\ell+1, N_{\tau_{\ell+1}}} 
			- G^{k,n}(\mu_{\tau_{k-1}+}) \Bigg] + G^{k, n}(\mu_t)
			\\
			& \qquad \qquad \qquad \qquad \qquad + \sum_{\ell=1}^{k-1} \Gamma^{G, \ell, N_{\tau_{\ell}}}(\tau_{\ell}) \sigma^{k, n}_{\ell+1, N_{\tau_{\ell+1}}} + \Gamma^{G, k, n}(t).
		\end{align*}
		The last two terms are $EG(t)$ of \eqref{eq : GR add}; using the identity $V^{\varphi, 0, N_0}(0) = G^0(\mu_0)$ and rearranging the terms inside the biggest square brackets yield the result \eqref{eq : V varphi decomposition}.
	\end{proof}
	
	\smallskip
	
	\subsubsection{Multiplicative generation}	\label{subsubsec : MGTS}
	
	In what follows, we also fix a piecewise generating function $G$ in $C^2(\U)$, but with an extra condition that the reciprocal $1/G^{k, n}(\mu_t)$ is locally bounded on every $(k, n)$-dissection set. Recalling the notations in Section~\ref{subsubsec : AGTS}, we shall recursively define two $\U$-valued process $\eta$ and $\psi$ as follows. At time $\tau_0 = 0$, let us set
	\begin{equation}
		\eta^{(0)} \equiv \psi^{(0)} \equiv \varphi^{(0)}		\label{def : psi 0}
	\end{equation}
	as in \eqref{def : varphi 0}. For each $k = 1, 2, \cdots, $ consider a $\mathcal{F}_{\tau_{k-1}}$-measurable random variable $\delta^{G, \psi, k, n}$
	\begin{equation}	\label{def : delta}
		\delta^{G, \psi, k, n} := \frac{V^{\psi, k-1, N_{\tau_{k-1}}}(\tau_{k-1})}{G^{k, n}(\mu_{\tau_{k-1}+})} \cdot \frac{\sum_{j=1}^{N_{\tau_{k-1}}}S_j(\tau_{k-1})}{\sum_{j=1}^n S_j(\tau_{k-1}+)},
	\end{equation}
	and a multiplicative measurable modification $\widehat{G}$ of $G$
	\begin{equation}	\label{eq : G modification2}
		\widehat{G}^0 = G^0, \qquad \widehat{G}^{k, n} = \delta^{G, \psi, k, n} G^{k, n}, \qquad \forall \, (k, n) \in \N^2.
	\end{equation}
	Let us recall from \eqref{def : Gamma k, n} the Gamma process $\Gamma^{G, k, n}$ on the $(k, n)$-dissection set and define
	\begin{equation}	\label{def : E k, n}
		E^{G, k, n}(t) := 0 + \mathcal{E} \bigg( \int_{\tau_{k-1}+}^t \frac{1}{G^{k, n}(\mu_{s})} \, d\Gamma^{G, k, n}(s) \bigg) \hat{\mathbbm{1}}_{\rrbracket \tau_{k-1}, \tau_k \rrbracket \cap (\R_+ \times \Omega^{k, n})}, \qquad t \ge 0,
	\end{equation}
	where $\mathcal{E}\big(L(t)\big)$ denotes the stochastic exponential of a finite variation process $L$ with $L(\tau_{k-1}+) = 0$ defined on the $(k, n)$-dissection set:
	\begin{equation}	\label{def : stochastic exponential}
		\mathcal{E}\big(L(t)\big) := \exp \big(L(t)\big) \prod_{\tau_{k-1} < s \le t} \Big(1 + \Delta L(s) \Big) \exp \Big( -\Delta L(s)\Big)	
	\end{equation}
	such that $d\mathcal{E}(L)(t) = \mathcal{E}(L)(t-) dL(t)$ holds. Thanks to the assumption on $G$ that the reciprocal $1/G^{k, n}(\mu_t)$ is locally bounded, the process $E^{G, k, n}(t)$ is of finite variation on the $(k, n)$-dissection set. We continue to define for $k = 1, 2, \cdots$, and any $n \in \N$
	\begin{align}
		\eta^{(k, n)}(t) := 0^{(n)} &+ \delta^{G, \psi, k, n} E^{G, k, n}(t-) \nabla G^{k, n} (\mu_{t-}) \hat{\mathbbm{1}}_{\rrbracket \tau_{k-1}, \tau_k \rrbracket \cap (\R_+ \times \Omega^{k, n})}	         \nonumber
		\\
		= 0^{(n)} &+ E^{G, k, n}(t-) \nabla \widehat{G}^{k, n} (\mu_{t-}) \hat{\mathbbm{1}}_{\rrbracket \tau_{k-1}, \tau_k \rrbracket \cap (\R_+ \times \Omega^{k, n})}.								\label{def : eta k, n2}
	\end{align}
	Furthermore, a $\mathcal{F}_{\tau_{k-1}}$-measurable random variable $C^{\widehat{G}, \eta, k, n}$ and the defect of self-financibility $Q^{\eta, \mu, k, n}$ for $\eta$ are defined as before in \eqref{def : Q dissection}, \eqref{def : C, G, mu, k, n}, and an $n$-dimensional vector process $\psi^{(k, n)}$ is given by
	\begin{equation}	\label{def : psi k, n}
		\psi^{(k, n)}_i(\cdot) := 0 + \Big( \eta^{(k, n)}_i(\cdot) - Q^{\eta, \mu, k, n}(\cdot) - C^{\widehat{G}, \eta, k, n} \Big) \hat{\mathbbm{1}}_{\rrbracket \tau_{k-1}, \tau_k \rrbracket \cap (\R_+ \times \Omega^{k, n})}, \quad i = 1, \cdots, n.
	\end{equation}
	Finally, we construct a $\U$-valued process $\psi$ collecting the dissections
	\begin{equation}	\label{def : psi}
		\psi := \psi^{(0)} \hat{\mathbbm{1}}_{\{0\}\times \Omega}  + \sum_{k=1}^{\infty} \sum_{n=1}^{\infty} \psi^{(k, n)} \hat{\mathbbm{1}}_{\rrbracket \tau_{k-1}, \tau_k \rrbracket \cap (\R_+ \times \Omega^{k, n})}.
	\end{equation}
	The next result shows that $\psi$ is indeed a trading strategy and we call it \textit{multiplicatively generated} from the generating function $G$.
	
	\smallskip
	
	\begin{prop} [Multiplicative generation]	\label{prop : MGTS}
		For a piecewise function $G \in C^2(\U)$ of $\mu$ such that the reciprocal $1/G^{k, n}$ is locally bounded for every $(k, n) \in \N^2$, the process $\psi$ in \eqref{def : psi} is a trading strategy and its relative wealth process $V^{\psi}$ is given by
		\begin{equation}	\label{eq : psi wealth}
			V^{\psi}(t) 
			= \widehat{G}^0(\mu_0) \hat{\mathbbm{1}}_{\{0\}\times \Omega}
			+ \sum_{k=1}^{\infty} \sum_{n=1}^{\infty} \widehat{G}^{k, n}(\mu_t) E^{G, k, n}(t) \hat{\mathbbm{1}}_{\rrbracket \tau_{k-1}, \tau_k \rrbracket \cap (\R_+ \times \Omega^{k, n})}.
		\end{equation}
	\end{prop}
	
	\begin{proof}
		As in the proof of Proposition~\ref{prop : AGTS}, it is straightforward to check the predictability and $\mu^{k, n}$-integrability of $\eta^{(k, n)}$ and $\psi^{(k, n)}$, thanks to the local boundedness assumption on $1/G^{k, n}$ for each $(k, n) \in \N^2$. This shows $\psi \in \mathcal{L}(\mu)$. Moreover, we can prove in the same manner that $\psi$ satisfies the self-financing condition \eqref{eq : combined self-financing} and conclude that $\psi$ is a trading strategy.
		
		We fix an arbitrary pair $(k, n) \in \N^2$ and note that the process $E^{G, k, n}(\cdot)$ is of finite variation and the identity $E^{\widehat{G}, k, n}(\cdot) = E^{G, k, n}(\cdot)$ holds. By applying the product rule~(Theorem~I.4.49 of \cite{JacodS}) and recalling the definition \eqref{def : Gamma k, n}, we derive on each $(k, n)$-dissection set, i.e., $\forall \, (t, \omega) \in \, \rrbracket \tau_{k-1}, \tau_k \rrbracket \cap (\R_+ \times \Omega^{k, n})$:
		\begin{align}
			\widehat{G}^{k, n}(\mu_t)E^{G, k, n}(t) &- \widehat{G}^{k, n}(\mu_{\tau_{k-1}+})E^{G, k, n}(\tau_{k-1}+)					\nonumber
			\\
			& = \int_{\tau_{k-1}+}^t E^{G, k, n}(s-) \, d\widehat{G}^{k, n}(\mu_s) +\int_{\tau_{k-1}+}^t \widehat{G}^{k, n}(\mu_s) \, dE^{G, k, n}(s)	\nonumber
			\\
			& = \int_{\tau_{k-1}+}^t E^{G, k, n}(s-) \Big( d\widehat{G}^{k, n}(\mu_s) + d\Gamma^{\widehat{G}, k, n}(s) \Big)		\nonumber
			\\
			&= \int_{\tau_{k-1}+}^t E^{G, k, n}(s-) \sum_{i=1}^n \partial_i \widehat{G}^{k, n}(\mu_{s-}) \, d\mu^{k, n}_i(s)
			= I_{\mu^{k, n}}(\eta^{(k, n)})(t)					\label{eq : V psi diff}
			\\
			&= I_{\mu^{k, n}}(\psi^{(k, n)})(t) = V^{\psi, k, n}(t) - V^{\psi, k, n}(\tau_{k-1}+).		\nonumber
		\end{align}
		Here, the second-last equality is from the property \eqref{eq : increments add up to zero}, and the last identity follows from the self-financing condition \eqref{eq : self-financing} for $\psi^{(k, n)}$. Furthermore, we derive from $Q^{\eta, \mu, k, n}(\tau_{k-1}+) = 0$ and $E^{G, k, n}(\tau_{k-1}+) = 1$ that
		\begin{equation}	\label{eq : V psi at tau k-1+}
			V^{\psi, k, n}(\tau_{k-1}+)
			= \sum_{i=1}^n \eta^{(k, n)}_i(\tau_{k-1}+) \mu^{(k, n)}_i(\tau_{k-1}+) - C^{\widehat{G}, \eta, k, n}
			= \widehat{G}^{k, n}(\mu_{\tau_{k-1}+})E^{G, k, n}(\tau_{k-1}+).
		\end{equation}
		From these derivations, we conclude that
		\begin{equation}	\label{eq : V psi equal to GE}
			V^{\psi, k, n}(\cdot) \equiv \widehat{G}^{k, n}(\mu_{\cdot})E^{G, k, n}(\cdot)	
		\end{equation}
		on the $(k, n)$-dissection set. When $t=0$, it is trivial to check $V^{\psi}(0) = G^0(\mu_0)$, and the result \eqref{eq : psi wealth} follows.
	\end{proof}
	
	As in Theorem~\ref{thm : AGTS alternative}, we now derive the decomposition of the log-relative wealth $V^{\psi}(t)$ for any given $t > 0$.
	
	\begin{thm} [Multiplicative generation]	\label{thm : MGTS alternative}
		For an arbitrary pair $(t, \omega)$ with $t > 0$, belonging to a $(k, n)$-dissection set for $(k, n) \in \N^2$, the log relative wealth $\log V^{\psi}(t)$ of the multiplicatively generated trading strategy $\psi$ in Proposition~\ref{prop : MGTS} can be decomposed as the sum of the generating function, the excess growth~(EG), and the correction term~(C) which arises from dimensional jumps up to time $t$:
		\begin{equation}   \label{eq : V psi decomposition}
			\log V^{\psi}(t) = \log G^{k, n}(\mu_t) + EG(t) + C(t), \qquad \text{where}
		\end{equation}
		\begin{align}
			EG(t) &:= \sum_{\ell=1}^{k-1} \Bigg[ \int_{\tau_{\ell-1}+}^{\tau_{\ell}} \frac{d\Gamma^{G, \ell, N_{\tau_{\ell}}, c}(s)}{G^{\ell, N_{\tau_{\ell}}}(\mu_{s})} + \sum_{\tau_{\ell-1} < s \le \tau_{\ell}} \log \bigg( 1+ \frac{\Delta \Gamma^{G, \ell, N_{\tau_{\ell}}}(s)}{G^{\ell, N_{\tau_{\ell}}}(\mu_s)} \bigg) \Bigg]	\label{eq : GR mul}
			\\
			& \qquad \qquad \qquad \qquad + \int_{\tau_{k-1}+}^t \frac{d\Gamma^{G, k, n, c}(s)}{G^{k, n}(\mu_{s})} + \sum_{\tau_{k-1} < s \le t} \log \bigg( 1+ \frac{\Delta \Gamma^{G, k, n}(s)}{G^{k, n}(\mu_s)} \bigg), \nonumber
			\\
			C(t) &:= \sum_{\ell=1}^k \log \bigg( \frac{\sigma^{\ell, N_{\tau_{\ell}}} \, G^{\ell-1, N_{\tau_{\ell-1}}}(\mu_{\tau_{\ell-1}})}{G^{\ell, N_{\tau_{\ell}}}(\mu_{\tau_{\ell-1}+})} \bigg), 		\label{eq : C mul}
		\end{align}
		with convention $N_{\tau_k} = n$. Here, $\Gamma^{G, k, n, c}$ denotes the continuous part of the Gamma process $\Gamma^{G, k, n}$, the first term on the right-hand side of \eqref{eq : Gamma k, n}.
	\end{thm}
	
	\begin{proof}
		We first introduce the notation
		\begin{equation} 	\label{def : xi G, k, n}
			\xi^{G, k, n} := \frac{\sum_{j=1}^{N_{\tau_{k-1}}}S_j(\tau_{k-1})}{G^{k, n}(\mu_{\tau_{k-1}+}) \sum_{j=1}^n S_j(\tau_{k-1}+)}
		\end{equation}
		such that 
		\begin{equation}	\label{eq : delta xi}
			\delta^{G, \psi, k, n} = V^{\psi, k-1, N_{\tau_{k-1}}}(\tau_{k-1}) \xi^{G, k, n}	
		\end{equation}
		holds from \eqref{def : delta} for each $(k, n) \in \N^2$.
		
		Since dissection sets are disjoint, we have the identity $\log V^{\psi}(t) = \log V^{\psi, k, n}(t)$ from \eqref{def : relative wealth w.r.t market} for the given pair $(t, \omega)$~(omitting $\omega$). We repeatedly apply \eqref{eq : V psi equal to GE} and \eqref{eq : delta xi} to derive
		\begin{align*}
			\log V^{\psi, k, n}(t) &= \log \delta^{G, \psi, k, n} + \log G^{k, n}(\mu_t) + \log E^{G, k, n}(t)
			\\
			&= \log V^{\psi, k-1, N_{\tau_{k-1}}}(\tau_{k-1}) + \log \xi^{G, k, n} + \log G^{k, n}(\mu_t) + \log E^{G, k, n}(t)
			\\
			&= \, \cdots
			\\
			&= \log V^{\psi, 0, N_0}(0) + \sum_{\ell=1}^{k-1} \bigg[ \log \xi^{G, \ell, N_{\tau_{\ell}}} + \log G^{\ell, N_{\tau_{\ell}}}(\mu_{\tau_{\ell}}) + \log E^{G, \ell, N_{\tau_{\ell}}}(\tau_{\ell}) \bigg]
			\\
			& \qquad \qquad \qquad \qquad \qquad + \bigg[ \log \xi^{G, k, n} + \log G^{k, n}(\mu_t) + \log E^{G, k, n}(t) \bigg].
		\end{align*}
		We rearrange some of the terms on the right-hand side by using the identity $V^{\psi, 0, N_0}(0) = G^0(\mu_0)$ and the definitions \eqref{def : xi G, k, n}, \eqref{def : sigma k, n} to obtain the correction term $C(t)$:
		\begin{align*}
			& \quad \log V^{\psi, 0, N_0}(0) + \sum_{\ell=1}^{k-1} \bigg[ \log \xi^{G, \ell, N_{\tau_{\ell}}} + \log G^{\ell, N_{\tau_{\ell}}}(\mu_{\tau_{\ell}}) \bigg] + \log \xi^{G, k, n}
			\\
			&= \sum_{\ell=1}^k \Bigg[ \log \bigg( \frac{\sum_{j=1}^{N_{\tau_{\ell-1}}} S_j(\tau_{\ell-1})}{\sum_{j=1}^{N_{\tau_{\ell}}}S_j(\tau_{\ell-1}+)} \bigg) - \log \bigg( \frac{G^{\ell, N_{\tau_{\ell}}}(\mu_{\tau_{\ell-1}+})}{G^{\ell-1, N_{\tau_{\ell-1}}}(\mu_{\tau_{\ell-1}})} \bigg) \Bigg]
			=\sum_{\ell=1}^k \log \bigg( \frac{\sigma^{\ell, N_{\tau_{\ell}}} \, G^{\ell-1, N_{\tau_{\ell-1}}}(\mu_{\tau_{\ell-1}})}{G^{\ell, N_{\tau_{\ell}}}(\mu_{\tau_{\ell-1}+})} \bigg).
		\end{align*}
		For the remaining terms, we use \eqref{def : E k, n}, \eqref{def : stochastic exponential}, and Theorem II.13 of \cite{Protter}, to derive
		\begin{equation}   \label{eq : log E dissection}
			\log E^{G, k, n}(t) = \int_{\tau_{k-1}+}^t \frac{d\Gamma^{G, k, n, c}(s)}{G^{k, n}(\mu_{s})} + \sum_{\tau_{k-1} < s \le t} \log \bigg( 1+ \frac{\Delta \Gamma^{G, k, n}(s)}{G^{k, n}(\mu_s)} \bigg),
		\end{equation}
		and similar expressions for $\log E^{G, \ell, N_{\tau_{\ell}}}(\tau_{\ell})$ for $\ell = 1, \cdots, k-1$. Thus, the result \eqref{eq : V psi decomposition} follows.
	\end{proof}
	
	\smallskip
	
	\subsubsection{Balance condition}  \label{subsubsec : balanced}
	
	We first note that the component $\varphi^{(k, n)}_i(t)$ of additively generated trading strategy $\varphi$ in \eqref{def : varphi k, n} has an alternative representation on the $(k, n)$-dissection set $\rrbracket \tau_{k-1}, \tau_k \rrbracket \cap (\R_+ \times \Omega^{k, n})$:
	\begin{equation}   \label{eq : varphi k, n alternative}
		\varphi^{(k, n)}_i(t) = \vartheta^{(k, n)}_i(t) - \sum_{j=1}^n \vartheta^{(k, n)}_j(t) \mu^{(k, n)}_j(t) + V^{\varphi, k, n}(t),
	\end{equation}
	from the identities \eqref{eq : jump relationships} and \eqref{eq : V varphi, k, n}. Moreover, the last term $V^{\varphi, k, n}(t)$ can be replaced with the right-hand side of \eqref{eq : V varphi decomposition}, so the strategy $\varphi^{(k, n)}(t)$ is expressed in terms of the original generating family $G$ and its gradient, the excess growth and the correction term, up to time $t$.
	
	Similarly, the component $\psi^{(k, n)}_i(t)$ of multiplicatively generated strategy $\psi$ in \eqref{def : psi k, n} also admits an alternative representation from the identities \eqref{eq : V psi diff} and \eqref{eq : V psi at tau k-1+}:
	\begin{equation}    \label{eq : psi k, n alternative}
		\psi^{(k, n)}_i(t) = \eta^{(k, n)}_i(t) - \sum_{j=1}^n \eta^{(k, n)}_j(t) \mu^{(k, n)}_j(t) + V^{\psi, k, n}(t),	
	\end{equation}
	and we can apply the decomposition \eqref{eq : V psi decomposition} to the last term of \eqref{eq : psi k, n alternative}. Furthermore, for $\psi$, there is a condition to impose on each generating function $G^{k, n}$ to further simplify the representation of the corresponding portfolio $\pi^{\psi}$.
	
	\begin{defn} [Balance condition]	\label{Def : balance}
		For each $n \in \N$, an $n$-dimensional differentiable function $f$ is called \textit{balanced}, if $f$ satisfies
		\begin{equation*}
			f(x) = \sum_{i=1}^{n} x_i \big(\partial_i f(x)\big) ~~\qquad \forall \, x \in \R^{n}.
		\end{equation*}
	\end{defn}
	
	This balance condition is known not only to simplify the representations of functionally generated trading strategies and corresponding portfolios~\citep{Karatzas:Ruf:2017, Karatzas:Kim}, but also to handle discontinuities of an additional process other than market weights when generating trading strategies~\citep{Kim:market-to-book}. The balance condition shall also be used in Section~\ref{subsec : open market FGP}, since it enables rank-based portfolios to invest only in the fixed number of large capitalization stocks in open markets, adopting the idea of \cite{Karatzas:Kim2}.
	
	When $G^{k, n}$ satisfies the balance condition, it is straightforward to check the identity
	\begin{equation*}
		\sum_{i=1}^n \eta^{(k, n)}_i(t) \mu^{(k, n)}_i(t-) = \widehat{G}^{k, n}(\mu_{t-}) E^{G, k, n}(t-)
	\end{equation*}
	for every $t \in \rrbracket \tau_{k-1}, \tau_k \rrbracket$. The last equality, together with \eqref{eq : psi k, n alternative} and \eqref{eq : psi wealth}, yields 
	\begin{equation}    \label{eq : balanced psi}
		\psi^{(k, n)}_i(t-) = \eta^{(k, n)}_i(t-) = \partial_i \widehat{G}^{k, n}(\mu_{t-}) E^{G, k, n}(t-).
	\end{equation}
	Recalling the definition~\eqref{def : portfolio weights of TS} of portfolio weights and the fact that $\psi^{(k, n)}$~(and its portfolio weight $\pi^{(k, n)}$) is left-continuous, we have the following result.
	
	\begin{cor}	\label{cor : MGTS balanced}
		If every generating function $G^{k, n}$ satisfies the balance condition, then the portfolio weight $\pi^{(k, n)}_i$ of multiplicatively generated trading strategy $\psi^{(k, n)}_i$ is expressed as
		\begin{equation}    \label{eq: portfolio weights of MTS balanced}
			\pi^{(k, n)}_i(t) = \frac{\partial_i G^{k, n}(\mu_{t-}) \mu^{(k, n)}_i(t-)}{\sum_{j=1}^n \partial_j G^{k, n}(\mu_{t-}) \mu^{(k, n)}_j(t-)}, \quad i = 1, \cdots, n, \quad \text{on each } (k, n) \text{-dissection set}.
		\end{equation}
	\end{cor}
	Note that the right-hand side depends only on the market weight process and on the original generating function $G$~(not its measurable modification $\widehat{G}$), even though the components of the trading strategy in \eqref{eq : balanced psi} depend on $\widetilde{G}$, or a normalizing random variable $\delta^{G, \psi, k, n}$.
	
	We shall see in Section~\ref{subsec : examples} that the generating functions of so-called \textit{diversity-weighted} and \textit{equally-weighted} portfolios~(Examples~\ref{ex : diversity weighted}, \ref{ex : equally weighted}) satisfy the balance condition.
	
	\smallskip
	
	\subsubsection{Self-financing market portfolio}  \label{subsubsec : Sf mu}
	
	The relative wealth $V^{\vartheta}(t)$ of the functionally generated trading strategy $\vartheta$, which appears in Theorems~\ref{thm : AGTS alternative} and \ref{thm : MGTS alternative}, measures the wealth of $\vartheta$ relative to the total market capitalization at time $t$, as defined in \eqref{def : relative wealth}. By its construction, the trading strategy $\vartheta$ is designed to be self-financing at all times, whereas the total market capitalization~(the denominator in \eqref{def : relative wealth}) can undergo a jump at each moment of dimensional change.
	
	Moreover, the correction term $C(t)$ in \eqref{eq : C mul} of a multiplicatively generated strategy can be further decomposed as
	\begin{equation}    \label{eq : decomposition of C}
		C(t) = \sum_{\ell=1}^k \log \sigma^{\ell, N_{\tau_{\ell}}} + \sum_{\ell=1}^k \log \bigg( \frac{G^{\ell-1, N_{\tau_{\ell-1}}}(\mu_{\tau_{\ell-1}})}{G^{\ell, N_{\tau_{\ell}}}(\mu_{\tau_{\ell-1}+})} \bigg) =: C_{TM}(t) + C_G(t).
	\end{equation}
	Note that the first term $C_{TM}(t)$ on the right-hand side is independent of the generating function $G$. It accumulates the (log values of) jumps in the total market capitalization, whereas the second term $C_G(t)$ captures the (log values of) jumps in the generating function, at all previous dimensional changes up to time $t$.
	
	These observations give rise to the following notion of self-financing market portfolio. Consider a family $\mathcal{G} := \{G^0\} \cup \{G^{k, n}\}_{(k, n) \in \mathbb{N}^2}$ of functions, given by $G^0(x) = \sum_{i=1}^{N_0} x_i$ and $G^{k, n}(x) = \sum_{i=1}^{n} x_i$. It is easy to check the two trading strategies, additively and multiplicatively generated from $\mathcal{G}$, coincide; let us denote it by $\xi$. Its corresponding portfolio $\pi \equiv \pi^{\xi}$ is computed as 
	\begin{equation}    \label{def : self-financing market portfolio}
		\pi^{(k, n)}(t) = \mu^{(k, n)}(t-)
	\end{equation}
	for every $(k, n) \in \mathbb{N}^2$ from Corollary~\ref{cor : MGTS balanced}.
	
	\begin{defn} [Self-financing market portfolio]  \label{Def : self-financing market portfolio}
		The trading strategy $\psi$ and its portfolio $\xi$ of \eqref{def : self-financing market portfolio} are called the \textit{self-financing market trading strategy} and \textit{self-financing market portfolio}, respectively. We denote them by $\xi_{sf}$ and $\mu_{sf}$.
	\end{defn}
	
	From Theorem~\ref{thm : MGTS alternative} and \eqref{eq : decomposition of C}, the log relative wealth of $\xi_{sf}$ only contains the correction term $C_{TM}(t)$, in other words, $\log G^{k, n}(\mu_{\cdot}) \equiv EG(\cdot) \equiv C_G(\cdot) \equiv 0$, and
	\begin{equation}    \label{eq : relative wealth of xi}
		\log V^{\xi_{sf}}(t) = C_{TM}(t) = \sum_{\ell=1}^k \log \sigma^{\ell, N_{\tau_{\ell}}}, \qquad \text{or} \qquad V^{\xi_{sf}}(t) = \sigma^{k, n}_{1, N_{\tau_1}},
	\end{equation}
	for any $(t, \omega)$ on the $(k, n)$-dissection set, recalling the notation \eqref{def : sigma k, n, i}. We emphasize again that the above term $C_{TM}(t)$ appears in the decomposition \eqref{eq : V psi decomposition} of every functionally generated trading strategy, independent of the generating function. Moreover, from the identity \eqref{eq : balanced psi}, we can easily deduce that all components of $\xi_{sf}^{(k, n)}(t)$ are equal to $V^{\xi_{sf}, k, n}(t) = \sigma^{k, n}_{1, N_{\tau_1}}$. Therefore, $\xi_{sf}$ is just a buy-and-hold trading strategy investing equal shares in all the extant stocks between dimensional jumps; at each dimensional jump, the number of shares holding for every stock is adjusted according to the jumps in total market capitalization, redistributing its wealth to the next constituents of the market in a self-financing way. This also explains how the portfolio $\pi$ in \eqref{def : self-financing market portfolio} distributes its wealth according to relative capitalizations of extant stocks at all times.
	
	If an investor believes that the self-financing market portfolio should be the baseline to compare the relative performance of the functionally generated trading strategy $\vartheta$, one can easily compute the ratio of $V^{\vartheta}$ to $V^{\xi_{sf}}$. We note here that $V^{\xi_{sf}}(t)$ takes positive values for every $t$, due to the positivity assumption on the capitalization process in Definition~\ref{Def : price process RCLL}. Let us denote
	\begin{equation}
		U^{\vartheta}(t) := \frac{V^{\vartheta}(t)}{V^{\xi_{sf}}(t)}, \qquad t \ge 0,
	\end{equation}
	the relative wealth of trading strategy $\vartheta$ with respect to the self-financing market portfolio. As in the manner of \eqref{def : relative wealth w.r.t market}, the process $U^{\vartheta}$ can be expressed as a collection of $U^{\vartheta, k, n}(t) := V^{\vartheta, k, n}(t) / V^{\xi_{sf}, k, n}(t)$ on each $(k, n)$-dissection set. In particular, for a multiplicatively generated trading strategy $\psi$ in Theorem~\ref{thm : MGTS alternative}, we have the decomposition of the log relative wealth of $\psi$ with respect to $\xi_{sf}$
	\begin{equation}   \label{eq : U psi decomposition}
		\log U^{\psi}(t) = \log G^{k, n}(\mu_t) + EG(t) + C_G(t),
	\end{equation}
	with the same excess growth term $EG$ of \eqref{eq : GR mul}, but with the correction term $C(t)$ replaced by $C_G(t)$ of \eqref{eq : decomposition of C}.
	
	\smallskip
	
	\subsubsection{Long-term growth of relative wealth processes} \label{subsubsec : long term growth}
	
	In this part, we shall compare the behavior of the excess growth term $EG$ in \eqref{eq : GR add} with that of \eqref{eq : GR mul}, and point out some disadvantages of the additively generated trading strategies over the multiplicatively generated ones.
	
	The excess growth terms $EG(t)$ of \eqref{eq : GR add} and \eqref{eq : GR mul} contribute to the long-term growth of the relative wealth processes $V^{\varphi}$ and $V^{\psi}$. We recall from the representation \eqref{eq : Gamma k, n} that each Gamma process $\Gamma^{G, \ell, N_{\tau_{\ell}}}(\cdot)$ is nondecreasing if each $G^{\ell, N_{\tau_{\ell}}}$ is a concave function. Therefore, both of the expressions \eqref{eq : GR add} and \eqref{eq : GR mul} are nonnegative, if every generating function in the family $\{G^{k, n}\}_{(k, n) \in \N^2}$ is concave. However, even with concave generating functions, the former excess growth term \eqref{eq : GR add} is generally not nondecreasing in time. For an additively generated strategy $\varphi$, the expression \eqref{eq : GR add} may decrease at the next dimensional change $\tau_{k}$, if the next ratio $\sigma^{k+1, \widetilde{n}}$ of total market capitalization~(where $\widetilde{n} \in \N$ is a next market dimension) is significantly less than $1$. This is because all the accumulations $\Gamma^{G, \ell, N_{\tau_{\ell}}}(\tau_{\ell})$ of Gamma processes for $\ell = 1, \cdots, k-1$ in the `past' epochs, are affected by the ratios of the total market capitalization at `future' dimensional changes $\tau_k, \tau_{k+1}, \cdots $, as the quantity $\sigma^{k, n}$ will be multiplied to the whole expression of \eqref{eq : GR add} at the moment $\tau_k$ of the next dimensional jump, and so on. In this sense, we can say that the excess growth of \eqref{eq : GR add} is only `piecewise nondecreasing' in time $t$ with a potential of occasional plummet at the moments of dimensional changes.
	
	In contrast, the excess growth \eqref{eq : GR mul} of the multiplicatively generated strategy $\psi$ is independent of the ratios $\sigma^{\ell, N_{\tau_{\ell}}}$. Therefore, every term of \eqref{eq : GR mul} is nonnegative and nondecreasing in time, if all the generating functions are positive and concave. This shows that the long-term growth of a multiplicatively generated strategy is not affected by the shocks that arise from dimensional jumps in the market.
	
	Moreover, the correction term $C$ of $\psi$ has a nice decomposition \eqref{eq : decomposition of C}, which separates the universal term $C_{TM}$ with the other term $C_G$, whereas that of $\varphi$ in \eqref{eq : C add} does not admit such decomposition. Thus, there is no simple representation of the relative wealth $U^{\varphi}$ of the additively generated strategy $\varphi$ with respect to the self-financing market portfolio; $U^{\varphi}$ should depend on the ratios of \eqref{def : sigma k, n}. However, $U^{\psi}$ in \eqref{eq : U psi decomposition} is free of such ratios, so the excess growth $EG(\cdot)$ contributes to the long-term outperformance of $\psi$ with respect to the self-financing market portfolio if we expect long-term stability of the remaining part $\log G(\mu_{\cdot}) + C_G(\cdot)$, which only depends on the generating function $G$.
	
	Because of the aforementioned drawbacks of the additively generated strategies, we shall provide in the next section examples of multiplicatively generated portfolios with empirical evolution of each term that appears in the decompositions \eqref{eq : V psi decomposition} and \eqref{eq : U psi decomposition}.

	\bigskip
	
	\section{Empirical analyses} \label{sec : empirical}
	
	Our main purpose of this section is to examine how 
	dimensional changes in the equity market affect portfolio performances. More specifically, using real stock market data, we would like to identify the correction terms $C$, $C_{TM}$, and $C_G$ in the equations \eqref{eq : V psi decomposition}, \eqref{eq : relative wealth of xi}, and \eqref{eq : U psi decomposition} of the log relative wealth processes $V^{\psi}$ and $U^{\psi}$ for some classical portfolios, all of which are multiplicatively generated.
	
	For the reason that we shall use daily stock price data, we first develop the previous theory of functional generation of portfolios in a discrete-time market model as a special case of the general theory we studied in Section~\ref{sec : FGP}, for a more precise comparison of the correction terms. When computing the excess growth term of \eqref{eq : GR mul}, we are subject to approximation errors for measuring the integral terms involving the Gamma processes from a given discrete-time dataset. Since we shall consider a long period of time~(40 years) for analyzing portfolios, the approximation errors tend to be accumulated over a long time which hinders the exact analysis of the correction terms.
	
	\medskip
	
	\subsection{Discrete-time model}	\label{subsec : discrete time}
	
	Since we allow jumps in the capitalization~(and market weight) process, all of the results in the previous sections can be easily reformulated in a discrete-time equity market model. For the purpose of introducing a discrete-time model, we shall fix in this subsection an index set $\mathbb{T} := (t_j)_{j=0}^{\infty}$ satisfying $0 = t_0 < t_1 < \cdots$. Moreover, for a given $\U$-valued discrete-time stochastic process $\{X(t_j) : t_j \in \mathbb{T}\}$ on a probability space $(\Omega, \mathcal{F}, \mathbb{P})$, let us consider a sequence of stopping times $(\tau_k)_{k \ge 0} \subset \mathbb{T}$ at which the dimension of $X$ changes, i.e., 
	\begin{equation*}
		\tau_0 := 0, \qquad 
		\tau_k := \inf \Big\{ t_j > \tau_{k-1} : \dim \big(X(t_j)\big) \neq \dim \big(X(t_{j-1})\big) \Big\}, \qquad  k \in \N.
	\end{equation*}
	Then, we can construct a $\U$-valued piecewise-constant process $\widetilde{X}$ by the recipe $\widetilde{X}(0) = X(0)$ and for $j = 1, 2, \cdots$
	\begin{align}
		\text{if } t_j \in (\tau_k)_{k \ge 1} : \qquad &\widetilde{X}(t) = X(t_{j-1}) \quad \text{for} \quad t_{j-1} < t \le t_j,       \label{def : X tilde}
		\\
		\text{if } t_j \notin (\tau_k)_{k \ge 1} : \qquad &\widetilde{X}(t) = X(t_{j-1}) \quad \text{for} \quad t_{j-1} < t < t_j, \quad \text{and} \quad \widetilde{X}(t_j) = X(t_j),      \nonumber
	\end{align}
	such that $\widetilde{X}$ is right-continuous between the dimensional jumps and left-continuous at the dimensional jumps. For the filtration, we consider a right-continuous extension $\mathcal{F}_t = \mathcal{F}^X_{t_{j-1}} := \sigma(X_{t_0}, \cdots, X_{t_{j-1}})$ for $t_{j-1} \le t < t_j$ of the natural filtration $\mathcal{F}^X$ of $X$. From this construction, it is straightforward to verify that $\widetilde{X}$ is a $\mathcal{F}_t$-progressive process and the sequence $(\tau_k)_{k \ge 0}$ is the (minimal) reset sequence of $\widetilde{X}$ satisfying the conditions of Definition~\ref{Def : reset sequence}, thus $\widetilde{X}$ is recognized as a piecewise RCLL semimartingale in the sense of Definition~\ref{Def : piecewise semimartingale}.
	
	Therefore, when observing capitalizations of an equity market (of stochastic dimension) only at the discrete times $\mathbb{T}$, such that the capitalization process $S$ is given as a $\U$-valued discrete stochastic process, we shall consider its piecewise-constant continuous-time version $\widetilde{S}$ with the corresponding market weight process $\widetilde{\mu}$ of $\widetilde{S}$, having the dimension process $N$, to apply the previous theory for generating trading strategies from $\widetilde{\mu}$. In other words, for any $\omega \in \Omega$, we shall identify the trajectory $\{\mu_{t_j}(\omega)\}_{t_j \in \mathbb{T}}$ of the discrete process with its continuous-time version $\{\widetilde{\mu}_t(\omega)\}_{t \ge 0}$.
	
	Now that $\widetilde{\mu}$ is piecewise-constant, the continuous part of every integral with the integrator involving (dissections of) $\widetilde{\mu}$ vanishes, and we can rewrite all the expressions in the earlier sections in terms of the original discrete-time process $\mu$ from the construction \eqref{def : X tilde}. In particular, for $(t_j, \omega) \in \rrbracket \tau_{k-1}, \tau_k \llbracket \cap (\R_+ \times \Omega^{k, n})$, the first self-financing condition \eqref{eq : self-financing} can be rewritten as
	\begin{equation*}
		\sum_{i=1}^n \vartheta_i(t_{j}) \mu_i(t_{j})
		= \sum_{i=1}^n \vartheta_i(t_{j-1}) \mu_i(t_{j}),
	\end{equation*}
	and each $(k, n)$-Gamma process of \eqref{eq : Gamma k, n} is represented as
	\begin{equation*}
		\Gamma^{G, k, n}(t_j) = 
		- \sum_{\tau_{k-1} < t_{q} \le t_j} d_{B, G^{k, n}}\big( \mu_{t_{q}}, \, \mu_{t_{q-1}}  \big).
	\end{equation*}
	The equation \eqref{eq : log E dissection} is now simplified to
	\begin{equation*}
		\log E^{G, k, n}(t_j) = \sum_{\tau_{k-1} < t_{q} \le t_j < \tau_k} \log \bigg( 1+ \frac{\Delta \Gamma^{G, k, n}(t_{q})}{G^{k, n}(\mu_{t_{q}})} \bigg)
		= \sum_{\tau_{k-1} < t_{q} \le t_j < \tau_k} \log \bigg( 1 - \frac{d_{B, G^{k, n}}\big( \mu_{t_{q}}, \, \mu_{t_{q-1}} \big)}{G^{k, n}(\mu_{t_{q}})} \bigg).
	\end{equation*}
	Therefore, Theorem~\ref{thm : MGTS alternative} holds with the decomposition \eqref{eq : V psi decomposition} (and \eqref{eq : decomposition of C}), rewritten as
	\begin{align}   
		\log V^{\psi}(t_j) &= \log G^{k, n}(\mu_{t_j}) + EG(t_j) + C_{TM}(t_j) + C_G(t_j), \qquad \text{where}      \label{eq : V psi decomposition discrete}
		\\
		EG(t_j) &= \sum_{\ell = 1}^{k-1} \sum_{\tau_{\ell-1} < t_q < \tau_{\ell}} \Big[ eg(\ell, N_{\tau_{\ell}}, t_q) \Big]
		+ \sum_{\tau_{k-1} < t_q \le t_j < \tau_k} \Big[ eg(k, n, t_q) \Big],        \label{def : cumul GR}
		\\
		C_{TM}(t_j) &= \sum_{\ell=1}^k \log  \sigma^{\ell, N_{\tau_{\ell}}}, \qquad
		C_G(t_j) = \sum_{\ell=1}^k \log \bigg( \frac{G^{\ell-1, N_{\tau_{\ell-1}}}(\mu_{\tau_{\ell-1}-})}{G^{\ell, N_{\tau_{\ell}}}(\mu_{\tau_{\ell-1}})} \bigg),  \label{def : cumul C}
	\end{align}
	with the convention $G^{0, N_0}(\mu_{\tau_0-}) = G^{1, N_{\tau_{1}}}(\mu_{\tau_0})$ and the notation 
	\begin{equation*}
		eg(\ell, N_{\tau_{\ell}}, t_q) := \log \bigg( 1-\frac{d_{B, G^{\ell, N_{\tau_{\ell}}}}\big( \mu_{t_{q}}, \, \mu_{t_{q-1}} \big)}{G^{\ell, N_{\tau_{\ell}}}(\mu_{t_q})} \bigg),
	\end{equation*}
	for a given pair $(t_j, \omega)$ in each $(k, n)$-dissection set for $(k, n) \in \N^2$ with $t_j < \tau_k$. Here and in what follows, $\tau_{k}-$ represents the last time index before $\tau_{k}$, whereas $\tau_{k}+$ is the next index after $\tau_{k}$, i.e., $\tau_{k}- = t_{j-1}$ and $\tau_k+ = t_{j+1}$ if $\tau_{k} = t_j$, with the convention $\tau_0- = \tau_0 = 0$. The expression $eg(\ell, N_{\tau_{\ell}}, t_q)$ represents the \textit{discrete excess growth} from $t_{q-1}$ to $t_q$, when the market is in the $\ell$-th epoch with dimension $N_{\tau_{\ell}}$, thus $EG(t_j)$ is the \textit{cumulative discrete excess growth} until time $t_j$.
	
	Since there is a dimensional jump between $\tau_{k}-$ and $\tau_k$, the last three terms of the equation \eqref{eq : V psi decomposition discrete} remains unchanged, i.e., $EG(\tau_k) = EG(\tau_k-)$, $C_{TM}(\mu_{\tau_k}) = C_{TM}(\mu_{\tau_k-})$, and $C_{G}(\mu_{\tau_k}) = C_{G}(\mu_{\tau_k-})$, due to its construction \eqref{def : X tilde}. However, during the next time window between $\tau_k$ and $\tau_k+$, the increment of the last term $C_G(\cdot)$ on the right-hand side of \eqref{eq : V psi decomposition discrete}, is offset by the previous increment of the first term $\log G(\mu_{\cdot})$:
	\begin{equation}    \label{eq : equal increments}
		C_G(\tau_{k}+) - C_G(\tau_{k}) 
		= - \Big\{ \log G^{k+1, N_{\tau_{k+1}}}(\mu_{\tau_{k}}) - \log G^{k, N_{\tau_{k}}}(\mu_{\tau_{k}-}) \Big\}.
	\end{equation}
	In other words, every summand of $C_{G}(\cdot)$ in \eqref{def : cumul C}, which occurs due to dimensional jump between $\tau_{\ell-1}-$ and $\tau_{\ell-1}$, is canceled out by the increment of $\log G(\mu_{\cdot})$ during the same period. Therefore, the decomposition \eqref{eq : V psi decomposition discrete} - \eqref{def : cumul C} can be rewritten as
	\begin{align*}
		\log V^{\psi}(t_j) = \sum_{\ell = 1}^{k-1} &\sum_{\tau_{\ell-1} < t_q < \tau_{\ell}} \bigg[ \log G^{\ell, N_{\ell}}(\mu_{t_q}) - \log G^{\ell, N_{\ell}}(\mu_{t_{q-1}}) \bigg] 
		\\
		+ &\sum_{\tau_{k-1} < t_q \le t_j < \tau_k} \bigg[ \log G^{k, n}(\mu_{t_q}) - \log G^{k, n} (\mu_{t_{q-1}}) \bigg] + EG(t_j) + C_{TM}(t_j),
	\end{align*}
	i.e., the evolution of the log generating function and the excess growth accumulated \textit{between} the past dimensional jumps and the universal correction term $C_{TM}$ accumulated \textit{at} those dimensional jumps affect the relative wealth $V^{\psi}$.
	
	
	For the log relative wealth $U^{\psi}$ in \eqref{eq : U psi decomposition}, we have the same representation in this discrete-time setting, except for the last correction term:
	\begin{align*}
		\log U^{\psi}(t_j) = \sum_{\ell = 1}^{k-1} &\sum_{\tau_{\ell-1} < t_q < \tau_{\ell}} \bigg[ \log G^{\ell, N_{\ell}}(\mu_{t_q}) - \log G^{\ell, N_{\ell}}(\mu_{t_{q-1}}) \bigg] 
		\\
		+ &\sum_{\tau_{k-1} < t_q \le t_j < \tau_k} \bigg[ \log G^{k, n}(\mu_{t_q}) - \log G^{k, n} (\mu_{t_{q-1}}) \bigg] + EG(t_j),
	\end{align*}
	The above derivations generalize the results of \cite{Wong:optimaltransport}, where the functional generation of portfolios is developed in a discrete-time market model of a fixed dimension. Here, we also refer to \cite{Campbell:Wong}, \cite{Pal:Wong:Geometry}, and \cite{Wong:Optimization}, for different aspects of the discrete-time setup of portfolio generation in a market of fixed dimension.
	
	\medskip
	
	\subsection{Examples}	\label{subsec : examples}
	We provide in this subsection some of the classical examples in the Stochastic portfolio theory under the discrete-time model. Empirical analyses of these examples will be given in the following subsections.
	
	When introducing a portfolio-generating function $G : \U \rightarrow \R$ as a piecewise function of the market weight process $\mu$ in the following examples, we shall only specify an $n$-dimensional `representative piece' $G^{k, n} : \R^n \rightarrow \R$, for a fixed pair $(k, n) \in \N^2$. All the other pieces can be easily inferred from this representative form; in fact, $G^{k, n} = G^{\ell, n}$ for any $\ell \in \N$. Moreover, we assume in the following that a general time index $t_j$ is in the $k$-th epoch with market dimension equal to $n$, i.e., the pair $(t_j, \omega)$ belongs to the $(k, n)$-dissection set $\rrbracket \tau_{k-1}, \tau_k \rrbracket \cap (\R_+ \times \Omega^{k, n})$.
	
	\smallskip
	
	\begin{example} [Diversity-weighted portfolio]  \label{ex : diversity weighted}
		For a fixed real number $p \in (0, 1]$, the function
		\begin{equation*}
			G^{k, n}(x) = \Big(\sum_{i=1}^n x_i^p \Big)^{\frac{1}{p}}
		\end{equation*}
		is balanced~(Definition~\ref{Def : balance}) and multiplicatively generates the portfolio from \eqref{eq: portfolio weights of MTS balanced}
		\begin{equation}    \label{eq : diversity pi}
			\pi_i(t_j) = \frac{\big(\mu_i(t_{j-1})\big)^p}{\sum_{\ell=1}^n \big(\mu_{\ell}(t_{j-1})\big)^p}, \qquad i = 1, \cdots, n.
		\end{equation}
		The parameter $p$ determines a measure of diversity; a portfolio with a smaller value of $p$ invests more wealth in smaller stocks. We also note that the case $p=1$ corresponds to the self-financing market portfolio in Section~\ref{subsubsec : Sf mu}. After some computation, we obtain from \eqref{eq : V psi decomposition discrete} the following decomposition of the log-relative wealth of the corresponding trading strategy $\psi$
		\begin{equation}    \label{eq : log V diversity}
			\log V^{\psi}(t_j) = \frac{1}{p} \log \Big( \sum_{i=1}^n \big(\mu_i(t_j)\big)^p \Big) + EG(t_j) + C_{TM}(t_j) + C_G(t_j), \qquad \text{where}
		\end{equation}
		the excess growth $EG(t_j)$ is given as \eqref{def : cumul GR} with
		\begin{align*}
			eg(\ell, N_{\tau_{\ell}}, t_q) &= \log \bigg( \sum_{i=1}^{N_{\tau_{\ell}}} \mu_i(t_q) \big( \mu_i(t_{q-1}) \big)^{p-1} \bigg)
			- \frac{1}{p} \Bigg[ \log \bigg( \sum_{i=1}^{N_{\tau_{\ell}}} \big( \mu_i(t_{q}) \big)^{p} \bigg) \bigg( \sum_{i=1}^{N_{\tau_{\ell}}} \big( \mu_i(t_{q-1}) \big)^{p} \bigg)^{p-1} \Bigg],
		\end{align*}
		and the correction term
		\begin{equation*}
			C_G(t_j) = \frac{1}{p} \sum_{\ell=1}^k 
			\log \Bigg( \frac{\sum_{i=1}^{N_{\tau_{\ell-1}}} \big( \mu_i(\tau_{\ell-1}-) \big)^p}{\sum_{i=1}^{N_{\tau_{\ell}}} \big( \mu_i(\tau_{\ell - 1}) \big)^p} \Bigg).
		\end{equation*}
	\end{example}
	
	\smallskip
	
	\begin{example} [Equal-weighted portfolio]  \label{ex : equally weighted}
		The following balanced function 
		\begin{equation*}
			G^{k, n}(x) = \prod_{i=1}^n (x_i)^{\frac{1}{n}}
		\end{equation*}
		multiplicatively generates the portfolio from \eqref{eq: portfolio weights of MTS balanced}
		\begin{equation*}
			\pi_i(t_j) = \frac{1}{n}, \qquad i = 1, \cdots, n,
		\end{equation*}
		which invests the same proportions of current wealth in all the existing stocks. We note that this portfolio is the limit of the diversity-weighted portfolio \eqref{eq : diversity pi} when $p \rightarrow 0$. The decomposition \eqref{eq : V psi decomposition discrete} of the log-relative wealth $\log V^{\psi}(t_j)$ is then computed as
		\begin{equation}    \label{eq : log V equal}
			\log V^{\psi}(t_j) = \frac{1}{n} \sum_{i=1}^n \log \mu_i(t_j) + EG(t_j) + C_{TM}(t_j) + C_G(t_j), \qquad \text{where}
		\end{equation}
		the excess growth $EG(t_j)$ is given as \eqref{def : cumul GR} with
		\begin{align*}
			eg(\ell, N_{\tau_{\ell}}, t_q) &= \frac{1}{N_{\tau_{\ell}}} \sum_{i=1}^{N_{\tau_{\ell}}} \log \frac{\mu_i(t_{q-1})}{\mu_i(t_{q})} + 
			\log \bigg( \frac{1}{N_{\tau_{\ell}}} \sum_{i=1}^{N_{\tau_{\ell}}} \frac{\mu_i(t_{q})}{\mu_i(t_{q-1})} \bigg),
		\end{align*}
		and the correction term equal to
		\begin{equation*}
			C_G(t_j) = \sum_{\ell=1}^k \Bigg[
			\frac{1}{N_{\tau_{\ell-1}}} \sum_{i=1}^{N_{\tau_{\ell-1}}} \log \mu_i(\tau_{\ell-1}-) - \frac{1}{N_{\tau_{\ell}}} \sum_{i=1}^{N_{\tau_{\ell}}} \log \mu_i(\tau_{\ell-1}) \Bigg].
		\end{equation*}
	\end{example}
	
	\smallskip
	
	\begin{example} [Entropy-weighted portfolio]    \label{ex : entropy weighted}
		The function
		\begin{equation*}
			G^{k, n}(x) = -\sum_{i=1}^n x_i \log x_i
		\end{equation*}
		multiplicatively generates the portfolio from \eqref{def : eta k, n2}, \eqref{eq : psi k, n alternative}, \eqref{eq : psi wealth}, and \eqref{def : portfolio weights of TS}
		\begin{equation*}
			\pi_i(t_j) = \frac{-\mu_i(t_{j-1}) \log \mu_i(t_{j-1})}{-\sum_{\ell=1}^n \mu_{\ell}(t_{j-1}) \log \mu_{\ell}(t_{j-1})}, \qquad i = 1, \cdots, n.
		\end{equation*}
		The log-relative wealth $\log V^{\psi}(t_j)$ of \eqref{eq : V psi decomposition discrete} is decomposed as
		\begin{equation}    \label{eq : log V entropy}
			\log V^{\psi}(t_j) = \log \Big(-\sum_{i=1}^n \mu_i(t_j) \log \mu_i(t_j) \Big) + EG(t_j) + C_{TM}(t_j) + C_G(t_j), \qquad \text{with}
		\end{equation}
		\begin{align*}
			eg(\ell, N_{\tau_{\ell}}, t_q) = &\log \Bigg( \frac{-\sum_{i=1}^{N_{\tau_{\ell}}} \mu_i(t_q) \log \mu_i(t_{q-1})}{-\sum_{i=1}^{N_{\tau_{\ell}}} \mu_i(t_q) \log \mu_i(t_q)} \Bigg),
			\\
			C_G(t_j) = \sum_{\ell=1}^k \Bigg[
			\log \bigg( -\sum_{i=1}^{N_{\tau_{\ell-1}}} \mu_i(\tau_{\ell-1}-) &\log \mu_i(\tau_{\ell-1}-) \bigg)
			- \log \bigg( -\sum_{i=1}^{N_{\tau_{\ell}}} \mu_i(\tau_{\ell-1}) \log \mu_i(\tau_{\ell-1}) \bigg) \Bigg].
		\end{align*}
	\end{example}
	
	\medskip
	
	\subsection{Data description}	\label{subsec : data}
	
	Our data contain daily closing prices of the stocks listed on the New York Stock Exchange (NYSE) and American Stock Exchange (AMEX) during 40 years (10086 trading days) between 1982 January 4th and 2021 December 31st. These data were obtained from the Center for Research in Security Prices \citep{CRSP} database, accessed via the Wharton research data services. We refer to \cite{Ruf:Github} for a more extensive description of the CRSP database.
	
\begin{figure}[!htb]
		\centering	
		\subfloat[Dimensional change of NYSE, AMEX]
		{\includegraphics[width=.49\linewidth]{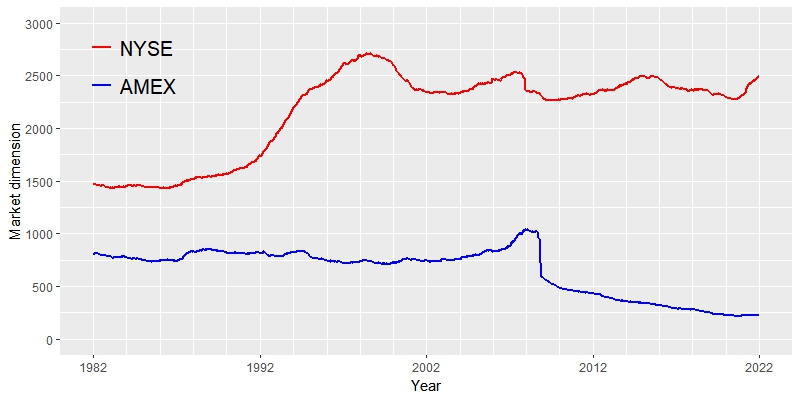}}
		\subfloat[$C_{TM}$ of self-financing market portfolio]
		{\includegraphics[width=.49\linewidth]{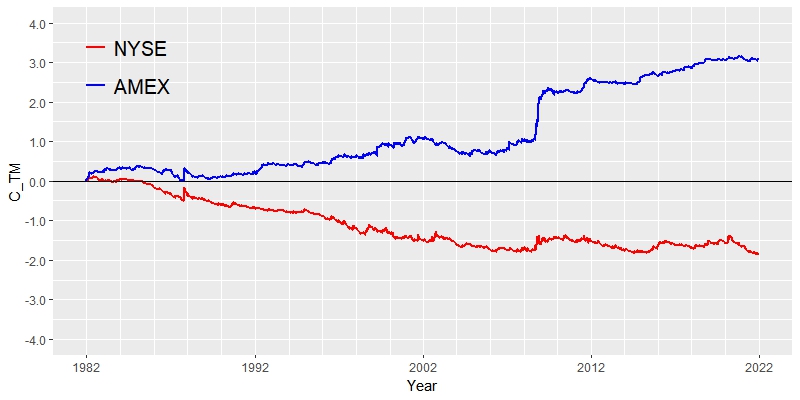}}
		\\
		\subfloat[Total market capitalization of NYSE]
		{\includegraphics[width=.49\linewidth]{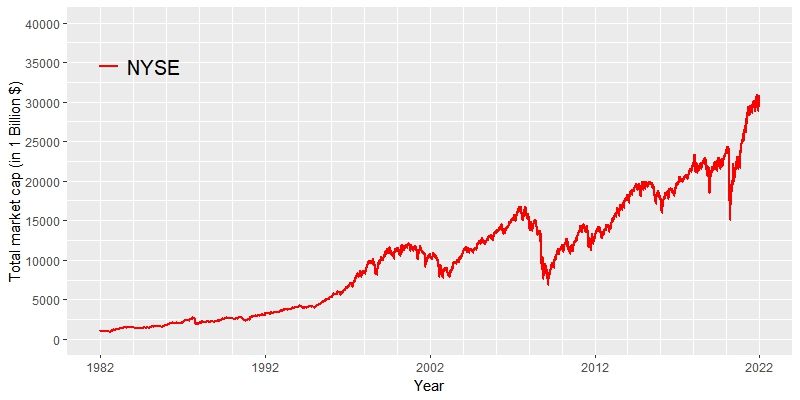}}
		\subfloat[Total market capitalization of AMEX]
		{\includegraphics[width=.49\linewidth]{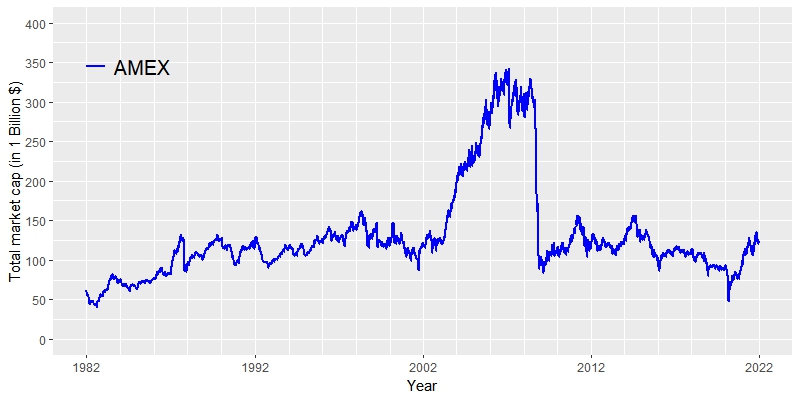}}
		\caption{Dimension and total market capitalization over time}
		\label{fig: overall market}
	\end{figure}

	In the discrete-time model introduced in Section~\ref{subsec : discrete time}, each element $t_j$ of the time index set $\mathbb{T} = (t_j = j)_{j=0}^{10085}$ represents the $j$-th trading day. Figure~\ref{fig: overall market} describes the evolutions of dimensional changes, the universal correction term $C_{TM}(t)$ in \eqref{eq : relative wealth of xi}, and the total market capitalizations of the two stock exchanges over $40$ years. In Figure~\ref{fig: overall market} (a), there were $6015$ and $3920$ jumps in dimension during $10086$ days for NYSE and AMEX respectively; in other words, on average, dimensional changes occurred every $1.68$ and $2.57$ trading days, respectively.
	
	Since the number of listed stocks on NYSE has increased over $40$ years, the correction term $C_{TM}$ of \eqref{eq : relative wealth of xi} for NYSE in Figure~\ref{fig: overall market} (b) exhibits overall decreasing movement. For the AMEX graph, we can observe the opposite tendency. A small hike near the year 1987 in the two graphs of Figure~\ref{fig: overall market} (b) is on account of the stock market crash in October 1987, known as Black Monday.
	
	In the graphs of AMEX in Figure~\ref{fig: overall market}, there was a significant delisting event in $2008$, which led to noticeable decreases in market dimension and total capitalization. The main reason is that NYSE Euronext, the multinational financial corporation which operates NYSE, acquired AMEX in $2008$, and AMEX is now known as the NYSE American. During this process of acquisition, many stocks are liquidated or dropped from the exchange, due to various reasons, e.g. failure to meet the new exchange’s financial guidelines for continued listing, or just company's request. The stock market crash in $2008$ is also responsible for the huge drop in total market capitalization of the two exchanges.
	
	We finally mention here that tradings of the portfolios in the next subsection are made every trading day~(daily rebalance) without any transaction costs, in order to precisely measure how much the correction terms impact the long-run relative performance of the portfolios in the decomposition identities. Since finding an outperforming portfolio~(or a relative arbitrage in Section~\ref{subsubsec : long term growth}) is not our goal, no transaction costs are considered in our empirical analysis. In the presence of transaction costs, trading frequency should be wisely determined by actual portfolio managers to maximize their profit, and we refer to \cite{Ruf:Xie:transaction} in this direction of study.
	
	\medskip
	
	\subsection{Empirical results}  \label{subsec : empirical results}
	
	We now provide the empirical results of individual terms in the decompositions \eqref{eq : log V diversity}, \eqref{eq : log V equal}, and \eqref{eq : log V entropy}, of the log relative wealth processes of the classical portfolios introduced in Section~\ref{subsec : examples}, using the dataset described in Section~\ref{subsec : data}. The evolutions of the $\log G$ term~(the first term on the right-hand side) and the excess growth ($EG$) will be illustrated in green and blue, respectively. The universal correction term $C_{TM}$, which is also depicted in Figure~\ref{fig: overall market} (b), will be in yellow, and the other correction term $C_G$ will be drawn in orange.
	
	Under the discrete-time market model, we note that the identities \eqref{eq : log V diversity}, \eqref{eq : log V equal}, and \eqref{eq : log V entropy} are exact, except at the moment when a stock is delisted from each of the two exchanges. In a delisting event, we use its delisting return which is available through our dataset from the CRSP database. However, our dataset contains a lot of missing delisting return variables. For those missing values, we shall consider two cases: (i) the most conservative case by setting all missing delisting returns equal to $-1$~(which means we lose all money which was allocated to the delisting stock), (ii) the most desirable case by setting the values equal to $0$~(which means we can liquidate the whole value of a stock before delisting from the exchange). In our dataset, there are some positive delisting returns, due to some reasons, but these are rare cases. When the $i$-th stock is delisted from the exchange with delisting return $\text{DLRET}_i$ at day $t_j$, we will accumulate the quantity $\log (1 + \pi_i(t_j) \times \text{DLRET}_i)$, which represents the change in the log return of $\pi$ due to the delisting, and we call this new term DLRET. For case (i), the DLRET graph will be drawn as a red solid line; for case (ii), it will be represented as a red dotted line. We recall here Remark~\ref{rem : delisting event} how our general continuous-time model in Section~\ref{sec : FGP} handles such delisting event.
	
	Finally, the two log relative wealth of the portfolios, with respect to the total market capitalization~($\log V^{\psi}$), and to the self-financing market portfolio~($\log U^{\psi}$) will be illustrated in black and purple, respectively. Therefore, from the identities \eqref{eq : V psi decomposition} and \eqref{eq : U psi decomposition}, we have the relationships in the following figures `black = green + blue + yellow + orange + red', and `purple = green + blue + orange + red'. Since there are solid and dotted red graphs, black and purple graphs also have two corresponding line types. We normalized the trading strategies at time $0$ such that every term in the decomposition takes an initial value of zero.
	
\begin{figure}[!htb]
		\centering	
		\subfloat[Diversity-weighted, $p = 0.75$]
		{\includegraphics[width=.49\linewidth]{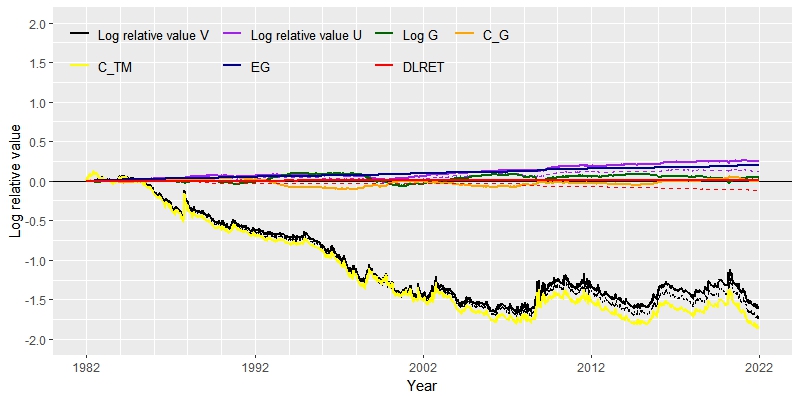}}
		\subfloat[Diversity-weighted, $p = 0.25$]
		{\includegraphics[width=.49\linewidth]{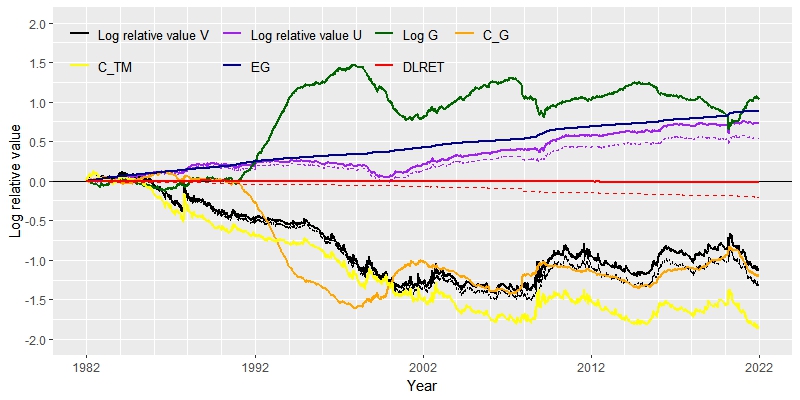}}
		\\
		\subfloat[Equally-weighted]
		{\includegraphics[width=.49\linewidth]{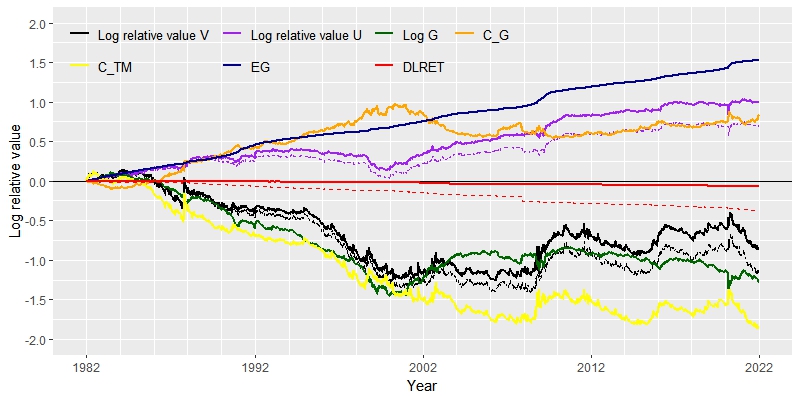}}
		\subfloat[Entropy-weighted]
		{\includegraphics[width=.49\linewidth]{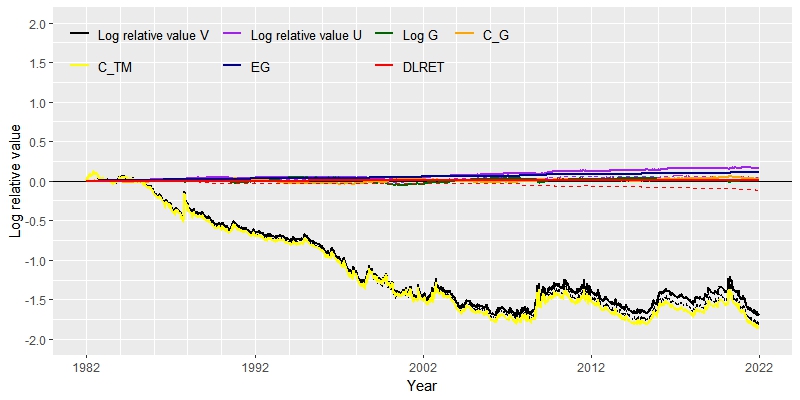}}
		\caption{Decomposition of the log-relative wealth of the portfolios on NYSE}
		\label{fig: portfolios NYSE}
	\end{figure}

\begin{figure}[!htb]
		\centering	
		\subfloat[Diversity-weighted, $p = 0.75$]
		{\includegraphics[width=.49\linewidth]{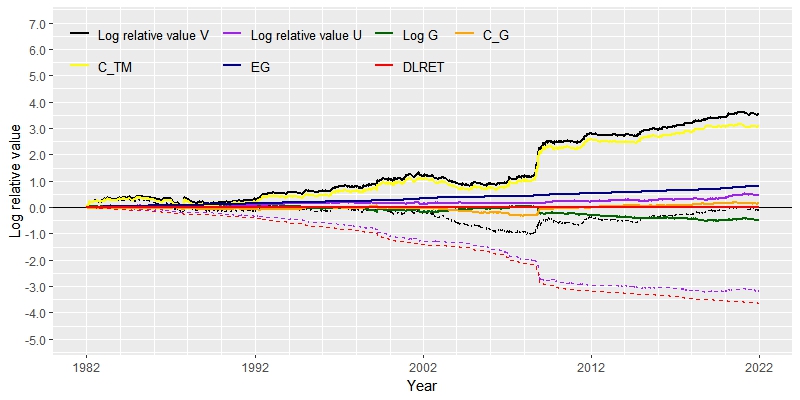}}
		\subfloat[Diversity-weighted, $p = 0.25$]
		{\includegraphics[width=.49\linewidth]{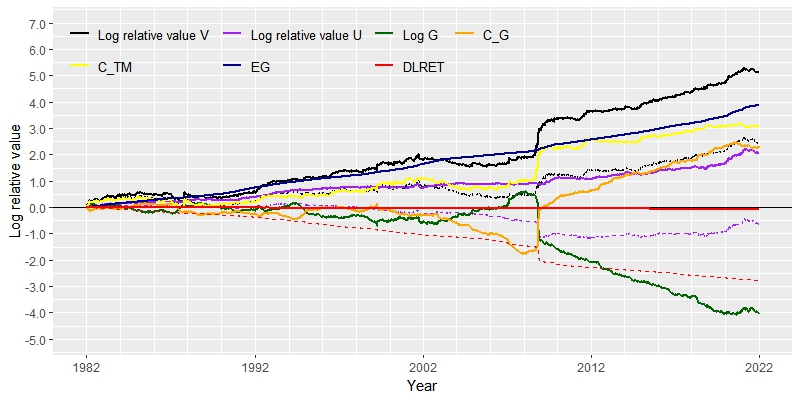}}
		\\
		\subfloat[Equally-weighted]
		{\includegraphics[width=.49\linewidth]{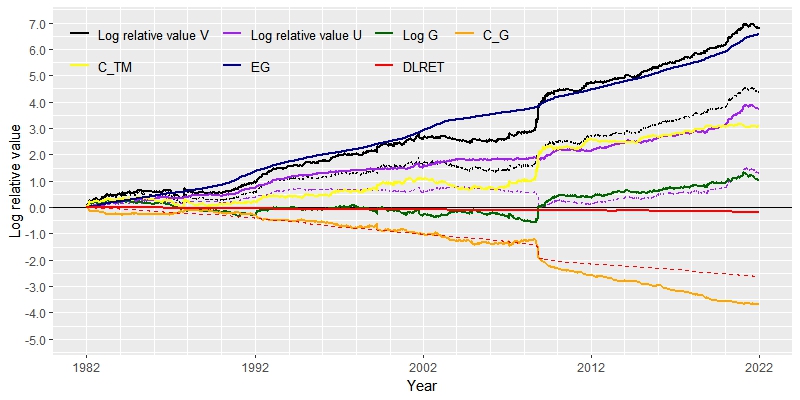}}
		\subfloat[Entropy-weighted]
		{\includegraphics[width=.49\linewidth]{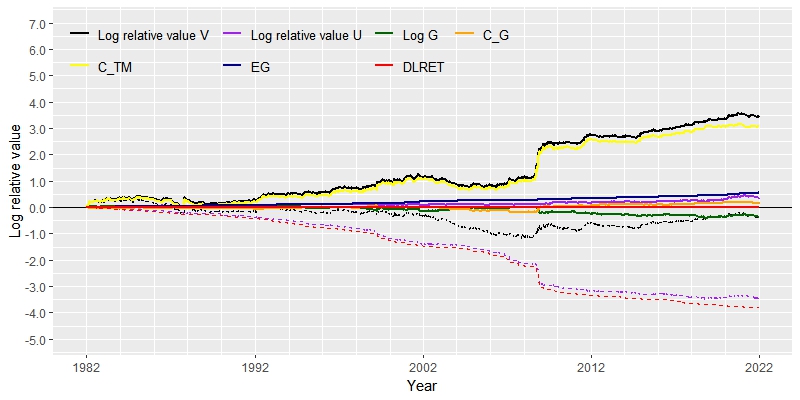}}
		\caption{Decomposition of the log-relative wealth of the portfolios on AMEX}
		\label{fig: portfolios AMEX}
	\end{figure}

	Figures~\ref{fig: portfolios NYSE} and \ref{fig: portfolios AMEX} show the aforementioned decompositions of the three portfolios, invested in the NYSE and AMEX, respectively.
	
	We now illustrate several important observations from Figures~\ref{fig: portfolios NYSE} and \ref{fig: portfolios AMEX}.
	\begin{enumerate} [(i)]
		\item The universal correction term $C_{TM}$, an inevitable term due to the jump in the total market capitalization, is quite huge; it is in fact a major factor of the relative wealth $V^{\psi}$ of the portfolios with respect to the total market capitalization, especially for the diversity-weighted portfolio with $p = 0.75$ and the entropy-weighted portfolio. Moreover, as mentioned in Section~\ref{subsec : data}, it influences the portfolio return on the two exchanges in the opposite way; for the portfolios on NYSE, it drags down the relative wealth $\log V^{\psi}$, whereas it contributes to the growth of $\log V^{\psi}$ for those on AMEX. This gives a reason for investors to use the self-financing market portfolio as the baseline when comparing the relative outperformance of their portfolios.
		
		\item For the diversity-weighted portfolios, smaller $p$ values~(or more concentration on the smaller stocks) result in higher excess growth ($EG$ term), which is well-known in the Stochastic portfolio theory (see, e.g. Example 3.4.4 of \cite{Fe}). Especially, in graphs (b) and (c), the excess growth significantly contributes to the growth in the log relative wealth $U^{\psi}$. However, such portfolios which overweight the smaller stocks, are subject to high turnover, thus should be carefully implemented in the presence of transaction costs.
		
		\item Furthermore, in those portfolios of (b) and (c), both the green graph ($\log G$ term) and the orange graph ($C_G$ term) fluctuate tremendously, but the directions of change are almost the opposite, like reflected mirror images of each other. This is expected from the fact that the dimensional change is quite frequent~(see Figure~\ref{fig: overall market} (a)), and from the relationship \eqref{eq : equal increments} of their increments at those moments of dimensional changes. Therefore, only the fluctuation of the generating function and the excess growth term \textit{between} two consecutive dimensional changes affect the evolution of relative wealth $U^{\psi}$ with respect to the self-financing market portfolio.
		
		\item As expected, the red graph (DLRET) negatively affects the portfolio performance, and its impact is larger for those on AMEX than on NYSE. Since there are more stocks on NYSE, a loss from a delisting of stock on NYSE (in the worst case of delisting return equal to $-1$) should have a smaller influence, on average, on the portfolio performance than that on AMEX. The actual evolution of the DLRET term should be somewhere in the middle of the red graphs, so these graphs provide at least bounds of the log relative wealth processes~(black and purple graphs).
	\end{enumerate}
	
	\bigskip
	
	\section{Conclusion} \label{sec : conclusion}
	
	By incorporating dimensional changes of an equity market in the functional generation of portfolios, this paper removes the assumption on the immutable size of the investable universe. We conclude by summarizing some directions for further extending the functional generation theory of portfolios from \cite{Karatzas:Kim}.
	\begin{enumerate} [(i)]
		\item When constructing portfolios, some observable, but non-tradable quantities $\lambda$ other than the market weights $\mu$ can be also used. Here, $\lambda$ is a $\U$-valued stochastic process of finite dimension with its own dimension process $\dim(\lambda)$~(which may coincide with $\dim(\mu)$ or not). The representative generating function $G^{k, n} : \U \times \U \rightarrow \mathbb{R}$ then takes two $\U$-valued processes $\mu$ and $\lambda$ as inputs.
		
		Examples of such additional process $\lambda$ include moving average, running maximum or minimum of the individual market weight $\mu^{(k, n)}_i(t)$ from its `birth' at time $\tau_{k-1}$, realized quadratic covariation $[\mu^{k, n}_i(t), \mu^{k, n}_j(t)]$, and stock fundamentals such as the book value (which we assume to have finite variation). We refer the reader to \cite{Kim:market-to-book, Mijatovic, Ruf:Xie, Schied:2016} in this direction of studies.
		
		\item Throughout the paper, we assume the portfolio-generating piecewise function $G$ to be in $C^2(\U)$~(Definition~\ref{def : piecewise function}), for the purpose of applying It\^o's formula. However, we can extend the class of portfolio-generating functions to less smooth functions when a generalized version of It\^o's formula~(or Tanaka's formula) is used.
		
		\item Even though the theory was developed in a probability space, we can remove any probabilistic assumptions and construct portfolios in a pathwise, probability-free setting. The capitalization process $S$ and the market weight vector $\mu$ can be modeled as $\U$-valued piecewise \textit{functions} instead of \textit{semimartingales}, in which case the reset sequence $(\tau_k)_{k=0}^{\infty}$ is just a sequence taking values in $[0, \infty]$, and we apply the pathwise It\^o-Tanaka theory when generating portfolios in a similar manner described in this paper. We refer to \cite{Model-free:Roughpath} for a more recent study in this direction using the rough path theory.
	\end{enumerate}
	
	Along with the above lists of applicable extensions of the theory from \cite{Karatzas:Kim}, we list a few more directions for future study in the context of our paper.
	\begin{enumerate} [(i)]
		\item In the examples of Section~\ref{sec : empirical}, we have used the `same' generating function for every epoch of the market even when its dimension changes. For example, in Section~\ref{subsec : empirical results}, the same parameter $p$ is used at all times for each diversity-weighted portfolio. However, we can choose different generating functions (or different parameters within a family of generating functions) for each epoch. Especially, if an investor can predict a certain trend in market diversity or expect a huge economic event (such as $2008$ stock market crash) for an upcoming epoch of the market, she can choose the generating function accordingly for the next epoch in order to maximize her profit. How the adoption of different generating functions for each epoch influences the portfolio return would be a practically interesting topic.
		
		\item We may extend some theoretical results in SPT under the market model of changing dimension. For example, \cite{Cuchiero_Scha_Wong} connects SPT with Cover's universal portfolio theory \citep{Cover_1991}. It would be interesting to find out whether a similar result holds when the dimension of the market fluctuates over time.
	\end{enumerate}

	\bigskip
	
	\subsection*{Acknowledgments}
	The authors greatly appreciate Ioannis Karatzas for detailed reading and feedback which improved this paper.
	
	\bigskip
	
	\subsection*{Funding}
	E. Bayraktar is supported in part by the National Science Foundation under grant DMS-2106556 and by the Susan M. Smith Professorship.
	
	\bigskip
	
	\subsection*{Data availability statement}
	The data that support the findings of this study are available from CRSP. Restrictions apply to the availability of these data, which were used under license for this study. Data are available from https://www.crsp.org/ with the permission of CRSP.
	
	\bigskip
	
	\renewcommand{\bibname}{References}
	\bibliography{aa_bib}

\begin{thebibliography}{}

\bibitem [\protect \citeauthoryear {%
Allan%
, Cuchiero%
, Liu%
\BCBL {}\ \BBA {} Prömel%
}{%
Allan%
\ \protect \BOthers {.}}{%
{\protect \APACyear {2023}}%
}]{%
Model-free:Roughpath}
\APACinsertmetastar {%
Model-free:Roughpath}%
\begin{APACrefauthors}%
Allan, A\BPBI L.%
, Cuchiero, C.%
, Liu, C.%
\BCBL {}\ \BBA {} Prömel, D\BPBI J.%
\end{APACrefauthors}%
\unskip\
\newblock
\APACrefYearMonthDay{2023}{}{}.
\newblock
{\BBOQ}\APACrefatitle {Model-free portfolio theory: A rough path approach}
  {Model-free portfolio theory: A rough path approach}.{\BBCQ}
\newblock
\APACjournalVolNumPages{Mathematical Finance}{To appear}{}{}.
\PrintBackRefs{\CurrentBib}

\bibitem [\protect \citeauthoryear {%
Bayraktar%
, Kim%
\BCBL {}\ \BBA {} Tilva%
}{%
Bayraktar%
\ \protect \BOthers {.}}{%
{\protect \APACyear {2022}}%
}]{%
BKT:arbitrage}
\APACinsertmetastar {%
BKT:arbitrage}%
\begin{APACrefauthors}%
Bayraktar, E.%
, Kim, D.%
\BCBL {}\ \BBA {} Tilva, A.%
\end{APACrefauthors}%
\unskip\
\newblock
\APACrefYearMonthDay{2022}{}{}.
\newblock
\APACrefbtitle {Arbitrage theory in a market of stochastic dimension.}
  {Arbitrage theory in a market of stochastic dimension.}
\newblock
\APACrefnote{Preprint, arXiv:2212.04623}
\PrintBackRefs{\CurrentBib}

\bibitem [\protect \citeauthoryear {%
Campbell%
\ \BBA {} Wong%
}{%
Campbell%
\ \BBA {} Wong%
}{%
{\protect \APACyear {2022}}%
}]{%
Campbell:Wong}
\APACinsertmetastar {%
Campbell:Wong}%
\begin{APACrefauthors}%
Campbell, S.%
\BCBT {}\ \BBA {} Wong, T\BHBI K\BPBI L.%
\end{APACrefauthors}%
\unskip\
\newblock
\APACrefYearMonthDay{2022}{}{}.
\newblock
{\BBOQ}\APACrefatitle {Functional Portfolio Optimization in Stochastic
  Portfolio Theory} {Functional portfolio optimization in stochastic portfolio
  theory}.{\BBCQ}
\newblock
\APACjournalVolNumPages{SIAM Journal on Financial Mathematics}{13}{2}{576-618}.
\PrintBackRefs{\CurrentBib}

\bibitem [\protect \citeauthoryear {%
Cover%
}{%
Cover%
}{%
{\protect \APACyear {1991}}%
}]{%
Cover_1991}
\APACinsertmetastar {%
Cover_1991}%
\begin{APACrefauthors}%
Cover, T.%
\end{APACrefauthors}%
\unskip\
\newblock
\APACrefYearMonthDay{1991}{}{}.
\newblock
{\BBOQ}\APACrefatitle {Universal portfolios} {Universal portfolios}.{\BBCQ}
\newblock
\APACjournalVolNumPages{Mathematical Finance}{1}{1}{1-29}.
\PrintBackRefs{\CurrentBib}

\bibitem [\protect \citeauthoryear {%
CRSP%
}{%
CRSP%
}{%
{\protect \APACyear {2023}}%
}]{%
CRSP}
\APACinsertmetastar {%
CRSP}%
\begin{APACrefauthors}%
CRSP.%
\end{APACrefauthors}%
\unskip\
\newblock
\APACrefYearMonthDay{2023}{}{}.
\newblock
\APACrefbtitle {{C}{R}{S}{P} {U}{S} stock database.} {{C}{R}{S}{P} {U}{S} stock
  database.}
\newblock
\APACrefnote{https://www.crsp.org, accessed via Wharton research data services}
\PrintBackRefs{\CurrentBib}

\bibitem [\protect \citeauthoryear {%
Cuchiero%
, Schachermayer%
\BCBL {}\ \BBA {} Wong%
}{%
Cuchiero%
\ \protect \BOthers {.}}{%
{\protect \APACyear {2019}}%
}]{%
Cuchiero_Scha_Wong}
\APACinsertmetastar {%
Cuchiero_Scha_Wong}%
\begin{APACrefauthors}%
Cuchiero, C.%
, Schachermayer, W.%
\BCBL {}\ \BBA {} Wong, T\BHBI K\BPBI L.%
\end{APACrefauthors}%
\unskip\
\newblock
\APACrefYearMonthDay{2019}{}{}.
\newblock
{\BBOQ}\APACrefatitle {Cover's universal portfolio, stochastic portfolio
  theory, and the num{\'e}raire portfolio} {Cover's universal portfolio,
  stochastic portfolio theory, and the num{\'e}raire portfolio}.{\BBCQ}
\newblock
\APACjournalVolNumPages{Mathematical Finance}{29}{3}{773-803}.
\PrintBackRefs{\CurrentBib}

\bibitem [\protect \citeauthoryear {%
E\BPBI R.~Fernholz%
}{%
E\BPBI R.~Fernholz%
}{%
{\protect \APACyear {2002}}%
}]{%
Fe}
\APACinsertmetastar {%
Fe}%
\begin{APACrefauthors}%
Fernholz, E\BPBI R.%
\end{APACrefauthors}%
\unskip\
\newblock
\APACrefYear{2002}.
\newblock
\APACrefbtitle {Stochastic Portfolio Theory} {Stochastic portfolio theory}\
  (\BVOL~48).
\newblock
\APACaddressPublisher{}{Springer-Verlag, New York}.
\newblock
\APACrefnote{Stochastic Modelling and Applied Probability}
\PrintBackRefs{\CurrentBib}

\bibitem [\protect \citeauthoryear {%
R.~Fernholz%
}{%
R.~Fernholz%
}{%
{\protect \APACyear {1999}}%
}]{%
F_generating}
\APACinsertmetastar {%
F_generating}%
\begin{APACrefauthors}%
Fernholz, R.%
\end{APACrefauthors}%
\unskip\
\newblock
\APACrefYearMonthDay{1999}{}{}.
\newblock
{\BBOQ}\APACrefatitle {Portfolio generating functions} {Portfolio generating
  functions}.{\BBCQ}
\newblock
\BIn{} M.~Avellaneda\ (\BED), \APACrefbtitle {Quantitative Analysis in
  Financial Markets.} {Quantitative analysis in financial markets.}
\newblock
\APACaddressPublisher{}{World Scientific}.
\PrintBackRefs{\CurrentBib}

\bibitem [\protect \citeauthoryear {%
R.~Fernholz%
}{%
R.~Fernholz%
}{%
{\protect \APACyear {2018}}%
}]{%
Fernholz:2018}
\APACinsertmetastar {%
Fernholz:2018}%
\begin{APACrefauthors}%
Fernholz, R.%
\end{APACrefauthors}%
\unskip\
\newblock
\APACrefYearMonthDay{2018}{}{}.
\newblock
\APACrefbtitle {Numeraire markets.} {Numeraire markets.}
\newblock
\APACrefnote{Preprint, arXiv:1801.07309}
\PrintBackRefs{\CurrentBib}

\bibitem [\protect \citeauthoryear {%
R.~Fernholz%
\ \BBA {} Karatzas%
}{%
R.~Fernholz%
\ \BBA {} Karatzas%
}{%
{\protect \APACyear {2009}}%
}]{%
FK_survey}
\APACinsertmetastar {%
FK_survey}%
\begin{APACrefauthors}%
Fernholz, R.%
\BCBT {}\ \BBA {} Karatzas, I.%
\end{APACrefauthors}%
\unskip\
\newblock
\APACrefYearMonthDay{2009}{}{}.
\newblock
{\BBOQ}\APACrefatitle {Stochastic {P}ortfolio {T}heory: an overview}
  {Stochastic {P}ortfolio {T}heory: an overview}.{\BBCQ}
\newblock
\BIn{} \APACrefbtitle {Handbook of Numerical Analysis} {Handbook of numerical
  analysis}\ (\BVOL\ Mathematical Modeling and Numerical Methods in Finance).
\newblock
\APACaddressPublisher{}{Elsevier}.
\PrintBackRefs{\CurrentBib}

\bibitem [\protect \citeauthoryear {%
Ghomrasni%
\ \BBA {} Pamen%
}{%
Ghomrasni%
\ \BBA {} Pamen%
}{%
{\protect \APACyear {2010}}%
}]{%
Ghomrasni:Pamen}
\APACinsertmetastar {%
Ghomrasni:Pamen}%
\begin{APACrefauthors}%
Ghomrasni, R.%
\BCBT {}\ \BBA {} Pamen, O\BPBI M.%
\end{APACrefauthors}%
\unskip\
\newblock
\APACrefYearMonthDay{2010}{}{}.
\newblock
{\BBOQ}\APACrefatitle {Decomposition of Order Statistics of Semimartingales
  Using Local Times} {Decomposition of order statistics of semimartingales
  using local times}.{\BBCQ}
\newblock
\APACjournalVolNumPages{Stochastic Analysis and Applications}{28}{3}{467-479}.
\PrintBackRefs{\CurrentBib}

\bibitem [\protect \citeauthoryear {%
Herdegen%
}{%
Herdegen%
}{%
{\protect \APACyear {2017}}%
}]{%
Herdegen:2015}
\APACinsertmetastar {%
Herdegen:2015}%
\begin{APACrefauthors}%
Herdegen, M.%
\end{APACrefauthors}%
\unskip\
\newblock
\APACrefYearMonthDay{2017}{}{}.
\newblock
{\BBOQ}\APACrefatitle {No-arbitrage in a num\'eraire-independent modelling
  framework} {No-arbitrage in a num\'eraire-independent modelling
  framework}.{\BBCQ}
\newblock
\APACjournalVolNumPages{Mathematical Finance}{27}{2}{568-603}.
\PrintBackRefs{\CurrentBib}

\bibitem [\protect \citeauthoryear {%
Itkin%
\ \BBA {} Larsson%
}{%
Itkin%
\ \BBA {} Larsson%
}{%
{\protect \APACyear {2021}}%
}]{%
Itkin:Larsson:Open}
\APACinsertmetastar {%
Itkin:Larsson:Open}%
\begin{APACrefauthors}%
Itkin, D.%
\BCBT {}\ \BBA {} Larsson, M.%
\end{APACrefauthors}%
\unskip\
\newblock
\APACrefYearMonthDay{2021}{}{}.
\newblock
\APACrefbtitle {Open Markets and Hybrid {J}acobi Processes.} {Open markets and
  hybrid {J}acobi processes.}
\newblock
\APACrefnote{Preprint, arXiv:2110.14046}
\PrintBackRefs{\CurrentBib}

\bibitem [\protect \citeauthoryear {%
Jacod%
\ \BBA {} Shiryaev%
}{%
Jacod%
\ \BBA {} Shiryaev%
}{%
{\protect \APACyear {2003}}%
}]{%
JacodS}
\APACinsertmetastar {%
JacodS}%
\begin{APACrefauthors}%
Jacod, J.%
\BCBT {}\ \BBA {} Shiryaev, A\BPBI N.%
\end{APACrefauthors}%
\unskip\
\newblock
\APACrefYear{2003}.
\newblock
\APACrefbtitle {Limit Theorems for Stochastic Processes} {Limit theorems for
  stochastic processes}\ (\PrintOrdinal{2nd}\ \BEd).
\newblock
\APACaddressPublisher{Berlin}{Springer}.
\PrintBackRefs{\CurrentBib}

\bibitem [\protect \citeauthoryear {%
Karatzas%
\ \BBA {} Kim%
}{%
Karatzas%
\ \BBA {} Kim%
}{%
{\protect \APACyear {2020}}%
}]{%
Karatzas:Kim}
\APACinsertmetastar {%
Karatzas:Kim}%
\begin{APACrefauthors}%
Karatzas, I.%
\BCBT {}\ \BBA {} Kim, D.%
\end{APACrefauthors}%
\unskip\
\newblock
\APACrefYearMonthDay{2020}{}{}.
\newblock
{\BBOQ}\APACrefatitle {Trading strategies generated pathwise by functions of
  market weights} {Trading strategies generated pathwise by functions of market
  weights}.{\BBCQ}
\newblock
\APACjournalVolNumPages{Finance and Stochastics}{24}{2}{423-463}.
\PrintBackRefs{\CurrentBib}

\bibitem [\protect \citeauthoryear {%
Karatzas%
\ \BBA {} Kim%
}{%
Karatzas%
\ \BBA {} Kim%
}{%
{\protect \APACyear {2021}}%
}]{%
Karatzas:Kim2}
\APACinsertmetastar {%
Karatzas:Kim2}%
\begin{APACrefauthors}%
Karatzas, I.%
\BCBT {}\ \BBA {} Kim, D.%
\end{APACrefauthors}%
\unskip\
\newblock
\APACrefYearMonthDay{2021}{}{}.
\newblock
{\BBOQ}\APACrefatitle {Open markets} {Open markets}.{\BBCQ}
\newblock
\APACjournalVolNumPages{Mathematical Finance}{31}{4}{1111-1161}.
\PrintBackRefs{\CurrentBib}

\bibitem [\protect \citeauthoryear {%
Karatzas%
\ \BBA {} Ruf%
}{%
Karatzas%
\ \BBA {} Ruf%
}{%
{\protect \APACyear {2017}}%
}]{%
Karatzas:Ruf:2017}
\APACinsertmetastar {%
Karatzas:Ruf:2017}%
\begin{APACrefauthors}%
Karatzas, I.%
\BCBT {}\ \BBA {} Ruf, J.%
\end{APACrefauthors}%
\unskip\
\newblock
\APACrefYearMonthDay{2017}{}{}.
\newblock
{\BBOQ}\APACrefatitle {Trading strategies generated by {L}yapunov functions}
  {Trading strategies generated by {L}yapunov functions}.{\BBCQ}
\newblock
\APACjournalVolNumPages{Finance and Stochastics}{21}{3}{753-787}.
\PrintBackRefs{\CurrentBib}

\bibitem [\protect \citeauthoryear {%
Karatzas%
\ \BBA {} Sarantsev%
}{%
Karatzas%
\ \BBA {} Sarantsev%
}{%
{\protect \APACyear {2016}}%
}]{%
Karatzas:Sarantsev}
\APACinsertmetastar {%
Karatzas:Sarantsev}%
\begin{APACrefauthors}%
Karatzas, I.%
\BCBT {}\ \BBA {} Sarantsev, A.%
\end{APACrefauthors}%
\unskip\
\newblock
\APACrefYearMonthDay{2016}{06}{}.
\newblock
{\BBOQ}\APACrefatitle {Diverse market models of competing Brownian particles
  with splits and mergers} {Diverse market models of competing brownian
  particles with splits and mergers}.{\BBCQ}
\newblock
\APACjournalVolNumPages{Ann. Appl. Probab.}{26}{3}{1329--1361}.
\PrintBackRefs{\CurrentBib}

\bibitem [\protect \citeauthoryear {%
Karatzas%
\ \BBA {} Shreve%
}{%
Karatzas%
\ \BBA {} Shreve%
}{%
{\protect \APACyear {1991}}%
}]{%
KS1}
\APACinsertmetastar {%
KS1}%
\begin{APACrefauthors}%
Karatzas, I.%
\BCBT {}\ \BBA {} Shreve, S\BPBI E.%
\end{APACrefauthors}%
\unskip\
\newblock
\APACrefYear{1991}.
\newblock
\APACrefbtitle {Brownian Motion and Stochastic Calculus} {Brownian motion and
  stochastic calculus}\ (\PrintOrdinal{Second}\ \BEd, \BVOL~113).
\newblock
\APACaddressPublisher{}{Springer-Verlag, New York}.
\PrintBackRefs{\CurrentBib}

\bibitem [\protect \citeauthoryear {%
Kim%
}{%
Kim%
}{%
{\protect \APACyear {2023}}%
}]{%
Kim:market-to-book}
\APACinsertmetastar {%
Kim:market-to-book}%
\begin{APACrefauthors}%
Kim, D.%
\end{APACrefauthors}%
\unskip\
\newblock
\APACrefYearMonthDay{2023}{}{}.
\newblock
{\BBOQ}\APACrefatitle {Market-to-book ratio in stochastic portfolio theory}
  {Market-to-book ratio in stochastic portfolio theory}.{\BBCQ}
\newblock
\APACjournalVolNumPages{Finance and Stochastics}{27}{2}{401-434}.
\PrintBackRefs{\CurrentBib}

\bibitem [\protect \citeauthoryear {%
Mijatovic%
}{%
Mijatovic%
}{%
{\protect \APACyear {2021}}%
}]{%
Mijatovic}
\APACinsertmetastar {%
Mijatovic}%
\begin{APACrefauthors}%
Mijatovic, P.%
\end{APACrefauthors}%
\unskip\
\newblock
\APACrefYearMonthDay{2021}{}{}.
\newblock
\APACrefbtitle {Beating the Market with Generalized Generating Portfolios.}
  {Beating the market with generalized generating portfolios.}
\newblock
\APACrefnote{Preprint, arXiv:2101.07084}
\PrintBackRefs{\CurrentBib}

\bibitem [\protect \citeauthoryear {%
Pal%
\ \BBA {} Wong%
}{%
Pal%
\ \BBA {} Wong%
}{%
{\protect \APACyear {2016}}%
}]{%
Pal:Wong:Geometry}
\APACinsertmetastar {%
Pal:Wong:Geometry}%
\begin{APACrefauthors}%
Pal, S.%
\BCBT {}\ \BBA {} Wong, T\BHBI K\BPBI L.%
\end{APACrefauthors}%
\unskip\
\newblock
\APACrefYearMonthDay{2016}{}{}.
\newblock
{\BBOQ}\APACrefatitle {The geometry of relative arbitrage} {The geometry of
  relative arbitrage}.{\BBCQ}
\newblock
\APACjournalVolNumPages{Mathematics and Financial Economics}{10}{}{263-293}.
\PrintBackRefs{\CurrentBib}

\bibitem [\protect \citeauthoryear {%
Protter%
}{%
Protter%
}{%
{\protect \APACyear {2003}}%
}]{%
Protter}
\APACinsertmetastar {%
Protter}%
\begin{APACrefauthors}%
Protter, P\BPBI E.%
\end{APACrefauthors}%
\unskip\
\newblock
\APACrefYear{2003}.
\newblock
\APACrefbtitle {Stochastic Integration and Differential Equations} {Stochastic
  integration and differential equations}\ (\PrintOrdinal{2nd}\ \BEd).
\newblock
\APACaddressPublisher{New York}{Springer}.
\PrintBackRefs{\CurrentBib}

\bibitem [\protect \citeauthoryear {%
Ruf%
}{%
Ruf%
}{%
{\protect \APACyear {2023}}%
}]{%
Ruf:Github}
\APACinsertmetastar {%
Ruf:Github}%
\begin{APACrefauthors}%
Ruf, J.%
\end{APACrefauthors}%
\unskip\
\newblock
\APACrefYearMonthDay{2023}{}{}.
\newblock
\APACrefbtitle {Empirical Finance with Equity Data ({P}h.{D}. Course),
  {L}{S}{E}.} {Empirical finance with equity data ({P}h.{D}. course),
  {L}{S}{E}.}
\newblock
\APACrefnote{Github https://github.com/johruf/CRSP\_on\_WRDS\_introduction}
\PrintBackRefs{\CurrentBib}

\bibitem [\protect \citeauthoryear {%
Ruf%
\ \BBA {} Xie%
}{%
Ruf%
\ \BBA {} Xie%
}{%
{\protect \APACyear {2019}}%
}]{%
Ruf:Xie}
\APACinsertmetastar {%
Ruf:Xie}%
\begin{APACrefauthors}%
Ruf, J.%
\BCBT {}\ \BBA {} Xie, K.%
\end{APACrefauthors}%
\unskip\
\newblock
\APACrefYearMonthDay{2019}{}{}.
\newblock
{\BBOQ}\APACrefatitle {Generalised {L}yapunov Functions and Functionally
  Generated Trading Strategies} {Generalised {L}yapunov functions and
  functionally generated trading strategies}.{\BBCQ}
\newblock
\APACjournalVolNumPages{Applied Mathematical Finance}{26}{4}{293-327}.
\PrintBackRefs{\CurrentBib}

\bibitem [\protect \citeauthoryear {%
Ruf%
\ \BBA {} Xie%
}{%
Ruf%
\ \BBA {} Xie%
}{%
{\protect \APACyear {2020}}%
}]{%
Ruf:Xie:transaction}
\APACinsertmetastar {%
Ruf:Xie:transaction}%
\begin{APACrefauthors}%
Ruf, J.%
\BCBT {}\ \BBA {} Xie, K.%
\end{APACrefauthors}%
\unskip\
\newblock
\APACrefYearMonthDay{2020}{}{}.
\newblock
{\BBOQ}\APACrefatitle {The Impact of Proportional Transaction Costs on
  Systematically Generated Portfolios} {The impact of proportional transaction
  costs on systematically generated portfolios}.{\BBCQ}
\newblock
\APACjournalVolNumPages{SIAM Journal on Financial Mathematics}{11}{3}{881-896}.
\PrintBackRefs{\CurrentBib}

\bibitem [\protect \citeauthoryear {%
Schied%
, Speiser%
\BCBL {}\ \BBA {} Voloshchenko%
}{%
Schied%
\ \protect \BOthers {.}}{%
{\protect \APACyear {2018}}%
}]{%
Schied:2016}
\APACinsertmetastar {%
Schied:2016}%
\begin{APACrefauthors}%
Schied, A.%
, Speiser, L.%
\BCBL {}\ \BBA {} Voloshchenko, I.%
\end{APACrefauthors}%
\unskip\
\newblock
\APACrefYearMonthDay{2018}{}{}.
\newblock
{\BBOQ}\APACrefatitle {Model-Free Portfolio Theory and Its Functional Master
  Formula} {Model-free portfolio theory and its functional master
  formula}.{\BBCQ}
\newblock
\APACjournalVolNumPages{SIAM Journal on Financial Mathematics}{9}{}{1074-1101}.
\PrintBackRefs{\CurrentBib}

\bibitem [\protect \citeauthoryear {%
Shiryaev%
\ \BBA {} Cherny%
}{%
Shiryaev%
\ \BBA {} Cherny%
}{%
{\protect \APACyear {2002}}%
}]{%
Shiryaev_vector}
\APACinsertmetastar {%
Shiryaev_vector}%
\begin{APACrefauthors}%
Shiryaev, A\BPBI N.%
\BCBT {}\ \BBA {} Cherny, A\BPBI S.%
\end{APACrefauthors}%
\unskip\
\newblock
\APACrefYearMonthDay{2002}{}{}.
\newblock
{\BBOQ}\APACrefatitle {Vector stochastic integrals and the {F}undamental
  {T}heorems of {A}sset {P}ricing} {Vector stochastic integrals and the
  {F}undamental {T}heorems of {A}sset {P}ricing}.{\BBCQ}
\newblock
\APACjournalVolNumPages{Proceedings of the Steklov Institute of
  Mathematics}{237}{}{6--49}.
\PrintBackRefs{\CurrentBib}

\bibitem [\protect \citeauthoryear {%
Strong%
}{%
Strong%
}{%
{\protect \APACyear {2014}}%
}]{%
Strong2}
\APACinsertmetastar {%
Strong2}%
\begin{APACrefauthors}%
Strong, W.%
\end{APACrefauthors}%
\unskip\
\newblock
\APACrefYearMonthDay{2014}{}{}.
\newblock
{\BBOQ}\APACrefatitle {Fundamental theorems of asset pricing for piecewise
  semimartingales of stochastic dimension} {Fundamental theorems of asset
  pricing for piecewise semimartingales of stochastic dimension}.{\BBCQ}
\newblock
\APACjournalVolNumPages{Finance and Stochastics}{18}{3}{487-514}.
\PrintBackRefs{\CurrentBib}

\bibitem [\protect \citeauthoryear {%
Wong%
}{%
Wong%
}{%
{\protect \APACyear {2015}}%
}]{%
Wong:Optimization}
\APACinsertmetastar {%
Wong:Optimization}%
\begin{APACrefauthors}%
Wong, T\BHBI K\BPBI L.%
\end{APACrefauthors}%
\unskip\
\newblock
\APACrefYearMonthDay{2015}{}{}.
\newblock
{\BBOQ}\APACrefatitle {Optimization of relative arbitrage} {Optimization of
  relative arbitrage}.{\BBCQ}
\newblock
\APACjournalVolNumPages{Annals of Finance}{11}{}{345-382}.
\PrintBackRefs{\CurrentBib}

\bibitem [\protect \citeauthoryear {%
Wong%
}{%
Wong%
}{%
{\protect \APACyear {2017}}%
}]{%
Wong:optimaltransport}
\APACinsertmetastar {%
Wong:optimaltransport}%
\begin{APACrefauthors}%
Wong, T\BHBI K\BPBI L.%
\end{APACrefauthors}%
\unskip\
\newblock
\APACrefYearMonthDay{2017}{}{}.
\newblock
\APACrefbtitle {On portfolios generated by optimal transport.} {On portfolios
  generated by optimal transport.}
\newblock
\APACrefnote{Preprint, arXiv:1709.03169}
\PrintBackRefs{\CurrentBib}

\end{thebibliography}

	\newpage
	
	\appendix
	
	\section{Rank based generation of portfolios}   \label{sec : rank FGP}
	
	Applying the similar method we developed in Section~\ref{sec : FGP}, we present in this Appendix generation of trading strategies depending on the ranks of companies, in terms of capitalization. We also study an open market embedded in the entire equity universe of stochastic dimension.
	
	\medskip
	
	\subsection{Rank based generation in the market of stochastic dimension}	\label{subsec : ranked FG}
	
	Let us define the $\ell$-th ranked components $v_{(\ell)}$ of an $n$-dimensional vector $v = (v_1, \cdots, v_n)$ for every $n \in \N$:
	\begin{equation*}
		v_{(\ell)} := \max_{1 \le i_1 \cdots \le i_{\ell} \le n} \min \{ v_{i_1}, \cdots, v_{i_{\ell}}\}, \qquad \ell \in [n] = \{1, \cdots, n\},
	\end{equation*}
	satisfying
	\begin{equation*}
		\max_{i=1, \cdots, n} v_i = v_{(1)} \ge v_{(2)} \ge \cdots \ge v_{(n)} = \min_{i=1, \cdots, n} v_i.
	\end{equation*}
	We shall use boldface symbols $\bm{v} := (v_{(1)}, v_{(2)}, \cdots, v_{(n)})$ to denote the vector arranged in a descending order.
	
	Moreover, for any given $\U$-valued process $X$, we denote 
	\begin{equation*}
		\bm{X}^{(k, n)}(\cdot) := \big(X^{(k, n)}_{(1)}(\cdot), \cdots, X^{(k, n)}_{(n)}(\cdot) \big)
	\end{equation*}
	the $n$-dimensional vector arranged in descending ranks of the dissection $X^{(k, n)}$ and construct the arranged $\U$-valued process $\bm{X}$ via dissection
	\begin{equation*}
		\bm{X}:= \bm{X}^{(0)} \hat{\mathbbm{1}}_{\{0\}\times \Omega}  + \sum_{k=1}^{\infty} \sum_{n=1}^{\infty} \bm{X}^{(k, n)} \hat{\mathbbm{1}}_{\rrbracket \tau_{k-1}, \tau_k \rrbracket \cap (\R_+ \times \Omega^{k, n})}.
	\end{equation*}
	In particular, if $X$ is a $\U$-valued piecewise semimartingale, then $\bm{X}$ is also a $\U$-valued piecewise semimartingale.
	
	We now consider the arranged process $\bm{\mu}$ of the market weight process $\mu$ of Definition~\ref{Def : market weights}. Recalling Remark~\ref{rem : two dissections}, we have two dissections $\bm{\mu}^{k, n}_{\ell} = \mu^{k, n}_{(\ell)}$ and $\bm{\mu}^{(k, n)}_{\ell} = \mu^{(k, n)}_{(\ell)}$ for $\ell \in [n]$ with the same increments, due to \eqref{eq : same increments}. However, the former dissection $\mu^{k, n}_{(\ell)}$ does not reflect the $\ell$-th ranked component of $\mu$ on $(k, n)$-dissection set, as it is reset to zero at $\tau_{k-1}$ by definition (see Figure~\ref{fig: example} (d)). Thus, throughout this Appendix, we shall use the latter dissection $\mu^{(k, n)}_{(\ell)}$ as an integrator, as every such integral will be considered solely on the $(k, n)$-dissection set.
	
	The generating function $G \in C^2(\U)$ now takes $\bm{\mu}$ as an input instead of $\mu$. On each $(k, n)$-dissection set, It\^o's formula gives
	\begin{align*}	
		&G^{k, n}(\bm{\mu}_t) = G^{k, n}(\bm{\mu}_{\tau_{k-1}+}) + I_{\bm{\mu}^{k, n}}\big(\nabla G^{k, n}(\bm{\mu}_-)\big)(t)
		\\
		& \qquad + \frac{1}{2} \int_{\tau_{k-1}+}^t \sum_{\ell=1}^n \sum_{\lambda=1}^n \partial^2_{\ell, \lambda} G^{k, n}(\bm{\mu}_{s-}) \, d[\mu^{(k, n), c}_{(\ell)}, \mu^{(k, n), c}_{(\lambda)}](s)
		+ \sum_{\tau_{k-1} < s \le t} d_{B, G^{k, n}}\big( \bm{\mu}_s, \, \bm{\mu}_{s-} \big),
	\end{align*}
	In the stochastic integral
	\begin{equation*}
		I_{\bm{\mu}^{k, n}}\big(\nabla G^{k, n}(\bm{\mu}_-)\big)(t)
		:= \int_{\tau_{k-1}+}^t \sum_{\ell=1}^n \partial_{\ell}G^{k, n}(\bm{\mu}_{s-}) \, d\mu^{(k, n)}_{(\ell)}(s),
	\end{equation*}
	we can replace the ranked integrators $d\mu^{(k, n)}_{(\ell)}$ with the original components $d\mu^{(k, n)}_i$ using Theorem~2.3 of \cite{Ghomrasni:Pamen}:
	\begin{align}
		& \quad G^{k, n}(\bm{\mu}_t) = G^{k, n}(\bm{\mu}_{\tau_{k-1}+}) 
		+ \int_{\tau_{k-1}+}^t \sum_{i=1}^n \sum_{\ell = 1}^n \partial_{\ell} G^{k, n}(\bm{\mu}_{s-}) \frac{\mathbbm{1}_{\{\mu^{(k, n)}_{(\ell)}(s-) = \mu^{(k, n)}_i(s-)\}}}{N^{k, n}_{\ell}(s-)} \, d\mu^{(k, n)}_i(s)	          \nonumber
		\\
		&+ \int_{\tau_{k-1}+}^t \sum_{\ell = 1}^n \sum_{j=\ell+1}^n \frac{\partial_{\ell} G^{k, n}(\bm{\mu}_{s-})}{N^{k, n}_{\ell}(s-)} \, d\mathcal{I}_s \big(\mu^{(k, n)}_{(\ell)} - \mu^{(k, n)}_{(j)}\big)	
		- \int_{\tau_{k-1}+}^t \sum_{\ell = 1}^n \sum_{j=1}^{\ell-1} \frac{\partial_{\ell} G^{k, n}(\bm{\mu}_{s-})}{N^{k, n}_{\ell}(s-)} \, d\mathcal{I}_s \big(\mu^{(k, n)}_{(j)} - \mu^{(k, n)}_{(\ell)}\big)			\nonumber
		\\
		&+ \frac{1}{2} \int_{\tau_{k-1}+}^t \sum_{\ell = 1}^n \sum_{\lambda = 1}^n \sum_{i=1}^n \sum_{j=1}^n \partial^2_{\ell, \lambda} G^{k, n}(\bm{\mu}_{s-}) \frac{\mathbbm{1}_{\{\mu^{(k, n)}_{(\ell)}(s-) = \mu^{(k, n)}_i(s-), \, \mu^{(k, n)}_{(\lambda)}(s-) = \mu^{(k, n)}_j(s-)\}}}{N^{k, n}_{\ell}(s-) \, N^{k, n}_{\lambda}(s-)} \, d[\mu^{(k, n), c}_i, \mu^{(k, n), c}_j](s)					\nonumber
		\\
		&+ \sum_{\tau_{k-1} < s \le t} d_{B, G^{k, n}}\big( \bm{\mu}_s, \, \bm{\mu}_{s-} \big).	        \label{eq : Ito ranked G}
	\end{align}
	Here, $N^{k, n}_{\ell}(t)$ is the number of components of $\bm{\mu}^{(k, n)}$ that are at rank $\ell$ at time $t$:
	\begin{equation*}
		N^{k, n}_{\ell}(t) := \big\vert \{ i: \mu^{(k, n)}_i(t) = \mu^{(k, n)}_{(\ell)}(t)\} \big\vert, 
	\end{equation*}
	and we denote $\mathcal{I}_t(Y) := L_t(Y) + \sum_{0 < s \le t} \mathbbm{1}_{\{Y(s-) = 0\}} \Delta Y(s)$, where $L_t(Y)$ is the local time accumulated at the origin by a semimartingale $Y$ up to time $t \ge 0$:
	\begin{equation}	\label{def : local time}
		2L_t(Y) := \vert Y(t) \vert - \vert Y(0) \vert - \int_0^t \text{sign}\big(Y(s-)\big) \, dY(s) - \sum_{0 < s \le t} \Big[ \vert Y(s) \vert - \vert Y(s-) \vert - \text{sign}\big(Y(s-)\big) \Delta Y(s) \Big],
	\end{equation}
	with the notation $\text{sign}(x) := 2 \times \mathbbm{1}_{(0, \infty)}(x) - 1$.
	
	On the right-hand side of \eqref{eq : Ito ranked G}, the expression
	\begin{equation}	\label{def : vartheta k, n ranked}
		\vartheta^{(k, n)}_i(s) := \sum_{\ell = 1}^n \partial_{\ell} G^{k, n}(\bm{\mu}_{s-}) \frac{\mathbbm{1}_{\{\mu^{(k, n)}_{(\ell)}(s-) = \mu^{(k, n)}_i(s-)\}}}{N^{k, n}_{\ell}(s-)}
	\end{equation}
	plays the role of integrand corresponding to the integrator $\mu^{k, n}_i$. We use this integrand $\vartheta^{(k, n)}$ to define the new $(k, n)$-Gamma process
	\begin{equation}	\label{def : Gamma k, n ranked}
		\Gamma^{G, k, n}(t) := 0 + \bigg[ G^{k, n}(\bm{\mu}_{\tau_{k-1}+}) - G^{k, n}(\bm{\mu}_t) + I_{\mu^{k, n}}(\vartheta^{(k, n)})(t) \bigg] \hat{\mathbbm{1}}_{\rrbracket \tau_{k-1}, \tau_k \rrbracket \cap (\R_+ \times \Omega^{k, n})}
	\end{equation}
	and apply \eqref{eq : Ito ranked G} to derive the alternative representation
	\begin{align}
		& \quad \Gamma^{G, k, n}(t) = 0 + \bigg[ -\int_{\tau_{k-1}+}^t \sum_{\ell = 1}^n \sum_{j=\ell+1}^n  \frac{\partial_{\ell} G^{k, n}(\bm{\mu}_{s-})}{N^{k, n}_{\ell}(s-)} \, d\mathcal{I}_s \big(\mu^{k, n}_{(\ell)} - \mu^{k, n}_{(j)}\big)	\nonumber
		\\
		& \qquad \qquad \qquad \qquad + \int_{\tau_{k-1}+}^t \sum_{\ell = 1}^n \sum_{j=1}^{\ell-1} \frac{\partial_{\ell} G^{k, n}(\bm{\mu}_{s-})}{N^{k, n}_{\ell}(s-)} \, d\mathcal{I}_s \big(\mu^{k, n}_{(j)} - \mu^{k, n}_{(\ell)}\big)			\nonumber
		\\
		& -\frac{1}{2} \int_{\tau_{k-1}+}^t \sum_{\ell, \lambda, i, j = 1}^n \partial^2_{\ell, \lambda} G^{k, n}(\bm{\mu}_{s-}) \frac{\mathbbm{1}_{\{\mu^{(k, n)}_{(\ell)}(s-) = \mu^{(k, n)}_i(s-), \, \mu^{(k, n)}_{(\lambda)}(s-) = \mu^{(k, n)}_j(s-)\}}}{N^{k, n}_{\ell}(s-) \, N^{k, n}_{\lambda}(s-)} \, d[\mu^{(k, n), c}_i, \mu^{(k, n), c}_j](s)		\nonumber
		\\
		& \qquad \qquad \qquad \qquad - \sum_{\tau_{k-1} < s \le t} d_{B, G^{k, n}}\big( \bm{\mu}_s, \, \bm{\mu}_{s-} \big) \Bigg] \hat{\mathbbm{1}}_{\rrbracket \tau_{k-1}, \tau_k \rrbracket \cap (\R_+ \times \Omega^{k, n})}.		\label{eq : Gamma k, n ranked}
	\end{align}
	
	In what follows, we shall recursively generate trading strategies as in Section~\ref{sec : FGP}, by making the vector $\vartheta^{(k, n)} := (\vartheta^{(k, n)}_1, \cdots, \vartheta^{(k, n)}_n)$ self-financing. By analogy with \eqref{def : vartheta 0} and \eqref{def : varphi 0}, we set two $N_0$-dimensional processes $\vartheta^{(0)}$, $\varphi^{(0)}$:
	\begin{alignat}{3}
		\vartheta^{(0)}_i(\cdot) &:= 0 + \sum_{\ell = 1}^{N_0} \partial_{\ell} G^0(\bm{\mu}_0) \frac{\mathbbm{1}_{\{\mu^{(k, n)}_{(\ell)}(0) = \mu^{(k, n)}_i(0)\}}}{N^{k, n}_{\ell}(0)} \hat{\mathbbm{1}}_{\{0\}\times \Omega}, \qquad && i = 1, \cdots, N_0,	\label{def : vartheta 0 ranked}
		\\
		\varphi^{(0)}_i(\cdot) &:= 0 + \big(\vartheta^{(0)}_i(\cdot) - C^{G, \vartheta, 0} \big) \hat{\mathbbm{1}}_{\{0\}\times \Omega}, \qquad && i = 1, \cdots, N_0,	\label{def : varphi ranked}
	\end{alignat}
	where $C^{G, \vartheta, 0} := \sum_{i=1}^{N_0} \vartheta^{(0)}_i(0) \mu_i(0) - G^0(\bm{\mu}_0)$. For $k = 1, 2, \cdots$, we define an $n$-dimensional process $\varphi^{(k, n)} = (\varphi^{(k, n)}_1, \cdots, \varphi^{(k, n)}_n)$ with components
	\begin{equation}	\label{def : varphi k, n ranked}
		\varphi^{(k, n)}_i(\cdot) := 0 + \Big( \vartheta^{(k, n)}_i(\cdot) - Q^{\vartheta, \mu, k, n}(\cdot) - C^{\widetilde{G}, \vartheta, k, n} \Big) \hat{\mathbbm{1}}_{\rrbracket \tau_{k-1}, \tau_k \rrbracket \cap (\R_+ \times \Omega^{k, n})},
	\end{equation}
	as in \eqref{def : varphi k, n} for each $n \in \N$. Here, $Q^{\vartheta, \mu, k, n}(\cdot)$, $\widetilde{G}$, and $C^{\widetilde{G}, \vartheta, k, n}$ are defined as in \eqref{def : Q dissection} - \eqref{def : C, G, mu, k, n}, with $\widetilde{G}^{k, n}(\mu_{\tau_{k-1}+})$ replaced by $\widetilde{G}^{k, n}(\bm{\mu}_{\tau_{k-1}+})$. Then, the $\U$-valued process $\varphi$, constructed from these dissections as in \eqref{def : varphi}, is a trading strategy satisfying the self-financing conditions.
	
	\begin{cor} [Additive generation]	\label{cor : AGTS ranked}
		The process $\varphi$ is a trading strategy and its relative wealth process $V^{\varphi}$ is given by
		\begin{equation}	\label{eq : varphi wealth ranked}	
			V^{\varphi}(t) 
			= \widetilde{G}(\bm{\mu}_t) + \sum_{k=1}^{\infty} \sum_{n=1}^{\infty} \Gamma^{G, k, n}(t) \hat{\mathbbm{1}}_{\rrbracket \tau_{k-1}, \tau_k \rrbracket \cap (\R_+ \times \Omega^{k, n})}
		\end{equation}
		for $t \ge 0$, where the $(k, n)$-Gamma process $\Gamma^{G, k, n}$ is given in \eqref{eq : Gamma k, n ranked}. Moreover, when the pair $(t, \omega)$ belongs to the $(k, n)$-dissection set for some $(k, n) \in \N^2$, the relative wealth process $V^{\varphi}(t)$ of \eqref{eq : varphi wealth ranked} has an alternative representation as in \eqref{eq : V varphi decomposition} - \eqref{eq : C add}, but $G(\mu)$ is replaced by $G(\bm{\mu})$ and the Gamma processes take the form of \eqref{eq : Gamma k, n ranked}.
	\end{cor}
	
	\begin{proof}
		We can prove that $\varphi$ is a trading strategy in the same manner as in the proof of Proposition~\ref{prop : AGTS}. We now derive as in \eqref{eq : additive ito} that
		\begin{equation*}
			V^{\varphi, k, n}(t) 
			= \widetilde{G}^{k, n}(\bm{\mu}_{\tau_{k-1}+}) + I_{\mu^{k, n}}(\vartheta^{(k, n)})(t)
		\end{equation*}
		holds for each $(k, n) \in \N^2$. Plugging \eqref{eq : Ito ranked G} into the last identity, along with \eqref{def : relative wealth w.r.t market}, yields the result \eqref{eq : varphi wealth ranked}. The last claim follows from the same argument in the proof of Theorem~\ref{thm : AGTS alternative}.
	\end{proof}
	
	\begin{rem}	\label{rem : AGTS alternative representation ranked}
		The component $\varphi^{(k, n)}_i(t)$ of $\varphi$ in Corollary~\ref{cor : AGTS ranked} admits the same representation as in \eqref{eq : varphi k, n alternative}:
		\begin{equation*}
			\varphi^{(k, n)}_i(t) 
			= \vartheta^{(k, n)}_i(t) - \sum_{j=1}^n \vartheta^{(k, n)}_j(t) \mu^{(k, n)}_j(t) + V^{\varphi, k, n}(t).
		\end{equation*}
	\end{rem}
	
	\smallskip
	
	For the multiplicative generation, we also impose the condition on $G \in C^2(\U)$ that the reciprocal $1/G^{k, n}(\mu_t)$ is locally bounded on every $(k, n)$-dissection set. Let us recall first $\varphi^{(0)}$ in \eqref{def : varphi ranked} and set 
	\begin{equation}	\label{def : psi 0 ranked}
		\eta^{(0)} \equiv \psi^{(0)} \equiv \varphi^{(0)}
	\end{equation}
	as before. We also consider the multiplicative measurable modification $\widehat{G}$ of $G$ as in \eqref{eq : G modification2} with $\mu$ replaced by $\bm{\mu}$ in the expression of $\delta^{G, \psi, k, n}$. 
	
	Furthermore, we define $E^{G, k, n}$ as in \eqref{def : E k, n} based on the new $(k, n)$-Gamma process in \eqref{def : Gamma k, n ranked}, where $\mu$ in the definition is replaced by $\bm{\mu}$. We next define two processes $\eta^{(k, n)}$ and $\psi^{(k, n)}$ in a recursive manner with $k = 1, 2, \cdots, $
	\begin{equation}	\label{def : eta k, n ranked}
		\eta^{(k, n)}(t) := 0^{(n)} + \delta^{G, \psi, k, n} E^{G, k, n}(t-) \vartheta^{(k, n)}(t) \hat{\mathbbm{1}}_{\rrbracket \tau_{k-1}, \tau_k \rrbracket \cap (\R_+ \times \Omega^{k, n})}, \qquad t \ge 0,
	\end{equation}
	similar to \eqref{def : eta k, n2}, and
	\begin{equation}	\label{def : psi k, n ranked}
		\psi^{(k, n)}_i(\cdot) := 0 + \Big( \eta^{(k, n)}_i(\cdot) - Q^{\eta, \mu, k, n}(\cdot) - C^{\widehat{G}, \eta, k, n} \Big) \hat{\mathbbm{1}}_{\rrbracket \tau_{k-1}, \tau_k \rrbracket \cap (\R_+ \times \Omega^{k, n})} , \quad i = 1, \cdots, n.
	\end{equation}
	as in \eqref{def : psi k, n} for each $n \in \N$. Then, the $\U$-valued process $\psi$ constructed as in \eqref{def : psi} is a rank based multiplicatively generated trading strategy.
	
	\begin{cor} [Multiplicative generation]	\label{cor : MGTS ranked}
		The process $\psi$ is a trading strategy and its relative wealth process $V^{\psi}$ is given by
		\begin{equation}	\label{eq : relative wealth psi ranked}
			V^{\psi}(t) 
			= \widehat{G}^0(\bm{\mu}_0) \hat{\mathbbm{1}}_{\{0\}\times \Omega}
			+ \sum_{k=1}^{\infty} \sum_{n=1}^{\infty} \widehat{G}^{k, n}(\bm{\mu}_t) E^{G, k, n}(t) \hat{\mathbbm{1}}_{\rrbracket \tau_{k-1}, \tau_k \rrbracket \cap (\R_+ \times \Omega^{k, n})},
		\end{equation}
		for $t \ge 0$ as in \eqref{eq : psi wealth}, but with the new Gamma processes of \eqref{eq : Gamma k, n ranked} in the expression of $E^{G, k, n}$. Furthermore, the log-relative wealth processes $\log V^{\psi}(t)$ and $\log U^{\psi}(t)$ can be represented as in \eqref{eq : V psi decomposition} - \eqref{eq : C mul}, and \eqref{eq : U psi decomposition}, respectively, when $(t, \omega)$ belongs to the $(k, n)$-dissection set, with $G(\bm{\mu})$ instead of $G(\mu)$ and the Gamma processes given in \eqref{eq : Gamma k, n ranked}.
	\end{cor}
	
	\begin{proof}
		Showing that $\psi$ is a trading strategy is again straightforward from the above construction. For the relative wealth process $V^{\psi}$, we derive just like the series of the identities of \eqref{eq : V psi diff}, but replacing $G(\mu)$ with $G(\bm{\mu})$ and \eqref{def : Gamma k, n} with \eqref{def : Gamma k, n ranked}, to prove $V^{\psi, k, n}(\cdot) \equiv \widehat{G}^{k, n}(\bm{\mu}_{\cdot})E^{G, k, n}(\cdot)$ on the $(k, n)$-dissection set. The last claim also follows from the proof of Theorem~\ref{thm : MGTS alternative}.
	\end{proof}
	
	\smallskip
	
	\begin{rem}		\label{rem : MGTS alternative representation ranked}
		Similar to \eqref{eq : psi k, n alternative}, the component $\psi^{(k, n)}_i(t)$ in Corollary~\ref{cor : MGTS ranked} also admits an alternative representation on the $(k, n)$-dissection set:
		\begin{equation*}
			\psi^{(k, n)}_i(t) = \eta^{(k, n)}_i(t) - \sum_{j=1}^n \eta^{(k, n)}_j(t) \mu^{(k, n)}_j(t) + V^{\psi, k, n}(t),
		\end{equation*}
		where the new Gamma process \eqref{eq : Gamma k, n ranked} is used in defining $E^{G, k, n}$. In particular, if every $G^{k, n}$ satisfies the balance condition, then we have a simpler representation
		\begin{equation} 	\label{eq : psi k, n balanced}
			\psi^{(k, n)}_i(t) = \eta^{(k, n)}_i(t)
		\end{equation}
		as in Corollary~\ref{cor : MGTS balanced}, since we have
		\begin{align*}
			\sum_{j=1}^n \eta^{(k, n)}_j(t) \mu^{(k, n)}_j(t) 
			&= \delta^{G, \psi, k, n} E^{G, k, n}(t-) \sum_{j=1}^n \vartheta^{(k, n)}_j(t) \mu^{(k, n)}_j(t) 
			\\
			&= \delta^{G, \psi, k, n} E^{G, k, n}(t-) \sum_{\ell=1}^n \partial_{\ell}G^{k, n}(\bm{\mu}_t) \mu^{(k, n)}_{(\ell)}(t) = \widehat{G}^{k, n}(\bm{\mu}_{t})E^{G, k, n}(t)	
		\end{align*}
		from \eqref{def : vartheta k, n ranked}.
	\end{rem}
	
	\medskip
	
	\subsection{Open market}	\label{subsec : open market FGP}
	
	The concept of an open market, consisting of a fixed number $m \in \mathbb{N}$, but fluctuating constituents of the largest capitalization stocks, has recently been studied in some literature \citep{Campbell:Wong, Fernholz:2018, Itkin:Larsson:Open, Karatzas:Kim2} to construct portfolios in a more realistic market model. In what follows, under some additional assumptions, we shall generate trading strategies in an open market embedded in an entire stock market with a stochastic dimension. 
	
	For a given $\U$-valued capitalization process $S$ of Definition~\ref{Def : price process RCLL} and corresponding market weight process $\mu$ of Definition~\ref{Def : market weights}, we shall fix a positive integer $m$ such that investors are only allowed to invest in the $m$ largest stocks in capitalization at all times. In other words, on each $(k, n)$-dissected market, we shall construct trading strategies that depend only on $\mu_{(1)}, \cdots, \mu_{(m)}$ and compare its performance relative to the total capitalization of the top $m$ index, that is, the sum of the $m$ largest companies' capitalization.
	
	We emphasize here that there is no restriction on the size $m$ of the open market. On the $(k, n)$-dissected market, in the case of $m < n$, one can only invest in the top $m$ stocks among $n$ extant stocks, whereas the top $m$ open market becomes the entire $(k, n)$-dissected market (thus investors can invest in any extant companies) in the other case of $m \ge n$. Thus, the results developed in the previous sections can be applied to the top $m$ open market in the latter case.
	
	In this subsection, we shall impose a special rule for breaking ties if two or more stocks have the same capitalization (e.g. a lexicographic rule which assigns a higher rank to a stock with a smaller index). Such a rule is necessary for an open market when choosing which stock to include in the top $m$ open market if there are multiple stocks having the same capitalization with rank $m$. For example, suppose that there are $N(t) = 3$ stocks extant in the market and we are considering the top $m=2$ open market. If the relationship $S_{(1)}(t) > S_{(2)}(t) = S_{(3)}(t)$ holds at the moment $t$, we need to choose one stock to include in the top $2$ open market among the two smallest stocks at time $t$. 
	
	Therefore, we define a process $[n] \times [0, \infty) \ni (i, t) \mapsto u^{k, n}_i(t) \in [n]$ for any $(k, n) \in \N^2$ such that each $u^{k, n}_i(\cdot)$ is predictable and satisfies
	\begin{align*}	
		S_i(t) &= S_{\big(u^{k, n}_i(t)\big)}(t), \qquad \text{on the dissection set } \rrbracket \tau_{k-1}, \tau_k \rrbracket \cap (\R_+ \times \Omega^{k, n}),
		\\
		u^{k, n}_i(t) &= i, \qquad \qquad \qquad ~~  \text{otherwise},
	\end{align*}
	for every $i \in [n]$. In other words, on the $(k, n)$-dissection set, $u^{k, n}_i(t)$ represents the rank of the $i$-th stock in terms of capitalization among $n$ stocks at time $t$. Since $u^{k, n}_i$ shall act only on the $(k, n)$-dissection set, the choice $u^{k, n}_i = i$ on the complement set, is not important. Because of the rule for breaking ties, each $u^{k, n}(\cdot)$ is a permutation process on the set $[n] = \{1, \cdots, n\}$. Moreover, since both index and rank of the stocks may shift due to the dimensional change~(at each reset sequence $\tau_k$), we assume that appropriate relabeling of index~(and corresponding rank) is done whenever necessary to inherit each company's dynamics after dimensional change.
	
	Since the discrimination rule breaks the symmetry among stocks having the same capitalization, the dynamics of the ranked semimartingales will be different from the one in Section~\ref{subsec : ranked FG}, and the equation \eqref{eq : Ito ranked G} is no longer relevant under the discrimination rule. To this end, we shall impose an additional assumption on the $\U$-valued market weight process $\mu$ to handle this rule for breaking ties.
	
	\begin{defn}
		A $\U$-valued piecewise semimartingale $X$ is called \textit{pathwise mutually nondegenerate}, if its $n$ components $X= (X_1, \cdots, X_n)$ on every $(k, n)$-dissection set satisfy the following conditions:
		\begin{enumerate} [(i)]
			\item each component $t \mapsto X_i(t)$ is continuous; 
			\item the set $\{ t : X_i(t) = X_j(t) \}$ has Lebesgue measure zero, $\mathbb{P}$-a.e., for all $i \neq j$;
			\item $L_t(X_{(k)}-X_{(\ell)}) \equiv 0$ holds $\mathbb{P}$-a.e., for all $|k-\ell| \geq 2$.
		\end{enumerate}
	\end{defn}
	
	Under the assumption that $\mu$ is pathwise mutually nondegenerate, every dissection $\mu^{(k, n)}$ is continuous on the $(k, n)$-dissection set. Moreover, Proposition~4.1.11 of \cite{Fe} proves that the ranked components $\mu^{k, n}_{(\ell)}(\cdot)$ are also continuous and have the dynamics on the $(k, n)$-dissection set:
	\begin{equation}	\label{eq : mu ranked dynamics simple}
		d\mu^{(k, n)}_{(\ell)}(t) = \sum_{i=1}^n \bm{1}_{\{u^{k, n}_i(t) = \ell\}} d\mu^{(k, n)}_i(t) + \frac{1}{2} dL_t(\mu^{(k, n)}_{(\ell)}-\mu^{(k, n)}_{(\ell+1)}) - \frac{1}{2} dL_t(\mu^{(k, n)}_{(\ell-1)}-\mu^{(k, n)}_{(\ell)}),
	\end{equation}
	for $\ell = 1, \cdots, n$, with conventions $L_t(\mu^{(k, n)}_{(0)}-\mu^{(k, n)}_{(1)}) \equiv L_t(\mu^{(k, n)}_{(N)}-\mu^{(k, n)}_{(N+1)}) \equiv 0$. Here, $L_t(Y)$ is the local time at the origin accumulated until time $t$ of the continuous semimartingale $Y$, without the last jump term in \eqref{def : local time}.
	
	For any $G \in C^2(\U)$ and arbitrary pair $(k, n) \in \N^2$, we apply It\^o's formula and the new dynamics \eqref{eq : mu ranked dynamics simple} to obtain
	\begin{align}
		& \quad G^{k, n}(\bm{\mu}_t) = G^{k, n}(\bm{\mu}_{\tau_{k-1}+}) 
		+ \int_{\tau_{k-1}+}^t \sum_{i=1}^n \sum_{\ell = 1}^n \partial_{\ell} G^{k, n}(\bm{\mu}_{s}) \mathbbm{1}_{\{u_i(s) = \ell\}} \, d\mu^{(k, n)}_i(s)	\label{eq : G Ito ranked simple}
		\\
		&+ \frac{1}{2} \int_{\tau_{k-1}+}^t \sum_{\ell = 1}^n \partial_{\ell} G^{k, n}(\bm{\mu}_{s}) \, dL_s \big(\mu^{(k, n)}_{(\ell)} - \mu^{(k, n)}_{(\ell+1)}\big)
		-\frac{1}{2} \int_{\tau_{k-1}+}^t \sum_{\ell = 1}^n \partial_{\ell} G^{k, n}(\bm{\mu}_{s}) \, dL_s \big(\mu^{(k, n)}_{(\ell-1)} - \mu^{(k, n)}_{(\ell)}\big)					\nonumber
		\\
		&+ \frac{1}{2} \int_{\tau_{k-1}+}^t \sum_{\ell = 1}^n \sum_{\lambda = 1}^n \sum_{i=1}^n \sum_{j=1}^n \partial^2_{\ell, \lambda} G^{k, n}(\bm{\mu}_{s}) \mathbbm{1}_{\{u^{k, n}_i(s) = \ell, \, u^{k, n}_j(s) = \lambda\}} \, d[\mu^{(k, n)}_i, \mu^{(k, n)}_j](s)							\nonumber
	\end{align}
	on the $(k, n)$-dissection set. We now denote
	\begin{align}	
		\vartheta^{(0)}_i(\cdot) & := 0 + \sum_{\ell = 1}^{N_0} \partial_{\ell} G^{0}(\bm{\mu}_{0}) \mathbbm{1}_{\{ u^{0}_i(0) = \ell \}} \hat{\mathbbm{1}}_{\{0\}\times \Omega}, \qquad \qquad \qquad \qquad i = 1, \cdots, N_0,	\label{def : vartheta 0 ranked simple}
		\\
		\vartheta^{(k, n)}_i(\cdot) & := 0 + \sum_{\ell = 1}^n \partial_{\ell} G^{k, n}(\bm{\mu}_{\cdot}) \mathbbm{1}_{\{ u^{k, n}_i(\cdot) = \ell \}} \hat{\mathbbm{1}}_{\rrbracket \tau_{k-1}, \tau_k \rrbracket \cap (\R_+ \times \Omega^{k, n})}, \qquad i = 1, \cdots, n,	\label{def : vartheta k, n ranked simple}
	\end{align}
	and use these dissections of $\vartheta$ to define the $(k, n)$-Gamma process $\Gamma^{G, k, n}$ as in \eqref{def : Gamma k, n ranked}. Then, the identity \eqref{eq : G Ito ranked simple} yields that $\Gamma^{G, k, n}(\cdot)$ is of finite variation with the alternative representation
	\begin{align}
		\Gamma^{G, k, n}(t) = 0 + \bigg[ &-\frac{1}{2} \int_{\tau_{k-1}+}^t \sum_{\ell = 1}^n \sum_{\lambda = 1}^n \sum_{i=1}^n \sum_{j=1}^n \partial^2_{\ell, \lambda} G^{k, n}(\bm{\mu}_{s}) \mathbbm{1}_{\{u^{k, n}_i(s) = \ell, \, u^{k, n}_j(s) = \lambda\}} \, d[\mu^{(k, n)}_i, \mu^{(k, n)}_j](s)						\nonumber
		\\
		& -\frac{1}{2} \int_{\tau_{k-1}+}^t \sum_{\ell = 1}^n \partial_{\ell} G^{k, n}(\bm{\mu}_{s}) \, dL_s \big(\mu^{(k, n)}_{(\ell)} - \mu^{(k, n)}_{(\ell+1)}\big)					\nonumber
		\\
		& + \frac{1}{2} \int_{\tau_{k-1}+}^t \sum_{\ell = 1}^n \partial_{\ell} G^{k, n}(\bm{\mu}_{s}) \, dL_s \big(\mu^{(k, n)}_{(\ell-1)} - \mu^{(k, n)}_{(\ell)}\big) \bigg] \hat{\mathbbm{1}}_{\rrbracket \tau_{k-1}, \tau_k \rrbracket \cap (\R_+ \times \Omega^{k, n})}.		\label{eq : Gamma k, n ranked simple}
	\end{align}
	
	Therefore, by replacing the basis $\vartheta^{(0)}$, $\vartheta^{(k, n)}$ in \eqref{def : vartheta 0 ranked}, \eqref{def : vartheta k, n ranked} and the corresponding Gamma process $\Gamma^{G, k, n}$ in \eqref{def : Gamma k, n ranked} with the new expressions defined in \eqref{def : vartheta 0 ranked simple}, \eqref{def : vartheta k, n ranked simple}, and \eqref{eq : Gamma k, n ranked simple}, respectively, we can construct $\varphi$ and $\psi$ from the new basis as in Section~\ref{subsec : ranked FG}, and obtain the same results as in Corollaries~\ref{cor : AGTS ranked}, \ref{cor : MGTS ranked}. The following corollary summarizes this construction.
	
	\begin{cor}	\label{cor : TS ranked simple}
		Let us assume that the market weight process $\mu$ is pathwise mutually nondegenerate and a special rule is used for breaking ties when ranking the stocks by their capitalization such that there is a one-to-one mapping between the ranks and the indices of the stocks at all times. Then, the process $\varphi$ constructed from $G \in C^2(\U)$ such that its dissections are recursively defined by \eqref{def : vartheta 0 ranked simple}, \eqref{def : vartheta k, n ranked simple}, \eqref{def : varphi ranked}, and \eqref{def : varphi k, n ranked}, is a trading strategy~(additively generated) and its relative wealth process is given by \eqref{eq : varphi wealth ranked}, or \eqref{eq : V varphi decomposition} - \eqref{eq : C add}, where the $(k, n)$-Gamma process $\Gamma^{G, k, n}$ is replaced by the one in \eqref{eq : Gamma k, n ranked simple}.
		
		Moreover, the process $\psi$ constructed from $G \in C^2(\U)$ such that $1/G^{k, n}$ is locally bounded for every $(k, n) \in \N^2$, via the recipe \eqref{def : vartheta 0 ranked simple}, \eqref{def : vartheta k, n ranked simple}, \eqref{def : psi 0 ranked}, \eqref{def : eta k, n ranked}, and \eqref{def : psi k, n ranked}, is a trading strategy~(multiplicatively generated) and its relative wealth process is given by \eqref{eq : relative wealth psi ranked}, or \eqref{eq : V psi decomposition} - \eqref{eq : C mul} and \eqref{eq : U psi decomposition}, where $\Gamma^{G, k, n}$ in the expression of $E^{G, k, n}$ is replaced by the one in \eqref{eq : Gamma k, n ranked simple}.
	\end{cor}
	
	Recalling that $m$ is the size of the open market, we construct the top $m$ open market portfolio in the following example.
	
	\begin{example}	[Top $m$ open market portfolio]	\label{ex : top m open market portfolio}
		Let us consider a specific example of generating function $G_m \in C^2(\U)$ defined as
		\begin{equation*}
			G^{k, n}_m (x) := \sum_{\ell=1}^{\min(m, n)} x_{\ell}, \qquad \forall \, (k, n) \in \N^2.
		\end{equation*}
		In each $(k, n)$-dissected market with $n \le m$, this function (either additively or multiplicatively) generates the self-financing market portfolio in Section~\ref{subsubsec : Sf mu}. Thus, let us consider the other case of $n > m$. Since each $G^{k, n}_m$ satisfies the balance condition of Definition~\ref{Def : balance}, the multiplicatively generated trading strategy $\psi_m$ of Corollary~\ref{cor : MGTS ranked} is given by
		\begin{equation*}
			\psi^{(k, n)}_{m, i}(\cdot) = E^{G_m, k, n}(\cdot) \vartheta^{(k, n)}_i(\cdot),
		\end{equation*}
		where
		\begin{equation*}
			\vartheta^{(k, n)}_i(\cdot) = \mathbbm{1}_{\{u_i(\cdot)\le m\}}, \qquad i = 1, \cdots, n, \qquad \text{and}
		\end{equation*}
		\begin{equation*}
			E^{G_m, k, n}(\cdot) = \exp \bigg( -\frac{1}{2} \int_{\tau_{k-1}+}^{\cdot} \frac{1}{\sum_{\ell=1}^m \mu^{(k, n)}_{(\ell)}(s)} dL_s(\mu^{(k, n)}_{(m)} - \mu^{(k, n)}_{(m+1)})\bigg),
		\end{equation*}
		from \eqref{eq : psi k, n balanced}, \eqref{def : vartheta k, n ranked simple}, \eqref{eq : Gamma k, n ranked simple}, and \eqref{def : E k, n}, on the $(k, n)$-dissection set. Its corresponding portfolio $\pi_m$ is computed from \eqref{def : portfolio weights of TS}
		\begin{equation}   \label{eq : top m market portfolio}
			\pi^{(k, n)}_{m, i}(\cdot) = \frac{\psi^{(k, n)}_{m, i}(\cdot) \mu^{(k, n)}_i(\cdot)}{\sum_{j=1}^n \psi^{(k, n)}_{m, j}(\cdot) \mu^{(k, n)}_j(\cdot)} 
			= \frac{ \sum_{\ell=1}^m \mathbbm{1}_{\{u_i(\cdot) = \ell\}} \mu^{(k, n)}_i(\cdot)}{ \sum_{j=1}^n \sum_{\ell=1}^m \mathbbm{1}_{\{u_j(\cdot) = \ell\}} \mu^{(k, n)}_j(\cdot)}
			= \frac{ \mathbbm{1}_{\{u_i(\cdot) \le m\}}  S_i(\cdot)}{\sum_{\ell=1}^m S_{(\ell)}(\cdot)}.
		\end{equation}
		Since the last denominator represents the total capitalization of the top $m$ open market, the portfolio weight $\pi^{(k, n)}_{m, i}$ for the $i$-th stock is equal to the relative capitalization of the $i$-th company with respect to the top $m$ open market whenever the company belongs to the top $m$ index, and the weight $\pi^{(k, n)}_{m, i}$ is zero otherwise. In other words, this portfolio distributes the wealth according to the relative capitalization of the stocks belonging to the top $m$ open market. Thus, we call this portfolio $\pi_m$, having $\pi^{(k, n)}_m$ as its $(k, n)$-dissection, the \textit{top $m$ open market portfolio}. The corresponding trading strategy $\psi_m$ above is called \textit{top $m$ open market trading strategy} and we denote it by $\xi_m \equiv \psi_m$, following the notation of Definition~\ref{Def : self-financing market portfolio}. Its log relative wealth with respect to the entire market capitalization is computed from Corollary~\ref{cor : MGTS ranked} (a version of \eqref{eq : V psi decomposition} - \eqref{eq : C mul} and \eqref{eq : decomposition of C}):
		\begin{align}
			\log V^{\xi_m}(t) 
			&= \log\Big(\sum_{i=1}^{\min (m, n)} \mu^{(k, n)}_{(i)}(t)\Big) + EG_m(t) + C_{TM}(t) + C_{G_m}(t), \qquad \text{where}		\label{eq : relative wealth top m market portfolio}
			\\
			EG_m(t) &= -\frac{1}{2} \sum_{\ell = 1}^{k-1}  \int_{\tau_{\ell-1}+}^{\tau_{\ell}} \frac{\mathbbm{1}_{\{N_{\tau_{\ell}} > m\}}}{\sum_{i=1}^{m} \mu^{(\ell, N_{\tau_{\ell}})}_{(i)}(s)} \, dL_s \big(\mu^{(\ell, N_{\tau_{\ell}})}_{(m)} - \mu^{(\ell, N_{\tau_{\ell}})}_{(m+1)}\big)    \nonumber
			\\
			& \qquad \qquad \qquad \qquad \qquad -\frac{1}{2} \int_{\tau_{k-1}+}^t \frac{\mathbbm{1}_{\{n > m\}}}{\sum_{i=1}^{m} \mu^{(k, n)}_{(i)}(s)} \, dL_s \big(\mu^{(k, n)}_{(m)} - \mu^{(k, n)}_{(m+1)}\big),	\nonumber
			\\
			C_{G_m}(t) &= \sum_{\ell = 1}^k \log \bigg( \frac{\sum_{i=1}^{\min(m, N_{\tau_{\ell-1}})} \mu^{(\ell-1, N_{\tau_{\ell-1}})}_{(i)}(\tau_{\ell-1})}{\sum_{i=1}^{\min(m, N_{\tau_{\ell}})} \mu^{(\ell, N_{\tau_{\ell}})}_{(i)}(\tau_{\ell-1}+)} \bigg)     \label{eq : C_G_m}
		\end{align}
		Here, the correction term $C_{TM}$ is the same as in \eqref{eq : decomposition of C}. The excess growth $EG_m(\cdot)$ is in fact nonincreasing in time, capturing the `leakage' of the top $m$ open market at the boundary, that is, a negative effect whenever a stock with the $m$-th rank is replaced by another stock which had the rank $m+1$ and dismissed from the top $m$ open market.
	\end{example}
	
	We now define a subset of trading strategies~(and corresponding portfolios) that invest only in the stocks belonging to the top $m$ open market at all times.
	
	\begin{defn} [Trading strategy among the top $m$ open market]
		A trading strategy $\vartheta$ is called \textit{trading strategy among the top $m$ open market}, if each dissection $\vartheta^{(k, n)}$ satisfies
		\begin{equation}	\label{con : TS in open market}
			\vartheta^{(k, n)}_i(\cdot) \mathbbm{1}_{\{u^{k, n}_i(\cdot) > m\}} \equiv 0, \qquad i = 1, \cdots, n,
		\end{equation}
		on the $(k, n)$-dissection set for every $(k, n) \in \N^2$.
	\end{defn}
	
	The condition \eqref{con : TS in open market} prohibits investing in the $i$-th stock whenever its rank is bigger than $m$. In what follows, we shall construct trading strategies among the top $m$ open market and compare its performance with respect to the top $m$ open market portfolio of \eqref{eq : top m market portfolio}, adopting the argument of \cite{Karatzas:Kim2}.
	
	The additively generated trading strategy $\varphi$ in Corollary~\ref{cor : TS ranked simple} takes the same representation as in Remark~\ref{rem : AGTS alternative representation ranked}; $\varphi^{(k, n)}_i$ depends on the Gamma process $\Gamma^{G, k, n}$ and this Gamma process given by \eqref{eq : Gamma k, n ranked simple} is nonzero unless the generating function $G$ is a constant function, i.e., $\partial_{\ell}G^{k, n} \equiv 0$ for every $\ell \in [n]$. Such dependence of $\varphi^{(k, n)}_i$ on the quantity $\Gamma^{G, k, n}$ is inevitable, even if the generating function $G^{k, n}$ is balanced. Thus, $\varphi^{(k, n)}$ may invest in the $i$-th company when the company fails to belong to the top $m$ open market. This is a characteristic of an additive generation; $\varphi$ distributes the cumulative earnings $\Gamma^{G, k, n}$ uniformly across all the stocks in the market. Therefore, additively generated trading strategies are not relevant for investing in open markets.
	
	On the other hand, the multiplicatively generated trading strategy $\psi$ of Corollary~\ref{cor : TS ranked simple} takes the simpler form of \eqref{eq : psi k, n balanced}, if the generating function $G^{k, n}$ is balanced:
	\begin{equation*}
		\psi^{(k, n)}_i(t) = \eta^{(k, n)}_i(t) = \delta^{G, \psi, k, n} E^{G, k, n}(t) \vartheta^{(k, n)}_i(t) = \sum_{\ell=1}^n \partial_{\ell}G^{k, n}(\bm{\mu}_t) \mathbbm{1}_{\{u^{k, n}_i(t) = \ell\}} \delta^{G, \psi, k, n} E^{G, k, n}(t).
	\end{equation*}
	Therefore, if the generating function $G^{k, n}$ depends only on the first $m$ components of $\bm{\mu}$ such that $\partial_{\ell}G^{k, n}(\bm{\mu}_{\cdot}) \equiv 0$ holds for $\ell > m$, then $\psi^{(k, n)}$ satisfies the condition \eqref{con : TS in open market}: $\psi^{(k, n)}_i(t) \mathbbm{1}_{\{u^{k, n}_i(t) > m\}} \equiv 0$ for every $i \in [n]$. We formulate this construction in the following result.
	
	\begin{cor}
		Suppose that a function $G \in C^2(\U)$ satisfies the following conditions:
		\begin{enumerate} [(i)]
			\item each $1/G^{k, n}$ is locally bounded;
			\item each $G^{k, n}$ is balanced in the sense of Definition~\ref{Def : balance};
			\item for every pair $(k, n) \in \N^2$ satisfying $m < n$, each $G^{k, n}$ depends only on the first $m$ components of its input such that $\partial_{\ell}G^{k, n} \equiv 0$ holds for every $\ell > m$.
		\end{enumerate}
		Then, the multiplicatively generated trading strategy $\psi$ from $G$ in Corollary~\ref{cor : TS ranked simple} is a trading strategy among the top $m$ stocks. Moreover, its log wealth process with respect to the top $m$ open market trading strategy is computed as
		\begin{align}
			& \log U^{\psi}_{\xi_m}(t) := \log \bigg( \frac{V^{\psi, k, n}(t)}{V^{\xi_m, k, n}(t)} \bigg) = \log V^{\psi, k, n}(t) - \log V^{\xi_m, k, n}(t)								\nonumber	
			\\
			= &\log G^{k, n}(\bm{\mu}_t) - \log \Big( \sum_{\ell=1}^{\min(m, n)} \mu^{(k, n)}_{(\ell)}(t) \Big)
			+ \Big[ EG(t) - EG_m(t) \Big] + \Big[ C_G(t) - C_{G_m}(t) \Big],			\label{eq : MGTS in the top open market relative wealth}
		\end{align}
		for an arbitrary pair $(t, \omega)$ belonging to a $(k, n)$-dissection set for $(k, n) \in \N^2$. Here, $EG_m$ and $C_{G_m}$ refer to the excess growth and the correction term of $\xi_m$, respectively, given in \eqref{eq : C_G_m}; $EG$ is from \eqref{eq : GR mul} without jump terms and the Gamma processes replaced by the ones in \eqref{eq : Gamma k, n ranked simple}; $C_G$ is from \eqref{eq : decomposition of C} with $G(\mu)$ replaced by $G(\bm{\mu})$.
	\end{cor}
	
	\begin{proof}
		We easily obtain the result \eqref{eq : MGTS in the top open market relative wealth} by comparing the log relative wealth $\log V^{\psi}$ in Corollary~\ref{cor : TS ranked simple} with \eqref{eq : relative wealth top m market portfolio}.
	\end{proof}

\end{document}